%% file: main.tex
\documentclass{svjour3}

\smartqed

\input{defs}

\journalname{Empirical Software Engineering}

\begin{document}

\title{An Empirical Study of Challenges in Machine Learning Asset Management}
%\subtitle{}
%\titlerunning{Short form of title}

\author{Zhimin~Zhao \and Yihao~Chen \and Abdul~Ali~Bangash \and Bram~Adams \and Ahmed~E.~Hassan}
%\authorrunning{Short form of author list}

\institute{Z. Zhao \at
    School of Computing, Queen's University, Kingston, ON, Canada\\
    \email{z.zhao@queensu.ca}
    \and
    Y. Chen \at
    School of Computing, Queen's University, Kingston, ON, Canada\\
    \email{yihao.chen@queensu.ca}
    \and
    A.A. Bangash \at 
    School of Computing, Queen's University, Kingston, ON, Canada\\
    \email{abdulali.b@queensu.ca}
    \and
    B. Adams \at 
    School of Computing, Queen's University, Kingston, ON, Canada\\
    \email{bram.adams@queensu.ca}
    \and
    A.E. Hassan \at 
    School of Computing, Queen's University, Kingston, ON, Canada\\
    \email{hassan@queensu.ca}
}

\date{Received: date / Accepted: date}

\maketitle

\begin{abstract}
\textit{Context:} In machine learning (ML) applications, assets include not only the ML models themselves, but also the datasets, algorithms, and deployment tools that are essential in the development, training, and implementation of these models. Efficient management of ML assets is critical to ensure optimal resource utilization, consistent model performance, and a streamlined ML development lifecycle. This practice contributes to faster iterations, adaptability, reduced time from model development to deployment, and the delivery of reliable and timely outputs.
\textit{Objective:} Despite research on ML asset management, there is still a significant knowledge gap on operational challenges, such as model versioning, data traceability, and collaboration issues, faced by asset management tool users. These challenges are crucial because they could directly impact the efficiency, reproducibility, and overall success of machine learning projects. Our study aims to bridge this empirical gap by analyzing user experience, feedback, and needs from Q\&A posts, shedding light on the real-world challenges they face and the solutions they have found. 
\textit{Methodology:} We examine $15,065$ Q\&A posts from multiple developer discussion platforms, including Stack Overflow, tool-specific forums, and GitHub/GitLab. Using a mixed-method approach, we classify the posts into knowledge inquiries and problem inquiries. We then apply BERTopic to extract challenge topics and compare their prevalence. Finally, we use the open card sorting approach to summarize solutions from solved inquiries, then cluster them with BERTopic, and analyze the relationship between challenges and solutions. 
\textit{Results:} We identify $133$ distinct topics in ML asset management-related inquiries, grouped into $16$ macro-topics, with software environment and dependency, model deployment and service, and model creation and training emerging as the most discussed. 
Additionally, we identify $79$ distinct solution topics, classified under $18$ macro-topics, with software environment and dependency, feature and component development, and file and directory management as the most proposed. \textit{Conclusion:} This study highlights critical areas within ML asset management that need further exploration, particularly around prevalent macro-topics identified as pain points for ML practitioners, emphasizing the need for collaborative efforts between academia, industry, and the broader research community. %For SE researchers, it is imperative to prioritize creating adaptive, intuitive environment management systems that integrate with various ML tools and platforms throughout the ML development lifecycle. SE educators should focus on enriching the curriculum in areas such as software environment and dependency management, as well as model deployment and serving, while also promoting the use of tool-specific forums for real-time, hands-on problem solving in ML asset management. Application developers need to improve their expertise in crucial areas such as software environment and dependency management to ensure seamless ML application development. Lastly, tool developers should pay attention to enhancing features tied to environment management and model deployment, making sure that tools are user-friendly, interoperable across platforms, and paired with comprehensive documentation.
%Lastly, we find that problem inquiries appear more frequently than knowledge inquiries related to machine learning asset management. Macro-topics concerning visualization and user interface, data preparation and development, and model deployment and serving show the highest unsolved rates. Meanwhile, topics on experiment tracking, source code tracking, and logs and monitoring take the longest median time to address. 
\end{abstract}


\input{sections/1-introduction}
\input{sections/2-background}
\input{sections/3-methodology}
\input{sections/4-rq1-results}
\input{sections/5-rq2-results}
\input{sections/6-rq3-results}
\input{sections/7-implications}
\input{sections/8-threats}
\input{sections/9-conclusion}
\input{sections/10-conflict}
\input{sections/11-availability}

% \begin{acknowledgements}
% \end{acknowledgements}

\bibliographystyle{spmpsci}
\bibliography{references}

\appendix
\input{sections/12-appendix}
\input{sections/13-appendix}

\end{document}

%% file: defs.tex
\usepackage{afterpage}
\usepackage{array}
\usepackage{booktabs}
\usepackage{cite}
\usepackage{enumitem}
\usepackage{float}
\usepackage{listings}
\usepackage{longtable}
\usepackage{multirow}
\usepackage{graphicx}
\usepackage{tcolorbox}
\usepackage{xcolor}
\usepackage{xspace}
\usepackage{xurl}
\usepackage{hyperref}
\usepackage[normalem]{ulem}

\tcbuselibrary{skins}

\newcolumntype{C}[1]{>{\centering\arraybackslash}p{#1}}

\newcommand{\code}[1]{{\small\texttt{#1}}\xspace}

\newcommand{\rpv}[1]{\textcolor{brown}{remove passive voice}}

\DeclareUnicodeCharacter{202F}{\,}
\extrafloats{500}

\makeatletter
\newcommand{\ie}{\emph{i.e.}\@ifnextchar.{\!\@gobble}{}}
\newcommand{\eg}{\emph{e.g.}\@ifnextchar.{\!\@gobble}{}}
\newcommand{\etc}{etc\@ifnextchar.{}{.\@}}
\makeatother

\newtcolorbox{mybox}[2][]{
top=0.15in,left=4pt,right=4pt,bottom=4pt,
fonttitle=\bfseries,
colbacktitle=gray,
colback=gray!5,
colframe=gray!40!black,
enhanced,
attach boxed title to top left={xshift=0em,yshift=-\tcboxedtitleheight/2},
boxed title style={size=small},
drop shadow={black!50!white},
title=#2,#1}

%% file: sections/1-introduction.tex
\section{Introduction}
\label{sec:introduction}
\vspace{-0.5em}

% \ali{This para introduce traditional asset management challenges}
From personalizing content recommendations on entertainment platforms~\cite{lapan2018deep} to enabling predictive diagnostics in healthcare~\cite{miotto2018deep}, and optimizing traffic flow in urban planning~\cite{nagy2018survey}, machine learning (ML) has transformed industries and our daily experiences. Despite these advances, the practical implementation of an ML-driven solution in a real-world scenario is fraught with challenges arising from the complexities of dynamic environments, such as changing data distributions, evolving requirements, and the need to constantly update and optimize models to maintain performance and accuracy. 

To help address these challenges, an intricate, yet essential task is asset management in ML-driven solutions. Asset management encompasses organizing, tracking, evaluating, and monitoring ML assets such as code, models, data sets, and other artifacts, which are crucial to the effective implementation of ML-driven solutions~\cite{idowu2021asset}. Although traditional software management tools, such as Git, are designed to handle software assets such as source code and documentat  ion, these tools lack native capabilities to effectively track ML-specific assets, such as model architectures, hyperparameters, and data splits~\cite{sculley2015hidden}. Therefore, it is imperative to have modern tools specifically designed to manage ML assets throughout the development lifecycle of ML-driven solutions.

%\ali{this paragraph talks about ML asset management tools}
In recent years, there has been a significant increase in the development of ML asset management tools. This growth can be attributed to joint efforts from academia and industry. In academia, tools such as CANDLE~\cite{wozniak2018candle}, DLHub~\cite{chard2019dlhub}, MLflow~\cite{zaharia2018accelerating}, ModelDB~\cite{vartak2018modeldb}, Pdmdims~\cite{peili2018deep}, and Runway~\cite{tsay2018runway} have emerged as prototypes for the management of machine learning experiments. Meanwhile, the industry has introduced tools such as Aim\footnote{\url{https://aimstack.io}}, ClearML\footnote{\url{https://clear.ml}}, DVC\footnote{\url{https://dvc.org}}, Kedro\footnote{\url{https://kedro.org}}, and Polyaxon\footnote{\url{https://polyaxon.com}}, among others. These innovative tools address and mitigate many of the challenges previously associated with traditional software management tools~\cite{idowu2021asset}.

%\ali{this paragraph talks about relevant research and study motivation}
However, several factors compromise the efficiency of ML asset management tools, leading to a myriad of challenges. For example, Kumar~et~al.~\cite{kumar2017data} offer an in-depth analysis of the challenges associated with data management in ML workloads. They further explore the primary techniques and systems formulated to address these challenges. Schlegel~et~al.~\cite{schlegel2023management} emphasize the complexities of ensuring the comparability, traceability, and reproducibility of ML artifacts. Idowu~et~al.~\cite{idowu2021asset} further stress that these challenges stem from the absence of tool interoperability and standardized management practices in asset management. While there is some preliminary research in this domain, the literature lacks comprehensive analysis across tools, examination of user challenges, and evaluation of the solutions to these challenges. This underscores the critical need for more extensive empirical studies to gain a holistic understanding from the tool users' perspective, with a focus on cross-tool analysis and solution evaluation.

%\yihao{This para talks about study methodology}
In this paper, we conduct a comprehensive study of developer discussion forums to identify the main challenges in ML asset management faced by ML practitioners and the strategies they adopt to address these challenges. We start by searching through multiple channels for tools that manage ML assets, then collect Q\&A posts about these tools from different developer discussion forums. This includes popular sites, such as Stack Overflow and GitHub, as well as forums set up by tool developers themselves (\ie, tool-specific forums). Based on the purpose of the inquiries, we sort these Q\&A posts into knowledge or problem types. Next, we use a topic modeling technique called BERTopic\footnote{\url{https://github.com/MaartenGr/BERTopic}} to extract the common topics in these inquiries. Then we manually group them into macro-topics, which are broad categories encompassing multiple related topics. We study the prevalence of these macro-topics to gain insights into the common challenges faced by ML practitioners. For inquiries marked as solved, we summarize the solutions provided in the content of the posts. We also use BERTopic to cluster these summaries to understand the patterns in the solutions. This analysis helps us to map the strategies that ML practitioners use to tackle the challenges in ML asset management. To our knowledge, this is the first study to use mixed methods to systematically explore user challenges in adopting ML asset management tools.

%\yihao{This para talks about findings and implications}
Our main findings and their implications include the following. \begin{itemize}
\item We identify $133$ distinctive challenge topics in ML asset management, which we subsequently group into $16$ macro-topics, \ie, high-level groupings of multiple interconnected topics with similar concerns. The most frequently discussed macro-topics are software environment and dependency ($18.89\%$), model deployment and serving ($10.59\%$), and model creation and training ($9\%$). The high prevalence indicates that these areas are of significant interest and concern for ML practitioners. Also, $55\%$ of the tools predominantly receive inquiries about the software environment and dependency topics. Furthermore, we observe different prevalent topics in each tool of the asset management ecosystem. %Domino\footnote{\url{https://domino.ai}} records a significant number of user inquiries ($38.46\%$) related to the topic of feature and component development. Furthermore, Optuna\footnote{\url{https://optuna.org}} ($32.73\%$) and Weights\&Biases ($20.28\%$) have the majority of inquiries about model creation and training. These patterns underscore that various tools in the ML asset management ecosystem address different challenges and requirements of ML practitioners. Lastly, there are more problem inquiries than knowledge ones in the Q\&A posts. This highlights the fact that ML practitioners encounter specific challenges or issues in ML asset management rather than just seeking general knowledge or information. 
\item We identify $79$ different solution topics in ML asset management, which we subsequently group into $18$ macro-topics. Among these, the most commonly proposed solutions are related to the software environment and dependencies ($23.31\%$), the development of features and components ($15.35\%$), and the file and directory ($9.64\%$). For knowledge inquiries, feature and component development methods are mainly used to solve challenges in source code management. Similarly, issues in experiment management are primarily solved using software environment and dependency-related solutions. On the other hand, for problem inquiries, a significant portion ($62.5\%$) of problems are resolved using software environment and dependency-related solutions. Additionally, feature and component development solutions are often applied to address challenges related to logging and monitoring. Moreover, problem inquiries tend to be solved more effectively using skills or knowledge from different domains, as opposed to knowledge inquiries. This observation highlights the necessity to employ varied approaches when addressing knowledge inquiries compared to problem ones. %We also observe that the ranking of solution macro-topics somewhat aligns with the challenge macro-topics that we have identified in RQ1. This is because challenges in one area often have solutions in the same area, which we call ``self-resolution'' in our study. This might be because ML tasks are specialized, so experts in one area are best suited to solve issues in that area. To understand the extent of self-resolution, we used a threshold of $25\%$ of in-between posts to indicate a substantial presence of self-resolution in a specific macro-topic. Our findings show that knowledge inquiries have a higher rate of self-resolution compared to problem inquiries. Specifically, $56.25\%$ of the macro-topics in knowledge inquiries experience self-resolution, compared to only $25\%$ in problem inquiries. On the other hand, we notice a pattern in which specific types of challenge are often addressed by solutions from different domains, a phenomenon we term ``counter-self-resolution''.
\item We find that Stack Overflow is the primary forum for practitioners seeking help with asset management, accounting for $48.82\%$ of all such inquiries. Tool-specific forums are next, contributing $34.19\%$, while repository-specific forums have the fewest, at only $17.16\%$. Of the tools examined, $25\%$ witness most user inquiries on tool-specific forums, $20\%$ on repository-specific forums, and $35\%$ on Stack Overflow. This trend implies a potential area of exploration for tool maintainers to enhance user engagement on their respective forums. Additionally, inquiries related to software environment and dependency are the most prevalent in both Stack Overflow, GitHub, and tool-specific forums. This suggests that regardless of where practitioners seek help, managing software environments and dependencies remains a top challenge in ML asset management. Furthermore, the frequency distribution of Q\&A posts varies significantly between different discussion forums. Developers can leverage these insights to customize content and discussions to align with the distinct characteristics and preferences of the audience in each forum.
%We also notice some interesting patterns when comparing topic difficulty on different forums. Tool-specific forums have the highest rate of unsolved questions at $78.32\%$. However, they also have the shortest median solution time, taking just $13.06$ hours. On the other hand, GitHub, even with its lower unsolved rate of $30.35\%$, takes much longer to resolve issues, with a median time of $217.51$ hours. Interestingly, Chi-square tests reveal no significant differences in unsolved rates across macro-topics between any pair of forums at the significance level of $0.05$. This implies that the difference in unsolved rates between forums is not because one forum is better at addressing a particular kind of topic. Instead, the pattern is consistent across all macro-topics. Furthermore, Mann-Whitney U tests show that user inquiries on tool-specific forums generally take significantly shorter time to resolve compared to GitHub and Stack Overflow at the significance level of $0.05$. Therefore, if users need quick responses, they should go to tool-specific forums. But for a wider range of solutions, GitHub and Stack Overflow might be better options. 
\end{itemize}

%\yihao{This para talks about paper structure}
%We organize the paper as follows. First, we highlight the related literature and provide essential background information in Section~\ref{sec:background}. Next, we detail our research methodology in Section~\ref{sec:methodology}. From Section~\ref{sec:rq1-results} to Section~\ref{sec:rq3-results}, we provide a comprehensive analysis of research questions RQ1 to RQ3. Then, we describe the key takeaways of our study for tool/application developers, researchers, and educators in the software engineering domain, respectively, in Section~\ref{sec:implications}. Following this, we discuss the potential threats to the validity of our research and mitigation strategies in Section~\ref{sec:threats}. Finally, we wrap up with conclusions and potential avenues for future research in Section~\ref{sec:conclusion}. For further reference, supplementary materials are made available in Section~\ref{sec:appendix}. Furthermore, we provide replication packages~\cite{replication_package} to ensure the reproducibility of our study. These packages include essential artifacts such as data collection, filtering, and processing scripts, topic information, and datasets.

%% file: sections/2-background.tex
\section{Background and Related Work}
\vspace{-0.5em}
\label{sec:background}

In this section, we provide a comprehensive overview of the role and challenges of asset management in the ML development lifecycle. %We first introduce the various stages of the lifecycle, emphasizing the iterative and dynamic nature of progressing from problem understanding to model monitoring. Therein, we highlight the emergence of ML assets and underscore their significance at each stage, leading to a detailed examination of machine learning asset management. We then provide an exhaustive exploration of the challenges inherent in managing ML assets and presenting academic and industrial responses in the form of specialized tools developed to mitigate these challenges. We underscore persistent challenges, including comparability, traceability, and reproducibility of ML assets, the complexity of lineage tracing, and the absence of standardized management methods, highlighting the ongoing need for innovations and solutions in this domain.

\subsection{Machine learning development lifecycle}
\vspace{-0.5em}

\begin{figure*}
\includegraphics[width=\columnwidth]{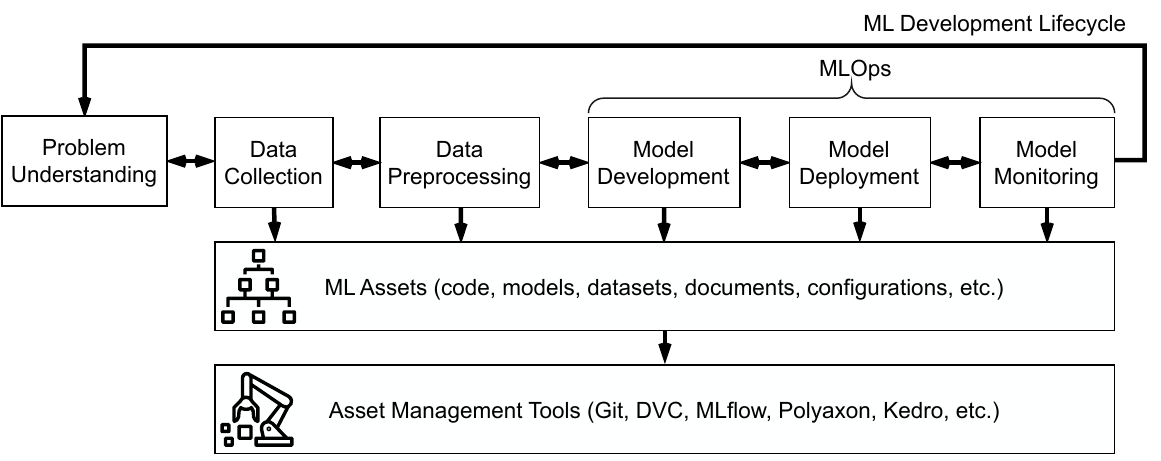}
\caption{Machine learning development lifecyle and asset management.}
\label{fig:lifecycle}
\vspace{-1.5em}
\end{figure*}

The ML development lifecycle, as shown in Figure~\ref{fig:lifecycle}, is a comprehensive process that encapsulates various stages, each integral to the development and deployment of ML-driven solutions. This lifecycle, as described by Amershi~et~al.~\cite{amershi2019software}, is not a strictly linear progression, but rather a dynamic flow with potential feedback loops. Each stage may require iterative refinement if certain criteria are not met, necessitating a review of previous stages.
\begin{itemize}
    \item \textit{Problem Understanding} stage is about understanding the problem at hand. It involves identifying potential features, understanding their significance in the context of existing or upcoming ML-driven solutions, and selecting an appropriate model architecture tailored to the problem.
    \item \textit{Data Collection} stage involves sourcing, integrating, and augmenting datasets, often using generic data for initial model training and specialized data for transfer learning.
    \item \textit{Data Preprocessing} stage involves cleaning, labeling, and transforming the datasets. It also emphasizes the extraction and selection of the most informative features that serve as input to the ML models.
    \item \textit{Model Development} stage involves training, evaluating, and testing models on selected features. The performance of these models is typically measured using predefined metrics in validation and test datasets.
    \item \textit{Model Deployment} stage involves integrating the trained model into real-world applications, allowing it to make inferences based on new data.
    \item \textit{Model Monitoring} stage is about tracking how the ML model performs over time. If any degradations in performance is observed, the current model might be re-trained with newly collected data.
\end{itemize}

\subsection{Machine learning asset management}
\vspace{-0.5em}

% The paragraph discusses the challenges in managing ML models
In the context of this lifecycle, ML assets emerge as the foundational elements employed throughout each stage. These assets, which encompass datasets, models, hyperparameters, configurations, and more, are significant in ensuring the success of ML operations (MLOps). This pivotal role necessitates the adoption of specialized machine learning asset management tools, such as MLflow. Whether accessed through intuitive web dashboards or command-line interfaces, these tools support the automation of operational tasks such as tracking, exploring, retrieving, collaborating on, archiving, and discarding ML assets~\cite{idowu2021asset}.
%The ML field has transformed from its early stages, characterized by basic algorithms, to a modern epoch where advanced systems are integral to production environments. This evolution is not only marked by the sophistication of algorithms and models but is also closely linked with advancements in the methods and tools designed to manage various types of ML assets~\cite{idowu2021asset}. Central to the ML ecosystem is the discipline of asset management, which provides the necessary frameworks and operations to navigate the multifaceted challenges associated with the development, deployment and operation of ML-driven solutions. 

However, managing these assets is fraught with a series of challenges. Schelter~et~al.~\cite{schelter2015challenges} categorize the challenges of managing ML models into three core categories: conceptual, data management, and engineering. They highlight issues such as model definition, multiple input handling, metric selection, data quality, and system compatibility in programming languages and hardware. Furthermore, Polyzotis~et~al.~\cite{polyzotis2018data} identify three main areas of focus in data lifecycle management for machine learning: understanding, validation/cleaning, and preparation. Some of the most important challenges include dealing with missing or inconsistent data, managing data from heterogeneous sources, ensuring data privacy and security, and keeping up with the latest developments in the ML community. Further expanding on this, Munappy~et~al.~\cite{munappy2022data} identify $20$ data management challenges encountered by deep learning practitioners, categorizing them into four groups based on the presence of solutions. These challenges cover issues ranging from data search, quality, and privacy to data versioning, labeling, and scalability. 

% The paragraph details how both academic and industrial sectors have developed specialized tools for machine learning asset management in response to emerging challenges.
In response to these intricate challenges, both research communities and companies have introduced specialized tools for asset management. Academically, tools such as ModelHub~\cite{miao2017modelhub}, DLHub~\cite{chard2019dlhub}, and Pdmdims~\cite{peili2018deep} have been introduced for model management. For experiment management, prototypes include Codalab~\cite{pavao2022codalab}, MLflow~\cite{zaharia2018accelerating}, Runway~\cite{tsay2018runway}, and ModelDB~\cite{vartak2018modeldb}. Furthermore, for comprehensive lifecycle management, solutions such as CANDLE~\cite{wozniak2018candle}, Ease.ML~\cite{aguilar2021ease}, VeML~\cite{le2023veml}, Scanflow~\cite{bravo2022scanflow}, TITAN~\cite{benitez2021titan}, Stratum~\cite{bhattacharjee2019stratum}, ProvDB~\cite{miao2017provdb}, and ModelOps~\cite{hummer2019modelops} have been proposed. In the industrial domain, commercial tools such as Aim\footnote{\url{https://aimstack.io}}, ClearML\footnote{\url{https://clear.ml}}, cnvrg.io\footnote{\url{https://cnvrg.io}}, Comet\footnote{\url{https://comet.com}}, Domino\footnote{\url{https://domino.ai}}, DVC\footnote{\url{https://dvc.org}}, Kedro\footnote{\url{https://kedro.org}}, Polyaxon\footnote{\url{https://polyaxon.com}}, and Weights\&Biases\footnote{\url{https://wandb.ai}} have been developed to address these challenges.

% The paragraph discusses the persistent challenges in managing ML assets despite advancements in tools
Despite these advances in tool support, managing ML assets using these tools still faces important challenges. Kumar~et~al.~\cite{kumar2017data} provide a comprehensive review of the data management-related challenges that arise in ML workloads and analyze the key techniques and systems that have been developed to address these challenges. Moreover, Schlegel~et~al.~\cite{schlegel2023management} highlight the difficulties in ensuring the comparability, traceability, and reproducibility of artifacts, such as models, data, logs, and metadata, throughout the ML development lifecycle, even with specialized tools. Furthermore, Melin~\cite{melin2023tackling} underscores the need for end-to-end versioning in the ML development lifecycle, highlights the complexities of lineage tracing due to framework integration issues, and advocates for clear guidelines in managing ML artifacts. In addition, Idowu~et~al.~\cite{idowu2021asset} highlight challenges such as limited interoperability between tools, tight library coupling, and the overhead of the necessary code instrumentation. They also point out the absence of standardized management methods for ML assets, forcing users into ad hoc solutions that hinder efficient model development. 

Compared to previous studies, our research explores the experience and feedback of tool users related to machine learning asset management, shedding light on the unexplored challenges and their practical solutions.

% \subsection{DevOps, DataOps, and MLOps}

%In light of this, three operational paradigms stand out: DevOps, MLOps, and DataOps. DevOps, as the name suggests, merges development and operations. It is a practice that streamlines software development to ensure faster and high-quality delivery~\cite{ebert2016devops}. By integrating software development (Dev) with IT operations (Ops), DevOps aims to reduce development time while ensuring continuous and quality software delivery. On the other hand, MLOps, short for Machine Learning Operations, focuses on simplifying the deployment and maintenance of machine learning models in real-world settings~\cite{kreuzberger2023machine}. It is a collaborative function that involves data scientists, DevOps engineers, and IT, focusing on aspects such as automation, workflow orchestration, reproducibility, continuous ML training, evaluation, and more. Lastly, DataOps, short for Data Operations, emphasizes data management through increased collaboration and automation~\cite{munappy2020ad}. Its primary goal is to speed up value delivery by ensuring consistent management and changes in data, data models, and associated artifacts. In summary, when these three paradigms are seamlessly combined, they provide a clear and comprehensive roadmap for handling the complexities of the machine learning development lifecycle.

%% file: sections/3-methodology.tex
\section{Methodology}
\label{sec:methodology}
\vspace{-0.5em}
In this section, we first outline our research goals and questions, followed by a detailed explanation of our study design.

\subsection{Goal and Research Questions}
\label{sec:method:subsec:goal}
\vspace{-0.5em}

%We formulate our study using the ``GQM'' (goal, question, metric)~\cite{Goal-Question-Metric} template as follows.

Our study aims to investigate the challenges associated with the management of ML assets. We aim to understand their prevalence and proposed solutions across developer discussion forums. Our research is specifically targeted at a diverse group of stakeholders, including data scientists, ML engineers, infrastructure architects, and other roles in the ML development lifecycle, such as operations managers. Collectively, we describe this group as ``ML practitioners''. The rationale behind focusing on these professionals is rooted in their profound engagement and proficiency in the ML development lifecycle. By understanding their perspectives and experiences, our research seeks to illuminate the pathways for more efficient and streamlined management of ML assets.

To achieve our stated goal, we investigate three research questions.
\begin{itemize}
    \item \textbf{RQ$_1$:} \textit{What topics are most frequently discussed related to machine learning asset management?} In RQ1, we identify common (macro-)topics in Q\&A posts on ML asset management. We then analyze and compare the frequency of different types of inquiry and macro-topics to determine the most common ones. Our goal is to gain a clear understanding of the key challenges that ML practitioners face in asset management.
    % \item \textbf{RQ$_2$}: \textit{Which topics regarding machine learning asset management are more challenging to address in developer forums?}
    \item \textbf{RQ$_2$}: \textit{What topics of solutions exist for the challenges related to machine learning asset management?} In RQ2, we identify common (macro-)topics found in solutions that address the challenges of ML asset management. We explore the interconnections between challenges and their solutions to uncover discernible trends or correlations in different domains. Our goal is to construct a comprehensive guide for ML practitioners, helping them navigate and overcome these challenges effectively by aligning them with proven solutions.
    \item \textbf{RQ$_3$:} \textit{What are the commonalities and differences between developer forums in their discussion related to machine learning asset management?} In RQ3, we investigate what is consistent and what varies between the different forums in terms of asset management discussion. Our goal is to uncover the unique strengths and patterns of each forum, helping ML practitioners navigate and utilize these forums effectively for their specific needs and interests.
\end{itemize}

\subsection{Study Design}
\label{sec:method:subsec:design}
\vspace{-0.5em} 

\begin{figure}[!t]
\centering
\includegraphics[width=\columnwidth]{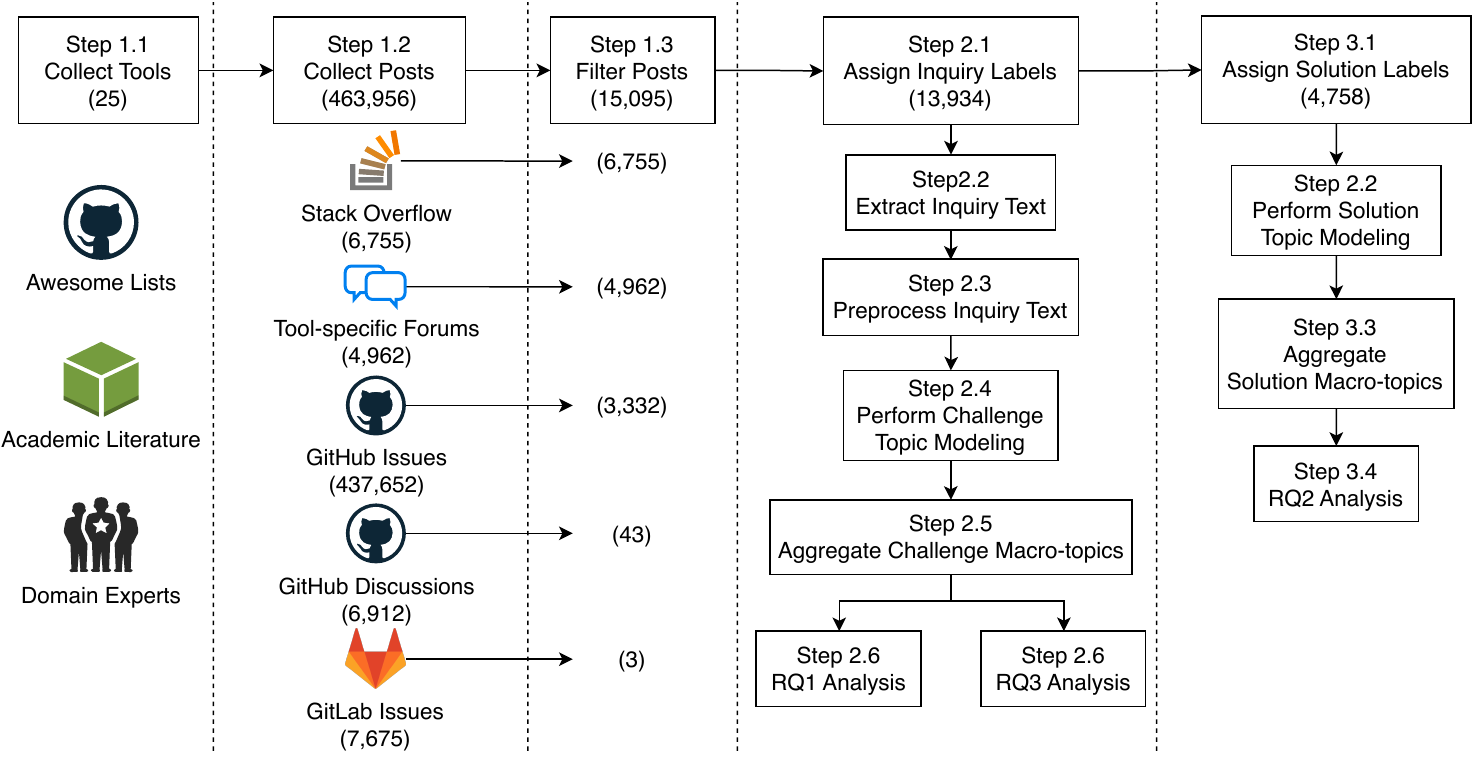}
\vspace{-2em}
\caption{Study workflow to analyze Q\&A posts related to ML asset management from developer discussion forums.}
\label{fig:study workflow}
\vspace{-1.5em}
\end{figure}

We present the workflow of our study in Figure~\ref{fig:study workflow}. Our study begins with the collection of Q\&A posts from developer discussion forums. Once collected, we classify these posts into two types: problem and knowledge. Following this classification, we preprocess the posts and employ the BERTopic technique to extract their topics. We then manually group them into broader categories, termed ``macro-topics''. Furthermore, we evaluate and compare the frequency of these macro-topics based on the type of inquiry and the specific discussion forums from which they are sourced. This evaluation helps to address our research questions, specifically RQ1 and RQ3. To wrap up our study, we manually label the solutions found in the posts and extract the topics associated with these solutions. This step sets the stage for the macro-topic mapping between challenges and solutions in RQ2.

\subsubsection{Phase 1: Data collection}
\label{sec:method:subsec:design:phase1}
\vspace{-1em}

In this phase, our goal is to collect Q\&A posts related to ML asset management from multiple developer discussion forums. To achieve this, we first start a multivocal literature review, aiming to compile an exhaustive list of tools that are specifically designed to manage ML assets. From this compilation, we derive a curated set of tool-specific keywords. Using these identified keywords, we then extract relevant Q\&A posts from multiple developer discussion forums.

\textbf{Step 1.1: Collect ML asset management-related tools.}

We begin by reviewing a range of literature, focusing on references to ML asset management tools. We formulate our initial search queries by refining the strategies used for literature retrieval, as introduced in the foundational work on asset management by Idowu~et~al.~\cite{idowu2021asset}. To expand our search, we include terms such as ``MLOps tools'' and ``machine learning experiment management tools'', among others related to asset management. Using these terms, we search Github Awesome Lists\footnote{\url{https://github.com/topics/awesome}} and Google Scholar\footnote{\url{https://scholar.google.com}} to find relevant literature, with the filtered results listed in Table~\ref{tab:retrieved literature} (Appendix~\ref{sec:appendix-2}). We follow a progressive stop method: If we do not find any relevant articles on three consecutive search pages, we conclude our search. Note that we ignore the literature that is similar to any work identified in the previous search. Furthermore, we employ a backward snowballing process~\cite{jalali2012systematic} to examine the references cited in the literature we review, identifying additional relevant works. This approach ensures that we consider all pertinent information and insights available in the existing body of knowledge.

Then we apply the following exclusion criteria to the tools that are referenced in the collected literature.
\begin{itemize}
    \item The tool is either restricted for in-house use, such as FBLearner Flow\footnote{\url{https://venturebeat.com/dev/facebook-details-its-company-wide-machine-learning-platform-fblearner-flow}}, or no longer available, such as Dot Science\footnote{\url{https://www.datanami.com/this-just-in/dotscience-is-shutting-down}}.
    \item The tool is hosted in an open-source repository, but it has earned less than $100$ stars. Although the number of stars in a repository may not be the only indicator of its quality or utility, it is often seen as a measure of trust and usage in the community~\cite{borges2018s}.
    \item The tool appears to be poorly maintained or not adequately documented. For example, its open-source repository has not received updates for more than a year. This metric has been used in the previous literature as an indicator to identify dormant open-source projects~\cite{coelho2017modern,izquierdo2017empirical,mens2014survivability,khondhu2013all}.
    \item The tool is specifically designed for a particular type of ML asset, such as a dataset or model, such as Dolt\footnote{\url{https://github.com/dolthub/dolt}}.
    \item The tool is a multipurpose asset management tool that is not specifically tailored for ML applications, such as Pachyderm\footnote{\url{https://github.com/pachyderm/pachyderm}}.
    \item The tool is in the research phase and has not yet been discussed in any developer forum or used in open-source repositories, such as VeML~\cite{le2023veml}.
\end{itemize}

We review more than $500$ tools and narrow them down to $25$ tools after applying the above exclusion criteria. We present the status (open source or proprietary), launch date, and website address of each tool in Table~\ref{tab:curated tools} (Appendix~\ref{sec:appendix-2}). In this table, the status indicates whether they are open-source or proprietary tools; the launch date indicates the date when the tool is accessible to the public; and the website shows the links to these tools. We find that 52\% of the tools (13 of 25) feature open source repositories, indicating a significant presence of open source options for machine learning asset management.

\textbf{Step 1.2: Scrape the tool-related posts from the developer discussion forums}

Having the list of tools, we scrape Q\&A posts from various discussion forums that mention the names of tools. Our focus is on forums predominantly used by developers, specifically those that cater to ML practitioners. In our study, we have chosen to analyze posts from three distinct types of forum: general forums including Stack Overflow, tool-specific forums (officially hosted by the tool providers for discussions related to their tools), such as the MLflow Google Group, and repository-specific forums (dedicated to discussions around specific repositories), such as GitHub Issues. In the following sections, we detail our methodology for collecting posts from each of these forums.

\textbf{General forums} 

For general developer discussion forums, we exclusively concentrate on Stack Overflow. Stack Overflow\footnote{\url{https://stackoverflow.com}} is a renowned platform where people, ranging from beginners to experienced professionals, engage in asking and answering questions about programming and software development. Building on this premise, it is worth noting that, instead of explicitly discussing ML asset management concepts, ML practitioners often allude to these concepts by referencing ML asset management tools in their posts. This can be either through direct mentions of the tool's name or, more commonly, via tags. Specifically, tags play a pivotal role in providing context to discussions about ML asset management tools. 

Given their significance, we prioritize tags as our primary filtering criterion for the tools under investigation. Thus, we employ an iterative and heuristic process to curate a relevant set of tags. For example, entering the keyword ``[*sagemaker*]'' into the Stack Overflow search bar retrieves several tags in the search results, such as \texttt{amazon-sagemaker}, \texttt{amazon-sagemaker-studio}, and \texttt{amazon-sagemaker-neo}, among others. In particular, some retrieved tags may not be directly related to asset management. Therefore, we meticulously review each tag in conjunction with its associated discussions to determine its relevance. Based on this evaluation, we compile a comprehensive list of tags that accurately represent the tools under study, as illustrated in Table~\ref{tab:stackoverflow tags} (Appendix~\ref{sec:appendix-2}). Using these tags, we formulate an SQL query, detailed in our replication package, to extract data from the Stack Exchange Data Explorer\footnote{\url{https://data.stackexchange.com}}. This query enables us to download $6,755$ unique Stack Overflow posts (up to $26^{th} of July, 2023$), ensuring a comprehensive and relevant dataset for our analysis. %Consequently, we avoid relying on keywords for filtering, as our observations indicate that untagged tool mentions often don't place the tool at the heart of the discussion.

\textbf{Tool-specific forums}

In addition to general developer discussion forums, we check the official website of each curated tool for information about any public discussion forum hosted by the tool developers. We provide the website addresses of the tool-specific discussion forums in Table~\ref{tab:tool-specific community websites} (Appendix~\ref{sec:appendix-2}). It is important to note that we have chosen to exclude official chat platforms such as Discord and Slack. This decision is based on the fact that chat data is typically held privately by tool organizations, and their policies prevent us from collecting this data for any public use. Since our primary focus is on Q\&A posts in these discussion forums, we have opted to exclude content, such as event announcements\footnote{\url{https://discuss.dvc.org/c/blog-discussions/5}}, which do not align with the objectives of our study. On the other hand, forums such as the MLflow Google Group do not differentiate between Q\&A posts and other types of content, such as product announcements. In these scenarios, we employ a manual filtering mechanism during our closed card sorting process, which is elaborated on in Step~2.1. Following this methodology, we have identified $4,962$ unique posts on tool-specific forums (up to $26^{th} July, 2023$). %For now, we decided to discard Slack and Discord data because 1) we are having difficulties with data collecting and 2) these data sources are not structured in a question-answer structure which makes it hard to differentiate questions from answers.

\textbf{Repository-specific forums} 

For repository-specific forums, we consider code collaboration platforms, such as GitHub\footnote{\url{https://github.com/features/issues}}, GitLab\footnote{\url{https://docs.gitlab.com/ee/user/project/issues}} and Bitbucket\footnote{\url{https://support.atlassian.com/bitbucket-cloud/docs/understand-bitbucket-issues}}. These platforms offer developers the opportunity to raise inquiries through repository issue channels. Although the discussion within these channels is tailored toward specific projects, they frequently underscore recurring challenges and community-proposed solutions related to ML asset management. Furthermore, GitHub provides an optional feature called ``GitHub Discussions'' for repositories, specifically catering to broader communication needs, such as Q\&A sessions\footnote{\url{https://github.com/features/discussions}}. We leverage these channels to extract insights relevant to our study.

First, we collect all repositories that have a dependency on the curated tools in their codebase. We observe that certain tools, such as DVC, possess GitHub dependency graphs. These graphs are essentially manifests detailing the ecosystems and packages upon which the tool relies. For such tools, we extract the necessary information from dependency repositories using an open source scraping tool\footnote{\url{https://pypi.org/project/github-dependents-info}}. For tools that lack dependency graphs, we rely on ``usage patterns'' to identify their presence in the repository codebase. A usage pattern is defined as a regular expression that represents a specific code snippet that indicates the use of a curated tool, such as importing a specific library. Our curated usage patterns are found in Table~\ref{tab:tool usage pattern} (Appendix~\ref{sec:appendix-2}). We derive these patterns by thoroughly examining the official documentation of each curated tool.

Then we resort to Sourcegraph to collect repository information that matches the usage pattern. Sourcegraph\footnote{\url{https://sourcegraph.com}} is a code search and navigation tool, facilitating developers to search, navigate, and understand their codebases on major code collaboration platforms such as GitHub, GitLab, and BitBucket. From this search, we have identified $37,782$ repositories on GitHub and $10$ on GitLab. Subsequently, we extracted issue posts from these repositories using the respective platform APIs. During this process, we have encountered some inaccessible repositories that return HTTP error status codes. These errors signify either a policy violation, as indicated by a $451$ error, or the nonexistence of the repository, represented by a $404$ error. Of the repositories that we can access, we download $437,652$ unique issues from GitHub and $7,675$ from GitLab (up to $26^{th} of July, 2023$).

To collect posts from the GitHub Discussions, we reuse the GitHub repositories during the aforementioned repository collection. We then narrow our focus to the ``Q\&A'' channels within each discussion forum, such as the one in Apache Airflow\footnote{\url{https://github.com/apache/airflow/discussions/categories/q-a}}, since only these channels contain Q\&A posts, distinguishing them from feature requests or polls. As a result, our dataset ends up with $6,912$ unique GitHub Discussions posts in repository-specific forums.

\textbf{Step 1.3:} \textit{Filter posts that are irrelevant to our study}

In this section, we filter out posts that are not relevant to our research objectives. For posts originating from Stack Overflow and tool-specific forums, we would manually filter them during closed card sorting detailed in Step~2.1.

For GitHub Discussions posts, we systematically curate a list of keywords pertinent to the tools being examined. First, we extract tool-specific keywords from the discussions heuristically. Subsequently, we assess the relevance of these keywords by searching for them within the posts to determine their effectiveness in identifying discussions pertinent to each tool. For example, our analysis of discussions related to Azure Machine Learning revealed that practitioners frequently use terms, such as ``aml'', ``azure machine learning'', ``azure ml'', ``azure-ml'', and ``azureml''. Table~\ref{tab:post title keywords} (Appendix~\ref{sec:appendix-2}) provides a comprehensive list of keywords identified for each tool evaluated. We then use these keywords to filter and retain tool-relevant posts, selecting any post whose title contained at least one of these keywords. Using this methodology, we identify $43$ relevant posts for further analysis from a total of $6,912$.

When analyzing issues on GitHub or GitLab, our initial step involves scrutinizing the 644 unique labels of all collected repositories. From our examination, we find that only four keywords – ``bug'', ``crash'', ``error'' and ``invalid'' – indicate a user inquiry-related issue. As a result, we exclude issues that do not contain these keywords in their labels, while keeping those with empty labels for additional review. Next, we further narrow our selection to issues whose titles contain any of the keywords outlined in Table~\ref{tab:post title keywords} (Appendix~\ref{sec:appendix-2}). After applying these filters, we are left with $3,332$ issues from the initial $437,652$ on GitHub and $3$ issues from the $7,675$ on GitLab. Following this, we select a random sample of $385$ posts from the refined post collection to achieve a confidence level of $95\%$ with a margin of error of $5\%$. Upon reviewing the content of these posts, we confirm that they are indeed discussing the specified tools, which validates our choice of keywords.

\subsubsection{Phase 2: Inquiry categorization}
\label{sec:method:subsec:design:phase2}
\vspace{-1em}

In this phase, our goal is two-fold in terms of inquiry categorization. First, we analyze the types of inquiry to understand their inherent characteristics. Second, we identify the common topics that emerge from these inquiries. To achieve this, we begin with closed card sorting~\cite{wood2008card} to categorize the types of inquiry from the collected posts. After categorizing, we preprocess the content of these posts and employ an appropriate topic modeling method. We then fine-tune the models to determine the optimal hyperparameters. Once the topics are identified, we manually group them into macro-topics. To address RQ1, we have evaluated the prevalence of each (macro-)topic, along with the type of inquiry. For RQ3, we similarly analyze prevalence patterns across discussion forums, drawing parallels with the methods used for RQ1.

\textbf{Step~2.1:} \textit{Closed card sorting on inquiries}

\begin{figure}
\fbox{\includegraphics[width=\columnwidth]{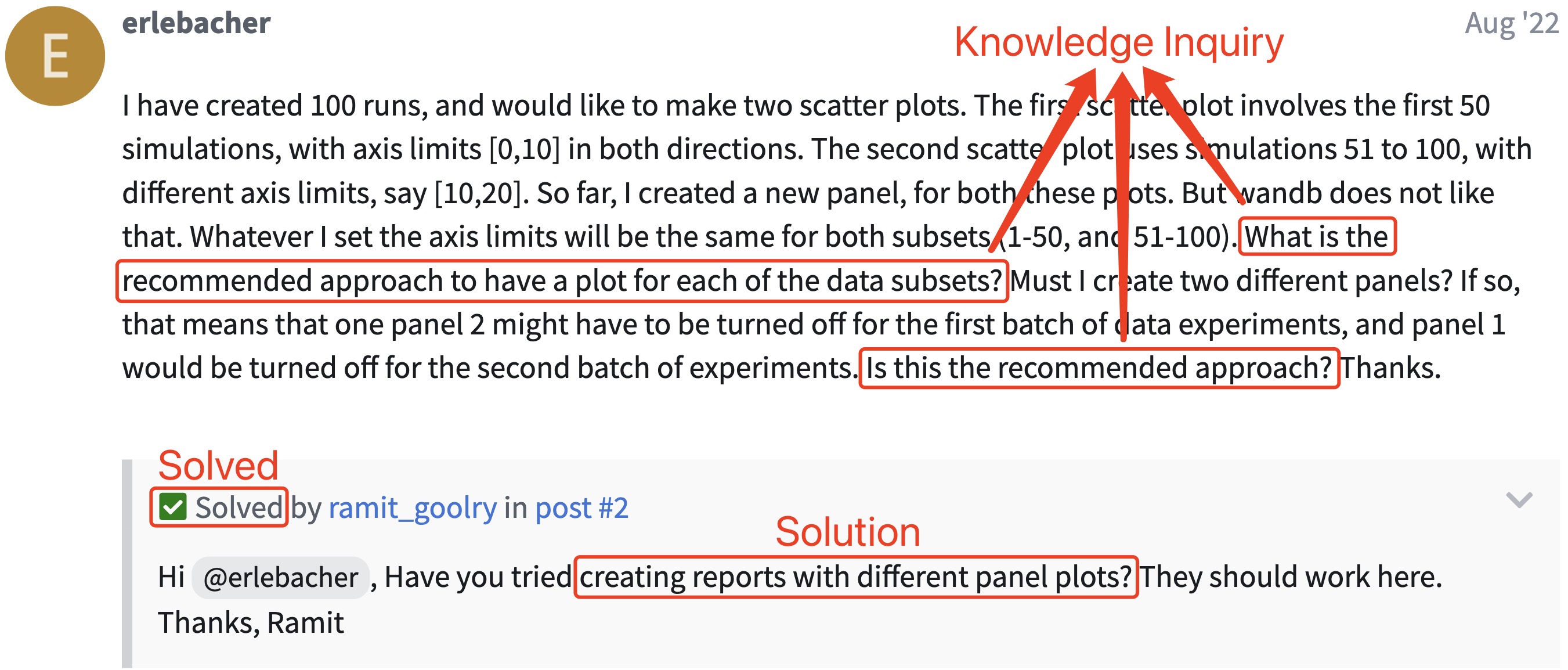}}
\vspace{-1.5em}
\caption{A solved knowledge inquiry from Weights\&Biases discussion forum.}
\label{fig:tool-specific}
\vspace{-1.3em}
\end{figure}

\begin{figure}
\fbox{\includegraphics[width=\columnwidth]{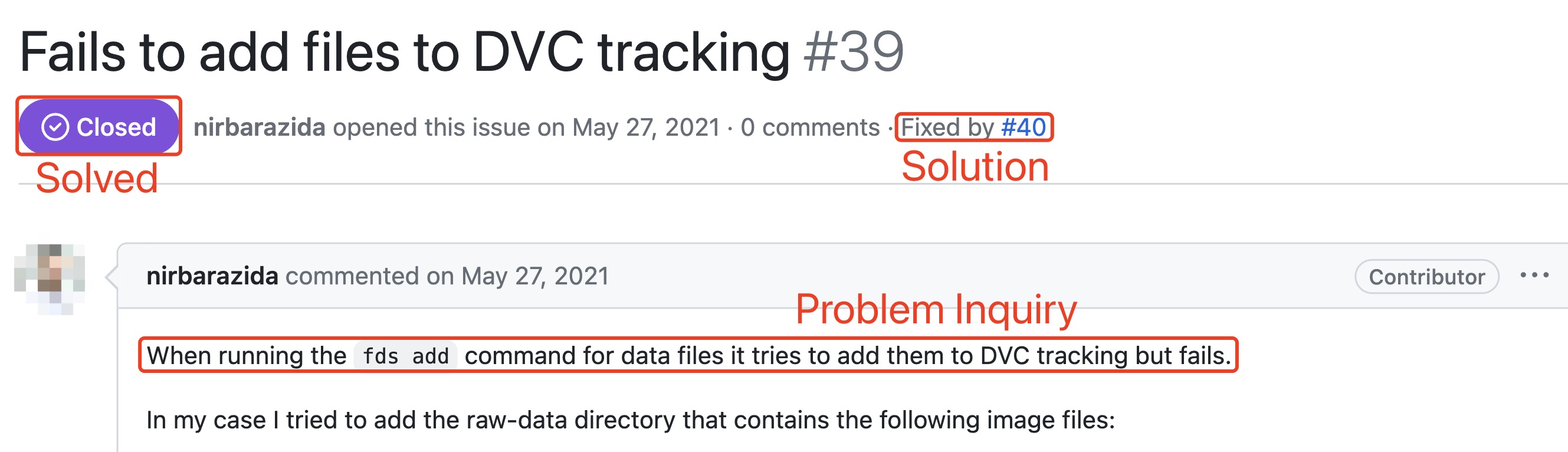}}
\vspace{-1.5em}
\caption{A closed problem inquiry in GitHub Issues.}
\label{fig:github issue-1}
\vspace{-1.5em}
\end{figure}

In the Q\&A context, inquiries primarily fall into two categories based on user intent: ``knowledge inquiry'' and ``problem inquiry'', as illustrated in Figures~\ref{fig:tool-specific} and Figure~\ref{fig:github issue-1}, respectively. Knowledge inquiry aims to understand concepts such as tool usage, technology comparison, and best practices, while problem inquiry aims to address specific issues such as bugs, errors, and performance degradation. This classification consolidates previous categorizations of ``conceptual'', ``knowledge-based'', and ``howto'' inquiries, as proposed by Chen~et~al.~\cite{chen2023practitioners} and Treude~et~al.~\cite{treude2011programmers}, under the umbrella of knowledge inquiries. For instance, a question such as ``What is the correct way to fine-tune a model?'' is labeled as a ``conceptual'' inquiry by Treude~et~al.~\cite{treude2011programmers}, whereas ``How to fine-tune a model'' is considered a ``howto'' inquiry by Chen~et~al.~\cite{chen2023practitioners}. Given the minor difference in verbal semantics, distinguishing between them in our research proves challenging and leads us to combine them into the knowledge inquiry category for clarity. This grouping reduces ambiguity among these types of inquiry, offering a clearer understanding of the common types of inquiry in discussion forums.

In addition to the problem and knowledge inquiries, posts that are non-inquiries\footnote{\url{https://github.com/awslabs/amazon-neptune-tools/issues/38}}, are mistagged\footnote{\url{https://stackoverflow.com/questions/63844663}}, or are irrelevant\footnote{\url{https://www.googlecloudcommunity.com/gc/AI-ML/What-you-think-about-CHATGPT/m-p/506958}} are labeled as ``not applicable'' (NA). Examples of non-inquiry posts encompass feature requests, event announcements, and user feedback surveys, to name a few. %feature requests, upcoming changes, event announcements, user feedback surveys, tutorials, advertisements, community updates, discussions on future plans, product launches, thank-you notes, usage claims, suggestions for improvements

The first and second authors opt to perform closed card sorting collaboratively, focusing on the three aforementioned types of inquiry. Closed card sorting is a method in which participants sort items into pre-defined categories on the collected posts~\cite{wood2008card}. From the total of $15,095$ posts, we randomly select $385$, providing a confidence level of $95\%$ and a margin of error of $5\%$. This sample size ensures an adequate representation of the general data~\cite{hastie2009elements}. The inter-rater reliability of this process, measured by Cohen's kappa~\cite{mchugh2012interrater}, is approximately $0.932$. This high value reflects a strong agreement between the authors, underscoring the consistency and reliability of our categorization process. In cases of disagreement, the authors engage in discussions to reconcile their perspectives. After filtering out posts with ``NA'' labels, the resulting dataset comprises $6,749$ Stack Overflow posts, $4,774$ tool-specific posts, and $2,403$ repository-specific posts. %In Table~\ref{post distribution:tool}, we show the distribution of Q\&A posts amongst various ML asset management tools. In particular, certain tools, such as \texttt{cnvrg.io}, do not have any Q\&A posts in our dataset. Note that the post number differs slightly from what was specified in the previous section, as some posts reference multiple curated tools together. 

% \begin{table}[!t]
% \centering
% \caption{The distribution of Q\&A posts across different ML asset management tools.}
% \begin{tabular}{lrrrrrrrrrrrrrrrrrrrr}
% \toprule
% \rotatebox{90}{Tools} & \rotatebox{90}{Amazon SageMaker} & \rotatebox{90}{Azure Machine Learning} & \rotatebox{90}{Weights \& Biases} & \rotatebox{90}{MLflow} & \rotatebox{90}{Vertex AI} & \rotatebox{90}{DVC} & \rotatebox{90}{Kedro} & \rotatebox{90}{Optuna} & \rotatebox{90}{Neptune} & \rotatebox{90}{Guild AI} & \rotatebox{90}{ClearML} & \rotatebox{90}{H2O AI Cloud} & \rotatebox{90}{MLRun} & \rotatebox{90}{Comet} & \rotatebox{90}{Polyaxon} & \rotatebox{90}{Sacred} & \rotatebox{90}{SigOpt} & \rotatebox{90}{Domino} & \rotatebox{90}{Aim} & \rotatebox{90}{Determined} \\
% \midrule
% count & 4325 & 4248 & 1378 & 1307 & 947 & 602 & 319 & 278 & 128 & 124 & 94 & 59 & 54 & 52 & 36 & 24 & 19 & 14 & 6 & 1 \\
% \bottomrule
% \end{tabular}
% \label{post distribution:tool}
% \end{table}

\textbf{Step~2.2:}\textit{Extract post content for topic modeling}

After categorizing the types of inquiry, we explore the topics of these inquiries using topic modeling, as described in detail in \textbf{Step~2.4}. To train a robust topic model, we opt for three distinct ways to extract content from the posts to achieve comprehensive coverage and the most accurate and relevant results.
\begin{enumerate}
    \item We extract the titles of the posts.%\cite{rahman2019snakes,allamanis2013and,rosen2016mobile}
    \item We extract the titles of the posts along with their bodies.
    \item We generate a refined title for each post using GPT-4\footnote{\url{https://platform.openai.com/docs/models/gpt-4}}, a large multimodal language model created by OpenAI. The parameter settings of GPT-4 are detailed in Table~\ref{tab:gpt-4} (Appendix~\ref{sec:appendix-2}). Our methodology involves a rigorous and iterative fine-tuning process of the input prompts, which includes approximately $20$ iterations. In every iteration, we meticulously adjust the input prompts to explore possible improvements in the coherence score associated with the identified topics. Subsequently, we conduct a manual assessment of the contextual relevance of each generated title. This iterative fine-tuning continues until we observe no further improvement in coherence scores, signifying the achievement of optimal prompt refinement. Consequently, the final optimized prompt, including the original content of the post, is sent to GPT-4 to generate the most coherent and contextually relevant refinement of the title of a given post.
    \begin{quote}
        \code{Refine the title of the following post to make it short and clear in simple English. \textbackslash n Title: [title text here] \textbackslash n Body: [body text here] \textbackslash n Refined Title:}
    \end{quote}
    We simplify and edit the content of $153$ posts that are too long to fit in GPT-4's $8,000$ token limit before feeding it to the model. This process involves removing verbose and non-essential sections, such as execution traces, to ensure that the text remains within the token constraints. 
    
    We opt to summarize the posts using GPT-4 due to observed inconsistencies in the quality of the titles and content of the posts. On the one hand, this decision comes from two main observations. First, the titles of the posts\footnote{\url{https://github.com/fastai/fastai/issues/3085}} can be vague, reducing the efficiency of our topic modeling efforts. This is a significant issue, as unclear titles can obscure the post's main subject, leading to less accurate topic classification. Second, the performance of topic modeling algorithms can suffer from repetition of words within posts~\cite{barde2017overview}. Specifically, excessive use of non-essential repeated words can compromise the algorithm's ability to accurately identify the core topics of discussion. Together, these factors necessitate a more sophisticated approach to post summarization to enhance our topic modeling process. On the other hand, recent research suggests that GPT-4 serves as a reliable and consistent substitute for human labeling in summarization tasks~\cite{gilardi2023chatgpt,cheng2023gpt}. In particular, a recent study shows that GPT-4 achieves a Spearman correlation of $0.514$ with human labelers in summarization tasks, surpassing all previous methods by a significant margin~\cite{liu2023gpteval}. %Empirical data further endorse the use of GPT-based models for text summarization in professional settings, as seen in articles by Towards Data Science\footnote{\url{https://towardsdatascience.com/choosing-the-right-language-model-for-your-nlp-use-case-1288ef3c4929}} and Quartz\footnote{\url{https://qz.com/how-chat-gpt-could-be-used-in-economics-research-1850114121}}.

    For evaluation purposes, we randomly sample $385$ posts from a total of $15,065$, with a confidence level of $95\%$ and a margin of error of $5\%$. After comparing these refined titles with the originals, we agree that the GPT-4 generated versions match or surpass the original titles in brevity, readability, and precision. To illustrate, consider the previous Fastai issue \#3085: the refined title, ``Broken Documentation Page for Neptune Callback'', is notably more descriptive than its original, \ie, ``https://docs.fast.ai/callback.neptune.html broken''.
\end{enumerate}

\textbf{Step~2.3:} \textit{Preprocess post content with NLP techniques}

After extracting or generating content using the three aforementioned methods, we preprocess it using basic natural language processing (NLP) techniques. This preprocessing is crucial for topic modeling, as it ensures that the data are clean and ready for accurate and meaningful results. The NLP preprocessing encompasses the following steps:
\begin{enumerate}
    \item Remove any escape character, punctuation mark, or numeric value.
    \item Remove any tool-specific keyword based on Table~\ref{tab:post title keywords} (Appendix~\ref{sec:appendix-2}).
    \item Keep nouns and verbs only.
    \item Remove any stop word. In our study, we categorize stop words into three distinct types. 
    \begin{enumerate}
        \item Everyday words in English, \eg, ``you'', ``like''.
        \item Discussion-specific words. Discussion-specific words are words that frequently occur in Q\&A posts, but have minimal relevance to the setting of asset management topics, \eg, ``discuss'', ``post'', ``difference''. In particular, we exclude error-related terms, such as ``error'', ``bug'', and ``crash'', which occur significantly within problem inquiries in our research, to enhance the clarity and interpretability of generated topics.
        \item Task-specific words. Task-specific words are words that are closely associated with ML tasks, but have limited relevance when it comes to the discussion of asset management. Such terms include ``regression'', ``forecasting'', ``segmentation'' and more. By filtering out task-specific stop words, we improve the relevance and clarity of our topics related to asset management. 
    \end{enumerate}
    We evaluate each stop word by checking the coherence score and reviewing the top 10 keywords for each topic. For a detailed list of each type of stop words, we refer to our replication packages~\cite{replication_package}. 
\end{enumerate}
Following these steps, the preprocessed text is ready for input into BERTopic for hyperparameter tuning.

\textbf{Step~2.4:}\textit{Perform topic modeling}

We use BERTopic\footnote{\url{https://github.com/MaartenGr/BERTopic}} to cluster topics from preprocessed content. Previous literature indicates that BERTopic consistently outperforms its competitors in various topic modeling tasks~\cite{grootendorst2022bertopic} and is frequently used in empirical software engineering research~\cite{diamantopoulos2023semantically,gu2023self,tao2023code}. BERT embeddings are adept at capturing semantic nuances, often outperforming sparse, high-dimensional bag-of-words, or n-gram models. Unlike LDA, which relies on word count, BERTopic captures the semantics of words beyond their frequency. For example, the function ``sort'' might be referred to as ``arrange'' or ``classify'' depending on the programming language or the preference of the developers. Thus, BERTopic has the unique capability to group words that may convey similar meanings but are phrased differently.

Since the default hyperparameters provided by BERTopic might not guarantee optimal performance, our goal is to define a comprehensive hyperparameter space for fine-tuning. This search space is initialized by a review of the official BERTopic documentation and insights from its GitHub issues\footnote{\url{https://github.com/MaartenGr/BERTopic/issues}}. As we navigate the search, we pay particular attention to variations in the C\_V score, a metric commonly used to measure the quality of topics produced by topic models~\cite{syed2017full}. If we observe that the most significant improvements occur at the boundaries of the current search range, we expand the hyperparameter search space heuristically. Conversely, if the model's performance begins to deteriorate, we cease expansion to ensure its stability and reliability. When certain hyperparameters do not lead to noticeable changes in performance, we regard them as irrelevant and exclude them from our search. %An example of our iterative approach can be seen in the fine-tuning of the \texttt{min\_cluster\_size} hyperparameter, which dictates the smallest permissible cluster size, thereby indirectly influencing the number of topics. A larger pool of topics tends to offer richer detail, but at the risk of topic overlap, a concern posed by Thomas~et~al.~\cite{thomas2013mining}. Our objective is to strike a balance between granularity and redundancy by determining an optimal topic count that augments our analysis while avoiding undue overlap. 

We show the range of hyperparameters used to fine-tune BERTopic for modeling challenges and solutions, respectively, in Table~\ref{tab:hyperparameter search space} (Appendix~\ref{sec:appendix-2}). The optimal hyperparameters are marked in \colorbox{green}{green}. Although coherence C\_V is our primary metric, we also consider other metrics when the coherence score of generated topic models is comparable. For instance, we evaluate the number of topics and outliers produced. Outliers refer to posts that do not fall into any of the created topics, suggesting weak semantic connections. Furthermore, we manually assess $5\%$ of the clustered posts for each topic to ensure their relevance and consistency. If the generated topics are difficult to interpret or have significant overlap, we would review and adjust the hyperparameters or go back to earlier steps. This could involve changing our preprocessing methods, including reviewing the list of stop words.

Upon identifying the most suitable topic model, we assign the most probable topic to each post. For outliers, we integrate them with their closest topic using the HDBSCAN technique. HDBSCAN, as described by McInnes~et~al.~\cite{mcinnes2017hdbscan}, is a soft clustering algorithm that identifies clusters of varying densities in the data without the need to predefine the number of clusters. %Having the topic summary, we exclude topics (\eg, ``team building'') that do not fall in our defined study scope. 

\textbf{Step~2.5:}\textit{Aggregate into macro-topics}

Upon reviewing the initial set of $133$ topics, we find them too numerous for an effective comparison. In addition, many of these topics exhibit strong interrelations. To streamline our analysis, we aggregate these topics into macro-topics. These macro-topics, which can be thought of as ``categories of topics'', represent high-level groupings of multiple interconnected topics with similar concerns. Previous studies~\cite{chen2020comprehensive,bagherzadeh2019going,Ahmed2018WhatDC} have employed the term ``category'' to denote such clusters, but we choose to use the term ``macro-topics'' for this aggregation. In the qualitative coding phase, we apply the ``negotiated agreement''~\cite{campbell2013coding} technique to improve the reliability of the coding. This technique involves collaborative coding by the authors to reach a consensus and refine the definitions of the codes, as seen in previous studies such as those by Chen~et~al.~\cite{chen2023practitioners}. Our consensus on coding is inspired by key ideas from seminal studies within the MLOps domain. We have been particularly influenced by the works of Idowu~et~al.~\cite{idowu2021asset}, Amershi~et~al.~\cite{amershi2019software}, and Schlegel~et~al.~\cite{schlegel2023management}.

\textbf{Step~2.6:}\textit{Collect prevalence metrics}

Next, we evaluate the macro-topics based on the ``prevalence'' metrics, a common practice in previous studies~\cite{openja2020analysis,Ahmed2018WhatDC,rosen2016mobile,venkatesh2016client,Yang2016WhatSQ}. Prevalence measures the frequency with which a topic emerges in our analyzed posts, expressed as a percentage of the entire dataset. %In contrast, ``difficulty'' is assessed through two metrics: the unsolved rate and the median solution time. The unsolved rate is a percentage indicating the number of posts on a topic that remain unsolved. Meanwhile, the median solution time quantifies the median duration, in hours, it takes to bring issues related to the topic to a solution. However, it is worth noting that solution does not always equate to an issue being genuinely addressed. %We provide a detailed explanation for each metric in Table~\ref{tab:term}.

\subsubsection{Phase 3: Solution categorization}
\label{sec:method:subsec:design:phase3}
\vspace{-1em}

In this phase, our objective is to understand and categorize the strategies that ML practitioners use to address inquiries, aiming to find prevalent solutions and explore their interconnection. We start by applying open card sorting to solved posts to summarize the solutions presented in them. With the labels generated from this process, we then use BERTopic for topic modeling to further categorize these solutions. After extracting the topics, we group them into macro-topics for a more holistic view. Then we investigate how frequently each solution macro-topic appears. Lastly, we analyze the correlation between the identified challenges and their corresponding solutions. %This streamlined approach ensures a logical progression and offers a clearer picture of the strategies employed by ML practitioners.

\textbf{Step 3.1:} \textit{Open card sorting on solutions}

It is crucial to understand how inquiries are addressed and concluded in different developer discussion forums. Platforms such as Stack Overflow, GitHub Discussions, and many tool-specific forums, such as Microsoft Q\&A\footnote{\url{https://learn.microsoft.com/en-us/answers/}}, allow users to mark any answer to an inquiry as ``accepted''. This acceptance is a clear indication that an inquiry is ``solved'', which generally encapsulates a ``solution'' that can effectively address the initial concerns. Figure~\ref{fig:tool-specific} shows a knowledge inquiry\footnote{\url{https://community.wandb.ai/t/axis-scales/2892}} in the Weights\&Biases discussion forum looking for best practices to visualize datasets. This inquiry is addressed by a suggestion to create reports using different panels. 

Unlike traditional Q\&A forums, some repository-specific discussion forums, such as GitHub Issues, do not employ an answer acceptance mechanism. Instead, they adopt a ``closure'' mechanism, indicating that the inquiry is solved, no longer considered active, or requires attention. Figure~\ref{fig:github issue-1} shows a closed problem inquiry\footnote{\url{https://github.com/DagsHub/fds/issues/39}} following a merge request. 

Following the above criteria, the first and second authors opt to perform open card sorting on the solved posts to categorize the solutions provided for user inquiries. Open card sorting is a method in which participants receive a set of cards, each containing a single piece of content or data, and are asked to group these cards into categories in a way that makes sense for them, without any predefined categories~\cite{wood2008card}. We choose open card sorting due to its ability to handle the versatility of solutions related to asset management. %Also, we decide against using GPT-4 for summarization purposes due to the potential presence of external links within the posts, leading to solutions , such as which could lead to solutions in the form of merge requests.
We randomly sample $385$ posts from $4,684$ solved posts, aiming for a $95\%$ confidence level and a $5\%$ margin of error. Our labels show a high level of agreement, confirmed by a Cohen's kappa of approximately $0.852$, indicating strong consistency in categorization. In cases of disagreement, the authors engage in discussions to reconcile their perspectives. 

To label posts, we use $2-5$ words that start with a verb and end with a noun. Although adjectives are selectively added for clarity, adverbs are used sparingly. This length is chosen to meet both the contextual demands of BERTopic for informative context and to resonate with human labeling practices for consistency. Stop words, punctuation, and non-UTF8 characters are excluded to ensure that only essential information is retained, avoiding any preprocessing that could result in the loss of important data. Following these criteria, the first author independently labels the remaining solved posts. Figure~\ref{fig:tool-specific} demonstrates a solution labeled ``create report''. 

During open card sorting, we notice that solved posts do not always have a satisfactory solution. We identify three types of abnormally solved inquiries, which we subsequently exclude from our analysis. We give the definition for each type of inquiry below.
\begin{itemize}
    \item Non-inquiry: This type refers to posts that seem to raise concerns, but upon closer inspection do not contain actual problems\footnote{\url{https://stackoverflow.com/questions/56046428}}. 
    \item Intermittent inquiry: In this category, the issues mentioned are inconsistent and cannot be replicated regularly\footnote{\url{https://stackoverflow.com/questions/72641789}}.
    \item Unsolved inquiry: These are inquiries that are marked as accepted or closed but still have unsolved issues\footnote{\url{https://stackoverflow.com/questions/65884046}}.
    % \item Normal inquiry: Any inquiry that does not fall into the above categories is considered normal.
\end{itemize}

\textbf{Step 3.2:}\textit{Perform topic modeling}

After the open card sorting process, we input our labels into BERTopic for topic modeling. Following a procedure similar to Step~2.4, we identify the optimal hyperparameters by optimizing the coherence score, minimizing the number of outliers, and reducing the overlap of topics. 

If the BERTopic performance falls short of our expectations, we undertake a manual adjustment of our labels. For example, in our initial setup, we labeled solutions related to Docker images with the term ``image''. However, we later realized that this term also encompassed solutions for ``image editing'', leading to incorrect grouping. To address this, we revised the label for Docker image solutions to ``docker-image'', while keeping the label ``image'' for solutions concerning image editing. Such refinements are essential to prevent future misclustering and ensure accurate categorization.

Similar to Step~2.4, once we identify the optimal topic model, we assign the most probable topic to each solved post using the HDBSCAN technique.

\textbf{Step 3.3:}\textit{Aggregate into macro-topics}

Given the observation that many of these topics ($79$ in total) share close relationships, the first three authors manually aggregate them into macro-topics using the negotiated agreement technique, similar to Step~2.5. If the generated topics are difficult to interpret or have significant overlap, we would review and adjust the hyperparameters or go back to the earlier steps. This could involve adjusting the open card sorting standard.

\textbf{Step 3.4:}\textit{Collect prevalence metrics}

In our study, we first collect metrics on the prevalence of each macro-topics. Using this data, we then create a heatmap that illustrates the number of posts between challenge and solution macro-topics. This approach not only helps us understand which solutions are commonly associated with specific challenges, but also sheds light on the potential causes leading to those challenges. 

%% file: sections/4-rq1-results.tex
\section{RQ1 Results: Prevalence of Challenge Topics}
\label{sec:rq1-results}
\vspace{-0.5em}

In RQ1, we investigate the challenges that ML practitioners face in asset management. To begin with, we categorize the collected posts into two primary types: knowledge and problem inquiries. Subsequently, we extract the topics of these inquiries, aggregate them into macro-topics, and clearly define each of these macro-topics along with illustrative examples. We also evaluate the prevalence of these macro-topics by measuring their frequency in the collected posts. Furthermore, we explore the intricate relationship between macro-topics and curated tools, emphasizing their correlation in the inquiries.

\subsection{Prevalence of posts based on type of inquiry}
\vspace{-0.5em} 

\textbf{We find that problem inquiries make up $59.42\%$ of the Q\&A posts, whereas knowledge inquiries account for $40.58\%$.} This shows a higher prevalence of problem inquiries compared to knowledge inquiries. This trend aligns with the findings of other studies in different domains, including those related to service mesh systems~\cite{chen2023practitioners}. However, this trend does not indicate that software engineers are typically more inclined to encounter technical issues than theoretical ones. For example, a study by Treude~et~al.~\cite{treude2011programmers} reports a higher prevalence of knowledge inquiries compared to other types of inquiry on Stack Overflow. %One possible explanation for the discrepancy could be the time gap between our dataset and that of Treude~et~al. The data from Treude~et~al. is collected nearly thirteen years ago. We hypothesize that the landscape of software development has evolved since then. As technology has advanced, software systems have become increasingly dependent on a myriad of internal and external components~\cite{belguidoum2007dependency}. This interdependency, while offering enhanced functionality and efficiency, also introduces a higher degree of complexity~\cite{decan2017empirical,kerzazi2014automated,merkel2014docker}. Practitioners are now navigating a landscape where actions in one component can significantly impact others, leading to unexpected challenges and difficulties. This intricate network of dependencies can make it easier for issues to arise, contributing to an increase in problem inquiries. In light of this, there is a clear need for more comparative studies. These will help us to understand the dynamics at play and test our hypothesis. 

\subsection{Macro-topics of the posts}
\vspace{-0.5em} 

\textbf{We identify $133$ distinct topics from the collected posts}, using the topic-modeling approach outlined in Step~2.5. Our optimal topic model achieves a coherence score of $0.7478$ and identifies $3,584$ posts as outliers. We have listed the names of all identified topics, each accompanied by an index, in Table~\ref{rq1:topic list} (Appendix~\ref{sec:appendix-2}). For a description of each topic, we refer to our replication package~\cite{replication_package}. Then we aggregate the topics into 16 macro-topics, which offer in-depth insights into the challenges of ML asset management, independent of specific frameworks and technologies. This study denotes macro-topics with the prefix $\hat{C}$ and topics with the prefix $C$. The subscript that follows each prefix indicates the (macro-)topic's index. In Table~\ref{rq1:macro-topic list}, we illustrate each macro-topic with its index, name, prevalence, and underlying topics.

\input{tables/rq1/rq1-macro-topic-list}

\begin{figure}[t]
\includegraphics[width=\columnwidth]{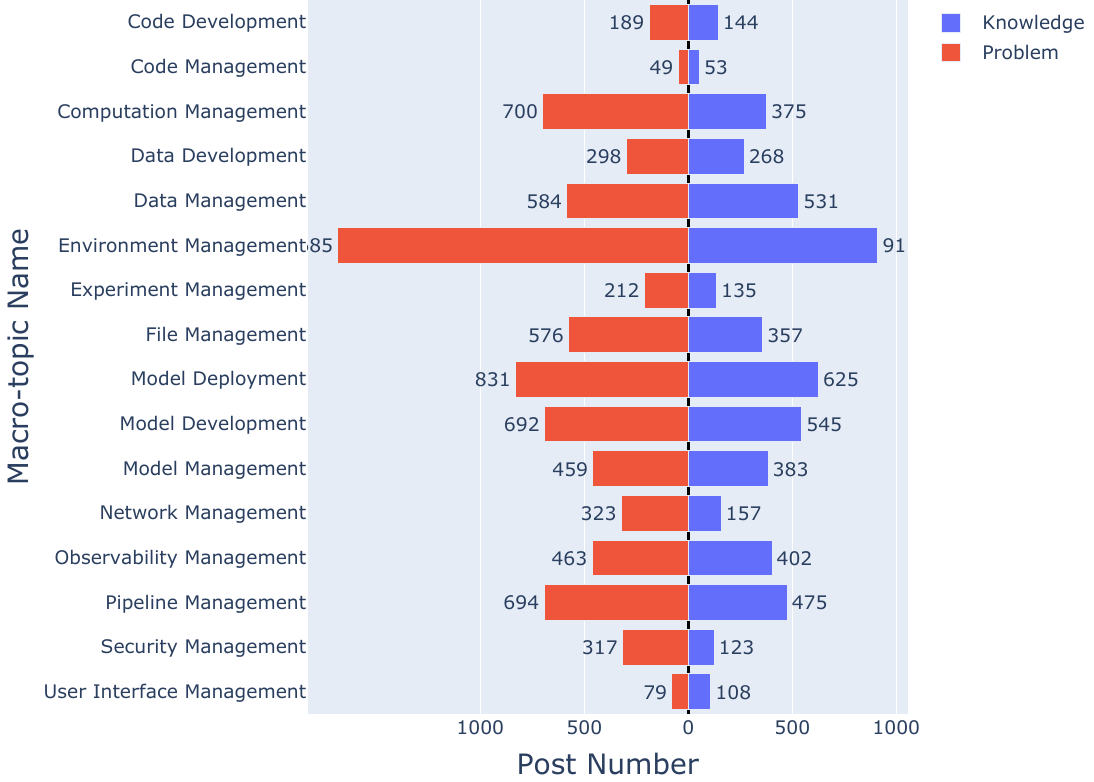}
\vspace{-2em}
\caption{A histogram representing the prevalence of each macro-topic in the knowledge and problem inquiries.}
\label{rq1:macro-topic frequency histogram}
\vspace{-1.5em}
\end{figure}

We compare the distribution of the macro-topics according to the type of inquiry and display the results in Figure~\ref{rq1:macro-topic frequency histogram}. In the subsequent sections, we explain each macro-topic by providing its definition, discussing underlying topics, and offering a snapshot of typical content through a sample post. For the sake of clarity in these samples, we abbreviate the verbose content in the posts: long descriptions are denoted by [TEXT], code and logs by [CODE], figures by [IMAGE], and hyperlinks by [URL].%Figure~\ref{rq1:macro-topic group frequency histogram} illustrates the frequency distribution of posts for each type of inquiry across these macro-topics.

% \begin{figure}[!t]
% \includegraphics[width=\columnwidth]{figures/rq1/Macro-topics group frequency histogram.pdf}
% \caption{Macro-topics group frequency histogram based on type of inquiry.}
% \label{rq1:macro-topic group frequency histogram}
% \end{figure}

\begin{longtable}{p{\linewidth}}
    \toprule
    $\hat{C}_{01}$ \textbf{Code Development} \\ 
    \midrule
    \textbf{Definition}: This macro-topic represents the posts related to creating and optimizing the scripts and code that govern the behavior of a machine learning system. It involves defining the attributes of an object ($C_{27}$), setting the parameters ($C_{28}$), and using APIs ($C_{57}$) to facilitate interoperations between different components of the software. Code development ensures a cohesive and functional codebase that can successfully drive machine learning operations and workflows. \\
    \textbf{Example}: The following example\footnote{\url{https://github.com/wandb/edu/issues/103}} is a problem inquiry from GitHub Issues. The user faces an attribute error when trying to access the \textit{WANDB\_PROJECT} attribute from the \textit{params} module in a Weights\&Biases run. \\
    $E_{01}$ \textit{When starting a new W\&B run, project=params.WANDB\_PROJECT returns AttributeError: module `params' has no attribute WANDB\_PROJECT} \\
    \\
    \toprule
    $\hat{C}_{02}$ \textbf{Code Management} \\ 
    \midrule
    \textbf{Definition}: This macro-topic represents the posts related to tracking, documenting, and monitoring changes and versions of codebases~\cite{rochkind1975source,loeliger2012version}. 
    It leverages tools such as Git ($C_{30}$), which serve as version control systems, to ensure consistency, collaboration, and traceability throughout the machine learning development lifecycle. \\ 
    \textbf{Example}: The following example\footnote{\url{https://stackoverflow.com/questions/75047065}} is a problem inquiry from Stack Overflow. The user encounters an error with DVC Git hooks when running \textit{git push}. \\ 
    $E_{02}$: \textit{DVC has Git hooks that are installed with the dvc install. The hooks were working fine, but after an error with dvc push and the DVC remote, I cannot git push because before git push gets executed, dvc push runs and generates an error. This means that I cannot push. How can I disable the DVC Git hooks so that I will no longer face the issue?} \\ 
    \\
    \toprule
    $\hat{C}_{03}$ \textbf{Computation Management}   \\ 
    \midrule
    \textbf{Definition}: This macro-topic represents the posts related to orchestration, allocation, and oversight of computational resources and services to ensure optimal and efficient operations in machine learning workflows~\cite{mustafa2015resource,manvi2014resource}. This includes monitoring instances ($C_{19}$), compute infrastructure ($C_{24}$), servers ($C_{37}$), compute clusters ($C_{42}$), system memory ($C_{44}$), specialized CUDA memory ($C_{61}$), handling situations where resource limits ($C_{34}$) or quota requests ($C_{102}$) are exceeded. Computation management ensures that computational resources are optimally allocated, scalable, and responsive to the demands of machine learning workflows, while maintaining system stability and efficiency. \\ 
    \textbf{Example}: The following example\footnote{\url{https://github.com/Azure/azureml-examples/issues/242}} is a problem inquiry from GitHub Issues. The user faces an error when setting up a Dask cluster following the tutorial. \\ 
    $E_{03}$: \textit{When I follow the using-dask tutorial to set up a Dask cluster, I fail to create a new cluster with an odd error that comes up: [Errno 113] No route to the host. I tried searching for it, but the results on Google will come with port and routing issues, which I don't know if are directly related to the problem here.} \\
    \\
    \toprule
    $\hat{C}_{04}$ \textbf{Data Development} \\ 
    \midrule
    \textbf{Definition}: This macro-topic represents the posts related to preparing, processing, and managing data for machine learning operations~\cite{sorokin2008utility,mckinney2011pandas,isah2019survey}. It encompasses tasks such as data annotation for training models ($C_{10}$), leveraging distributed computing systems such as Spark for large-scale data processing and analytics ($C_{13}$), manipulating the structure or values of dataframes ($C_{22}$), and using libraries such as pandas for versatile data manipulation and analysis ($C_{29}$). Data development ensures that data are properly curated, transformed and ready for subsequent machine learning tasks. \\ 
    \textbf{Example}: The following example\footnote{\url{https://github.com/aws/amazon-sagemaker-examples/issues/698}} is a problem inquiry from GitHub Issues. The user faces an error related to a mismatch between the number of labels in the multi-label data and the \textit{num\_classes} parameter. \\ 
    $E_{04}$: \textit{I am getting an error while running the sagemaker image classification multilabel example: Failed Reason: ClientError: Mismatch between the number of labels in the multilabel data and the num\_classes parameter.}  \\ 
    \\
    \toprule
    $\hat{C}_{05}$ \textbf{Data Management} \\ 
    \midrule
    \textbf{Definition}: This macro-topic represents the posts related to storing, orchestrating, retrieving, sharing, and connecting various data resources and artifacts throughout the machine learning lifecycle~\cite{rigby2009collaboration,barrak2021co}. It encompasses the use of buckets for storage ($C_{33}$), data factories ($C_{35}$), managing ML features in feature stores ($C_{38}$), downloading artifacts ($C_{54}$), using blob storage for unstructured data ($C_{67}$), exporting ($C_{75}$) and importing ($C_{117}$) data, organizing data in datasets ($C_{76}$), reading image formats ($C_{78}$), managing various artifacts produced during ML workflows ($C_{79}$), utilizing data stores for ML data ($C_{89}$) and connecting to relational database systems ($C_{131}$) while benefiting from shared cache storage ($C_{125}$). Data management ensures efficient storage, retrieval, and flow of data and artifacts, facilitating seamless machine learning operations and workflows. \\ 
    \textbf{Example}: The following example\footnote{\url{https://github.com/aws-samples/sagemaker-ssh-helper/issues/28}} is a problem inquiry from GitHub Issues. The user faces an invalid bucket name error when running \textit{sm-local} scripts and using the SageMaker defaults from a configuration file. \\ 
    $E_{05}$: \textit{Issue: Invalid bucket name ``sagemaker.config INFO - Fetched default config from location'': The name of the bucket must match the regex. This error message appears when a user executes `sm-local' scripts and uses SageMaker defaults to set values from the configuration file.}  \\ 
    \\
    \toprule
    $\hat{C}_{06}$ \textbf{Environment Management} \\ 
    \midrule
    \textbf{Definition}: This macro-topic represents the posts related to creating, configuring, deploying, and maintaining the various software and infrastructure elements necessary for the development, training, and deployment of machine learning models~\cite{belguidoum2007dependency,esparrachiari2018tracking}. It encompasses the management of notebook instances ($C_{02}$), Docker images ($C_{03}$), software versions ($C_{04}$), studios for ML development ($C_{07}$), RStudio IDE ($C_{14}$), workspaces ($C_{15}$), software configurations ($C_{39}$), initialization processes ($C_{40}$), Python interpreter ($C_{48}$), library imports ($C_{52}$), Terraform infrastructure creation ($C_{60}$), software configurations ($C_{64}$), module usage ($C_{87}$), deployments via Kubernetes ($C_{88}$), handling missing modules ($C_{96}$), package installations ($C_{97}$), notebook imports ($C_{99}$), software installations ($C_{101}$), training container configurations ($C_{105}$), container registries ($C_{107}$), global YAML configurations ($C_{110}$), software dependencies ($C_{115}$), and interactive Jupyter Notebook environments ($C_{123}$). Environment management ensures the integration and operation of tools, platforms, and configurations in the ML development lifecycle. \\ 
    \textbf{Example}: The following example\footnote{\url{https://stackoverflow.com/questions/50441181}} is a knowledge inquiry from Stack Overflow. The user inquires about the duties of ensuring consistency and updating software packages in a serverless compute environment such as SageMaker. \\ 
    $E_{06}$: \textit{I am learning AWS SageMaker which is supposed to be a serverless computing environment for machine learning. In this type of serverless compute environment, who is supposed to ensure the consistency of the software package and update the versions?} \\ 
    \\
    \toprule
    $\hat{C}_{07}$ \textbf{Experiment Management} \\ 
    \midrule
    \textbf{Definition}: This macro-topic represents the posts related to designing and executing controlled tests to collect data, evaluate models or hypotheses, and efficiently handle situations where multiple runs or executions overlap or nest in each other~\cite{idowu2021asset}. It encompasses the structured process of conducting controlled tests or investigations ($C_{18}$) and managing conditions where multiple runs or executions are intertwined ($C_{116}$). Experiment management streamlines the execution and evaluation of controlled tests while handling overlapping or nested runs in the ML development lifecycle. \\ 
    \textbf{Example}: The following example\footnote{\url{https://stackoverflow.com/questions/60088889}} is a problem inquiry from Stack Overflow. The user faces an error when creating a new experiment after deleting one with the same name. \\ 
    $E_{07}$: \textit{[TEXT] I am using Mlflow w/ backend Postgres db. Here is what I have run: [CODE] This deletes the experiment, but when I run a new experiment with the same name as the one I just deleted, it will return this error: [CODE]}  \\ 
    \\
    \toprule
    $\hat{C}_{08}$ \textbf{File Management} \\ 
    \midrule
    \textbf{Definition}: This macro-topic represents the posts related to the handling, storage, conversion, and manipulation of files~\cite{agrawal2007five,lu2013study}. It encompasses the organization of file directories ($C_{26}$), the conversion of file formats ($C_{36}$), the uploading ($C_{47}$) and downloading ($C_{72}$) files, the reading of files ($C_{55}$), the storage solutions ($C_{59}$), the serialization and deserialization of files ($C_{70}$), the management of input and output results ($C_{121}$) and the handling of training files ($C_{129}$). File management ensures efficient storage, retrieval, and manipulation of data and models throughout the machine learning lifecycle. \\ 
    \textbf{Example}: The following example\footnote{\url{https://stackoverflow.com/questions/56269391}} is a problem inquiry from Stack Overflow. The user cannot read the files into the DVC control when they are missing from the remote. \\ 
    $E_{08}$: \textit{I ran into a problem with DVC when some files are missing on the remote. [TEXT] How can I remove files from the DVC control and add them again?} \\ %For example, when I execute the DVC pull, I get the output: [CODE]. DVC status gives me: [CODE] I have access to the original mp4 files and I have tried to add them. I copy the mp4 files to the data folder and execute: [CODE] But there is no effect. 
    \\
    \toprule
    $\hat{C}_{09}$ \textbf{Model Deployment} \\ 
    \midrule
    \textbf{Definition}: This macro-topic represents the posts related to integrating a trained machine learning model into an application or system to make real-time predictions or decisions~\cite{paleyes2022challenges,baier2019challenges}. It encompasses prediction ($C_{09}$), deploying a model ($C_{21}$, $C_{53}$, $C_{80}$, $C_{94}$), offering it through web services ($C_{17}$), setting up inference ($C_{23}$, $C_{130}$) and model endpoints ($C_{43}$), deploying such endpoints ($C_{50}$, $C_{118}$), batch prediction ($C_{56}$), invoking endpoints ($C_{66}$, $C_{119}$), leveraging serverless platforms such as Endpoint Lambda ($C_{71}$) and ACI ($C_{111}$), and scoring ($C_{84}$) or serving the model ($C_{132}$) for production use. Model deployment ensures that trained machine learning models are readily accessible and functional in production environments for real-time or batch predictions. \\ 
    \textbf{Example}: The following example\footnote{\url{https://github.com/huggingface/transformers/issues/13111}} is a problem inquiry from GitHub Issues. The user faces an error when trying to deploy it to SageMaker endpoints. \\ 
    $E_{09}$: \textit{I have trained and saved a BertForSequenceClassification model to S3. I then used this notebook [URL] to deploy the model to SageMaker Endpoints. I ran: [CODE] I get the following error: [CODE]} \\ 
    \\
    \toprule
    $\hat{C}_{10}$ \textbf{Model Development} \\ 
    \midrule
    \textbf{Definition}: This macro-topic represents the posts related to training, tuning, and optimizing machine learning models~\cite{nguyen2019machine,wang2019various,bahrampour2015comparative}. It comprises the use of ML frameworks such as scikit-learn ($C_{114}$), PyTorch ($C_{08}$), and TensorFlow ($C_{12}$), paired with optimization methods including fine-tuning ($C_{133}$), hyperparameter adjustments ($C_{11}$), and sweeps ($C_{20}$), while undertaking tasks such as training jobs ($C_{31}$) and distributed training ($C_{46}$), and focusing on both general training ($C_{83}$, $C_{100}$) and specialized training of YOLO models ($C_{126}$). Model development ensures the creation of robust and efficient machine learning models that are trained, optimized, and validated using best practices and tools. \\ 
    \textbf{Example}: The following example\footnote{\url{https://github.com/aws/amazon-sagemaker-examples/issues/670}} is a problem inquiry from GitHub Issues. The user faces a \textit{No such file or directory} error when running a training job for a custom algorithm following the tutorial. \\ 
    $E_{10}$: \textit{I am trying to create a custom algorithm following the instructions given in this tutorial [URL]. When I am running the training job, it is failing with the error No such file or directory. As per the documentation, SageMaker should create these documents and copy the data and artifacts during the runtime. But this is not happening.} \\ 
    \\
    \toprule
    $\hat{C}_{11}$ \textbf{Model Management} \\ 
    \midrule
    \textbf{Definition}: This macro-topic represents the posts related to creating, saving, registering, loading, storing, compiling, and exchanging machine learning models, including their different formats, states, and associated metadata~\cite{jiang2023empirical,xiu2020exploratory}. It encompasses creation ($C_{82}$), storage ($C_{32}$), registration ($C_{62}$), loading ($C_{69}$), centralized storage in registries ($C_{77}$), compilation in NEO format ($C_{103}$), utilization of specialized frameworks such as Huggingface ($C_{106}$), exchange through formats such as ONNX ($C_{108}$), general use of trained or pre-trained models ($C_{109}$), distinction of trained models ($C_{112}$) and checkpointing of models during training ($C_{127}$). Model management ensures the systematic development, storage, optimization, and tracking of machine learning models throughout their lifecycle. \\ 
    \textbf{Example}: The following example\footnote{\url{https://stackoverflow.com/questions/74257398}} is a knowledge inquiry from Stack Overflow. The user asks about retrieving weights from the best performing Keras model in Optuna. \\ 
    $E_{11}$: \textit{Is there a way to pull the weights from the Keras model that performs the best and was created using an Optuna study? [TEXT]} \\ % The model with which I am working is a fully connected network with dense layers. The studies are called using the traditional method: [CODE] I can supply any additional code that might be necessary
    \\
    \toprule
    $\hat{C}_{12}$ \textbf{Network Management} \\ 
    \midrule
    \textbf{Definition}: This macro-topic represents the posts related to the establishment, monitoring, and optimization of communication links and interactions related to the endpoints of the machine learning model~\cite{xia2014survey,kreutz2014software}. It involves addressing broken links ($C_{73}$), managing endpoint content ($C_{81}$), conducting endpoint ping tests ($C_{85}$), establishing connections ($C_{104}$), utilizing web interfaces ($C_{113}$), receiving response endpoints ($C_{124}$), and overseeing endpoint transition ($C_{128}$). Network management ensures seamless integration, availability, and responsive interactions of the machine learning model endpoints in the ML development lifecycle. \\ 
    \textbf{Example}: The following example\footnote{\url{https://github.com/allegroai/clearml-server/issues/201}} is a knowledge inquiry from GitHub Issues. The user inquires about changing the default API server port in ClearML and updating the system with new credentials. \\ 
    $E_{12}$: \textit{I would like to change the default apiserver port from 8008 to 18008. I have changed it to docker-compose.yml. The ClearML app runs without problems. But when I create new credentials, it gives me an old port there. It would be great if I could change this port so it would show up when creating credentials. How to achieve it?} \\ 
    \\
    \toprule
    $\hat{C}_{13}$ \textbf{Observability Management} \\ 
    \midrule
    \textbf{Definition}: This macro-topic represents the posts related to tools, processes, and practices designed to record, monitor, compare, and visualize events, metrics, and performance related to machine learning models throughout their lifecycle~\cite{goniwada2022observability,klaise2020monitoring}. It encompasses the recording of events through logs ($C_{05}$), the usage of custom metrics ($C_{41}$), visualization with TensorBoard ($C_{49}$), logging performance metrics ($C_{63}$), benchmarking model performance ($C_{68}$), generating training reports ($C_{91}$), logging model specifics ($C_{92}$), continuous model monitoring ($C_{93}$) and integration of logs in Amazon CloudWatch ($C_{98}$). Observability management ensures continuous, transparent, and efficient monitoring and analysis of machine learning models, fostering improved performance, reliability, and insights. \\ 
    \textbf{Example}: The following example\footnote{\url{https://stackoverflow.com/questions/71505796}} is a knowledge inquiry from Stack Overflow. The user faces an issue with a binary classification metric evaluation function that leads to inaccurate recall metrics. \\ 
    $E_{13}$: \textit{I was trying to assess the metrics for binary classification and got a very strange result. It appears that eval\_and\_log\_metrics will not be able to tell which one is the true positive in the y\_test. So it just averages out true positive and true negative, which gave inflated recall metrics. How to pass the true positive value to the eval\_and\_log\_metrics?} \\ 
    \\
    \toprule
    $\hat{C}_{14}$ \textbf{Pipeline Management} \\ 
    \midrule
    \textbf{Definition}: This macro-topic represents the posts related to the orchestration and execution of tasks or steps in a machine learning process~\cite{schlegel2023management,xin2021production}. It orchestrates the comprehensive progression of a project through defined phases such as individual pipeline steps ($C_{01}$), handling large data through batch transform ($C_{16}$), mitigating issues such as Stuck Queue, incorporating inference pipeline strategies ($C_{65}$), improving efficiency through parallelization jobs ($C_{86}$), defining pipeline training steps ($C_{95}$), establishing lifecycle configuration ($C_{120}$), and integrating custom job tasks ($C_{122}$) to foster streamlined, effective and robust machine learning operations. Pipeline management ensures a structured and efficient approach to handling machine learning projects from start to finish. 
    \textbf{Example}: The following example\footnote{\url{https://github.com/Azure/MachineLearningNotebooks/issues/1927}} is a knowledge inquiry from GitHub Issues. The user asks about retrieving information from existing pipelines, including their scheduling and metadata, using the Azure Machine Learning Python SDK. \\ 
    $E_{14}$: \textit{I use Azure machine learning services to run an ml-model. When I go to the GUI of AML I can see all the existing pipelines, but I cannot see how they are scheduled. I need to obtain all published pipelines and the associated metadata. How can I get information about an existing pipeline with the Python SDK?} \\ 
    \\
    \toprule
    $\hat{C}_{15}$ \textbf{Security Management} \\ 
    \midrule
    \textbf{Definition}: This macro-topic represents the posts related to the control and protection of machine learning resources and services~\cite{soomro2016information,werlinger2009integrated}. It encompasses the structured utilization and control over user accounts ($C_{25}$), leveraging authentication processes ($C_{51}$) for verified access, ensuring secure API interactions through API keys ($C_{58}$), controlling access through permission authorizations ($C_{74}$) and handling instances of permission denial ($C_{90}$) to safeguard resources and services. Security management serves as the guardian in MLOps, orchestrating user account setups, verifying identities, and facilitating secure API accesses while meticulously granting permissions to prevent unauthorized operations and access. \\ 
    \textbf{Example}: The following example\footnote{\url{https://stackoverflow.com/questions/64039980}} is a problem inquiry from Stack Overflow. The user faces a permission error when deploying a TensorFlow model using Sagemaker. \\ 
    $E_{15}$: \textit{I can train my model using the Sagemaker TensorFlow container. Below is the [CODE]. But when I try to deploy the model, I get the following error: UnexpectedStatusException: Error hosting endpoint: Failed. Reason: Make sure that all images included in the model for the production variant AllTraffic exist and that the execution role used to create the model has permission to access them.} \\ 
    \\
    \toprule
    $\hat{C}_{16}$ \textbf{User Interface Management} \\ 
    \midrule
    \textbf{Definition}: This macro-topic represents the posts related to design, implementation, and maintenance of visual tools and dashboards that enable users to interact with and interpret machine learning systems, models, and data. It encompasses the creation and management of visual representations of data or information, such as plots ($C_{06}$). User interface management enhances user experience by leveraging visual tools to facilitate effective and intuitive data representation and analysis. \\ 
    \textbf{Example}: The following example\footnote{\url{https://community.wandb.ai/t/vega-code/4605}} illustrates a knowledge inquiry from Weights\&Biases discussion forum. The user inquires about the accessibility of the underlying source code for a Vega or Vega-lite visualization in a Weights\&Biases panel. \\ 
    $E_{16}$: \textit{Is it possible to see the underlying source code of Vega or Vega-lite for a particular panel on wandb?} \\ 
    % \bottomrule
\end{longtable}

% [$\hat{C}_{17}$] \textbf{Quality Assurance Management}: This macro-topic describes the process of evaluating or verifying the correctness, functionality, or performance of a system, model, or process. The following example illustrates a knowledge inquiry about the advantages of a GPU instance. The accepted response recommends that the poster perform further manual verification.
% $E_{17}$: \textit{Can H2o AutoML benefit from a GPU instance on Sagemaker platform? [TEXT]}

\subsection{Prevalence of posts based on macro-topics}
\vspace{-0.5em} 

We find that \textbf{Environment Management ($\hat{C}_{06}$) is the most prevalent macro-topic among the collected posts, constituting $18.89\%$ of the total posts.}  The high prevalence indicates that ensuring a consistent and controlled environment is challenging for ML practitioners throughout ML asset management. ML models are particularly sensitive to their design, training, and deployment environments\footnote{\url{https://neptune.ai/blog/ml-model-packaging}}. Even minor differences in framework versions, library dependencies, or hardware configurations can result in performance inconsistencies or failures~\cite{enck2022top}. 

Moreover, we find that \textbf{Model Deployment ($\hat{C}_{11}$) ranks as the second most prevalent macro-topic at $10.59\%$, with Model Development ($\hat{C}_{09}$) closely following as the third most prevalent macro-topic at $9\%$.} The prominence of Model Deployment ($\hat{C}_{11}$) indicates that a substantial part of the challenges and focus in ML asset management is on transitioning models from development or testing phases to production. The absence of well-defined routes from development to production exacerbates this issue, 
and is connected to an alarming AI project failure rate of nearly 50\%\footnote{\url{https://aws.amazon.com/blogs/startups/scaling-ai-ml-and-accelerating-ai-development-with-anyscale-and-aws/}}. Similarly, the notable prevalence of model development emphasizes the importance of core activities in the building of ML models. Activities such as selecting appropriate algorithms, feature engineering, and model training remain pivotal areas of discussion in the ML development lifecycle~\cite{giray2021software,baier2019challenges}.

Lastly, we observe that \textbf{Code Management ($\hat{C}_{02}$) remains the least prevalent macro-topic, accounting for only $0.74\%$ of the posts}. This may be due to the perception of source code management as well-established, along with the specialized nature of the forums related to ML asset management, \ie, discussions about code management are typically found on other platforms. For instance, the tag ``Git'' on Stack Overflow serves as a lively venue for discussions on this topic, underlining the trend of these dialogues migrating to spaces specifically dedicated to code management.

\subsection{Mapping between macro-topics and tools}
\vspace{-0.5em} 

\begin{figure}[!t]
\includegraphics[width=\columnwidth]{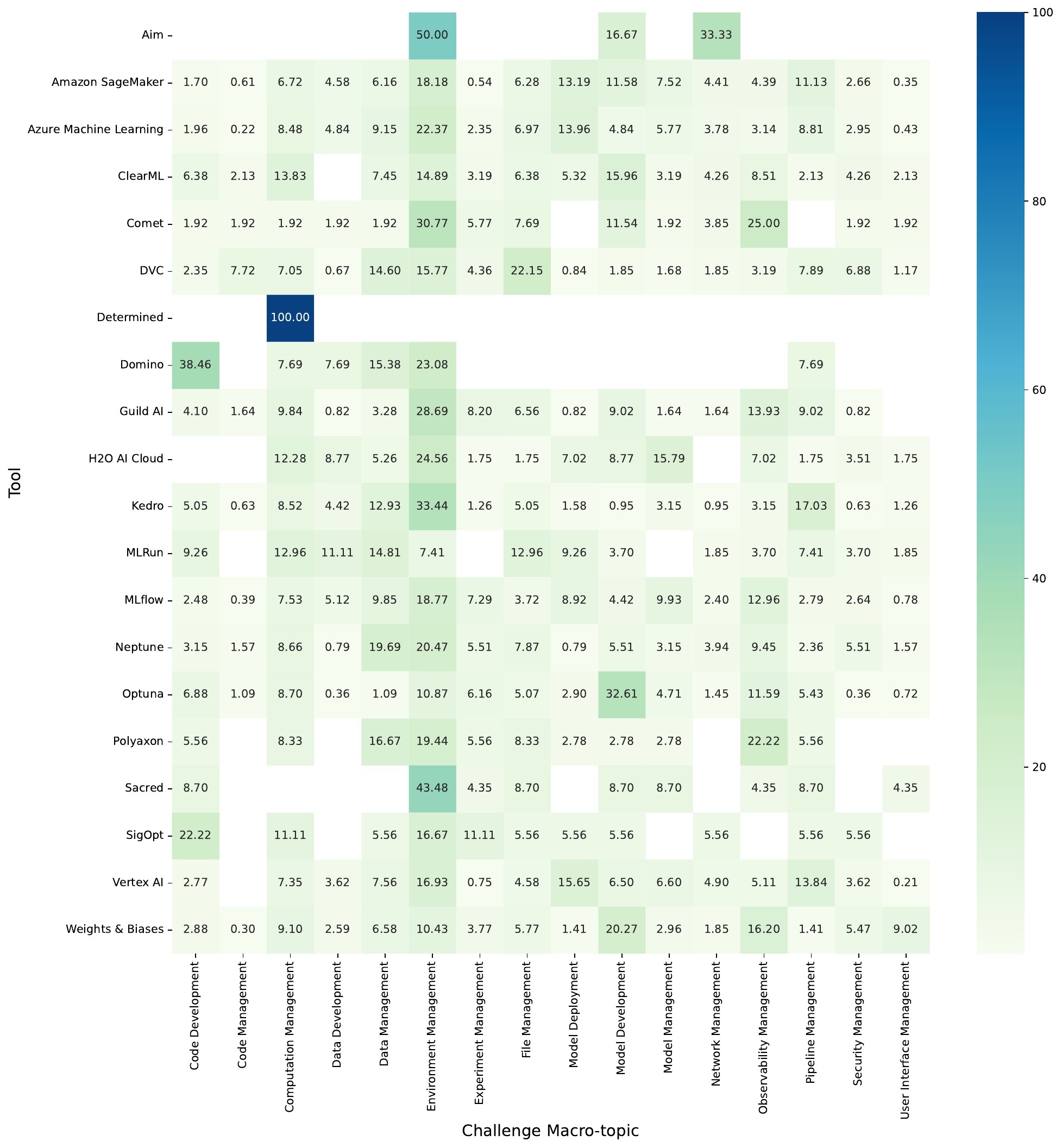}
\vspace{-2em}
\caption{Mapping between tools and challenge macro-topics based on the number of posts (Normalized across challenge macro-topics).}
\label{tool-challenge:tool}
\vspace{-1.5em}
\end{figure}

In Figure~\ref{tool-challenge:tool}, we present a heatmap showing the relationship between various tools and challenge macro-topics. Each block in the heatmap represents the percentage of posts associated with a specific tool to each of the challenge macro-topics. One key finding from the heatmap is that the \textbf{$55\%$ (11 out of 20) tools are primarily associated with inquiries related to Environment Management ($\hat{C}_{06}$).} This list includes Aim ($50.00\%$), AzureML ($22.37\%$), Comet ($30.77\%$), Guild AI ($28.69\%$), H20 AI Cloud ($24.56\%$), Kedro ($33.44\%$), MLflow ($18.77\%$), Neptune ($20.47\%$), Sacred ($43.48\%$), SageMaker ($18.18\%$) and Vertex AI ($16.93\%$). This trend suggests a strong focus of posts on managing the software environment and dependencies. The diverse ways in which different tools handle environment and dependency management underline the unique integration challenges each encounters. 

In contrast, \textbf{Domino stands out with a significant $38.46\%$ of user inquiries related to Code Development ($\hat{C}_{01}$).} This emphasizes the challenges Domino users face when creating ML-related components or features. Similarly, \textbf{Optuna ($32.61\%$) and Weights\&Biases ($20.27\%$) predominantly receive inquiries about Model Development ($\hat{C}_{10}$)}, pointing to the difficulties users face when developing ML models using these tools. An outlier is Determined, which is exclusively related to Computation Management ($\hat{C}_{03}$). However, this association might not be indicative of a broader trend, as only a single post corresponds to this tool in our collected posts. This suggests that challenges related to Determined might be underrepresented in our survey.

\begin{footnotesize}
\begin{mybox}{Summary of RQ1}
\begin{itemize}
    \item Problem inquiries (59.42\%) appear more frequently than knowledge inquiries (40.58\%) in Q\&A posts related to machine learning asset management.
    \item Environment Management ($18.89\%$), Model Deployment ($10.59\%$) and Model Development ($9\%$) are the most discussed topics in related posts to machine learning asset management, indicating a strong emphasis on the integration, training, and serving of machine learning models in practical applications.
    % \item For a significant $81.25\%$ of the macro-topics, the majority of inquiries are directed towards AzureML and SageMaker. This underscores their pivotal roles as holistic, end-to-end ML solution platforms, indicating their widespread use and perhaps areas where they might need further refinement or clearer documentation.
    \item $55\%$ of the tools receive the majority of their inquiries related to Environment Management. This pattern indicates a prevalent challenge in the ability of these tools to seamlessly integrate with other software platforms, highlighting an area for improvement and development.
\end{itemize}
\end{mybox}
\end{footnotesize}

%% file: tables/rq1/rq1-macro-topic-list.tex
\begin{table}[!t]
\centering
\caption{Macro-topic list of the identified challenges.}
\vspace{-0.5em}
\begin{tabular}{lllp{12em}}
\toprule
\textbf{Index} & \textbf{Name} & \textbf{Prevalence (\%)} & \textbf{Topic List} \\
% \cmidrule{1-2} \cmidrule{3-6}
\midrule
$\hat{C}_{01}$ & Code Development & 2.42 & [27, 28, 57] \\
\midrule
$\hat{C}_{02}$ & Code Management & 0.74 & [30] \\
\midrule
\multirow{2}{*}{$\hat{C}_{03}$} & \multirow{2}{*}{Computation Management} & \multirow{2}{*}{7.82} & [19, 24, 34, 37, 42, 44, 61, 102] \\
\midrule
$\hat{C}_{04}$ & Data Development & 4.12 & [10, 13, 22, 29] \\
\midrule
\multirow{2}{*}{$\hat{C}_{05}$} & \multirow{2}{*}{Data Management} & \multirow{2}{*}{8.11} & [33, 35, 38, 54, 67, 75, 76, 78, 79, 89, 117, 125, 131] \\
\midrule
\multirow{3}{*}{$\hat{C}_{06}$} & \multirow{3}{*}{Environment Management} & \multirow{3}{*}{18.89} & [2, 3, 4, 7, 14, 15, 39, 40, 48, 52, 60, 64, 87, 88, 96, 97, 99, 101, 105, 107, 110, 115, 123] \\
\midrule
$\hat{C}_{07}$ & Experiment Management & 2.52 & [18, 116] \\
\midrule
\multirow{2}{*}{$\hat{C}_{08}$} & \multirow{2}{*}{File Management} & \multirow{2}{*}{6.79} & [26, 36, 47, 55, 59, 70, 72, 121, 129] \\
\midrule
\multirow{3}{*}{$\hat{C}_{09}$} & \multirow{3}{*}{Model Deployment} & \multirow{3}{*}{10.59} & [9, 17, 21, 23, 43, 50, 53, 56, 66, 71, 80, 84, 94, 111, 118, 119, 130, 132] \\
\midrule
\multirow{2}{*}{$\hat{C}_{10}$} & \multirow{2}{*}{Model Development} & \multirow{2}{*}{9.00} & [8, 11, 12, 20, 31, 46, 83, 100, 114, 126, 133] \\
\midrule
\multirow{2}{*}{$\hat{C}_{11}$} & \multirow{2}{*}{Model Management} & \multirow{2}{*}{6.13} & [32, 62, 69, 77, 82, 103, 106, 108, 109, 112, 127] \\
\midrule
\multirow{2}{*}{$\hat{C}_{12}$} & \multirow{2}{*}{Network Management} & \multirow{2}{*}{3.49} & [73, 81, 85, 104, 113, 124, 128] \\
\midrule
\multirow{2}{*}{$\hat{C}_{13}$} & \multirow{2}{*}{Observability Management} & \multirow{2}{*}{6.29} & [5, 41, 49, 63, 68, 91, 92, 93, 98] \\
\midrule
\multirow{2}{*}{$\hat{C}_{14}$} & \multirow{2}{*}{Pipeline Management} & \multirow{2}{*}{8.51} & [1, 16, 45, 65, 86, 95, 120, 122] \\
\midrule
$\hat{C}_{15}$ & Security Management & 3.20 & [25, 51, 58, 74, 90] \\
\midrule
$\hat{C}_{16}$ & User Interface Management & 1.36 & [6] \\
\bottomrule
\end{tabular}
\label{rq1:macro-topic list}
\vspace{-1.3em}
\end{table}

%% file: sections/5-rq2-results.tex
\section{RQ2 Results: Prevalence of Solution Topics}
\label{sec:rq2-results}
\vspace{-0.5em}

In RQ2, we examine the solutions proposed by ML practitioners to address the challenges in ML asset management. We define these topics, offer illustrative examples, and compare their prevalence. In addition, we analyze the distribution of solutions by topic and type of inquiry. Lastly, linking challenges to their respective solution topics, we seek to find the prevailing patterns of solutions in ML asset management.

\subsection{Macro-topics of the solutions}
\label{sec:rq2-results:solution macro-topic}
\vspace{-0.5em} 

We use open card sorting to categorize solved inquiries, as detailed in Step~3.1. The frequency of each type of inquiry is presented in Table \ref{tab:types of solved inquiry}. In particular, $9.86\%$ ($159$ in total) GitHub issues are marked as ``closed'' without any solution found in their posts. This might indicate that these issues may be of low priority or may not have a significant impact on the software project.

\input{tables/rq2/types-of-solved-inquiry}

\textbf{We identify $79$ distinct topics from the solutions of normal inquiries} using the topic-modeling approach outlined in Step~3.3. Our optimal topic model achieves a coherence score of $0.9085$ and identifies $890$ posts as outliers. In our study, we use the prefix $\hat{R}$ to represent macro-topics and $R$ to represent regular topics. A detailed list of these topics, along with their indices and names, is provided in Table~\ref{rq2:topic list} (Appendix~\ref{sec:appendix-2}). For further descriptions of each topic, please refer to our replication packages~\cite{replication_package}.

Following the practice of RQ1, we formulate $18$ macro-topics to encapsulate the original topics. Table~\ref{rq2:macro-topic list} shows the macro-topics, their prevalence, and underlying topics. A notable observation is the naming overlap between certain macro-topics associated with solutions and those linked to challenges. This overlap signifies their shared relevance to specific aspects of asset management, underscoring the interconnectedness of challenges and solutions within the same thematic groups. %Additionally, we illustrate the frequency distribution of posts for each inquiry type across these macro-topics in Figure~\ref{rq2:macro-topic group frequency histogram}.

\input{tables/rq2/rq2-macro-topic-list}

In this section, we define each solution macro-topic and provide corresponding examples from our dataset of collected posts. Since the first $16$ solution macro-topics share identical names with their challenge counterparts, we focus exclusively on the two uniquely named solution macro-topics in this section, directing readers to the Appendix~\ref{sec:appendix-1} for the illustrations of the others.

\begin{longtable}{p{\linewidth}}
    \toprule
    $\hat{R}_{17}$ \textbf{Comparison \& Recommendation} \\ 
    \midrule
    \textbf{Definition}: Comparison \& Recommendation represents the posts related to the analytical process in which the efficacy and suitability of various tools, such as SDKs, APIs, databases, platforms, runtimes, versions, tasks, data or pipelines, are examined ($R_{82}$) and suggested ($R_{02}$) during the ML development lifecycle. Comparison and recommendation facilitate the selection of the most appropriate tools and resources regarding ML assets, thus promoting efficiency and superior outcomes through well-informed decisions in the production environment. \\
    \textbf{Example}: The following example\footnote{\url{https://stackoverflow.com/questions/72408785}} provides supported runtimes to help the user make an informed decision. \\
    $E_{17}$: \textit{Accepted Answer: MLRun has several different ways to run a piece of code. At this time, the following runtimes are supported: [TEXT]} \\
    \\
    \toprule
    $\hat{R}_{18}$ \textbf{Maintenance \& Support} \\ 
    \midrule
    \textbf{Definition}: Maintenance \& Support represents the posts related to the continuous process of updating and improving machine learning systems by implementing patches or fixes ($R_{08}$), and managing operating issues through support tickets ($R_{46}$)~\cite{grubb2003software,kitchenham1999towards}. It ensures the stability, performance, and effective communication of machine learning-based systems. \\
    \textbf{Example}: The following example\footnote{\url{https://stackoverflow.com/questions/73435172}} provides the development plan for the DVC-Hydra integration support. \\
    $E_{17}$: \textit{Accepted Answer: A DVC-Hydra integration is in development. You can see the proposal in [URL] and the development progress in [URL]. [TEXT]} \\
    % \bottomrule
\end{longtable}

% [$\hat{R}_{05}$] \textbf{Document Consultation} This macro-topic describes the writing, refining, and testing of the actual code that underpins ML models and related processes. The following example demonstrates a problem inquiry about verbose logs. The accepted answer implies that a hotfix has been implemented.
% $E_{17}$: \textit{Accepted Answer: I want to train YOLOv5 on aws sagemaker also deploy the model on sagemaker itself,need to know about entrypoint python script as well. how can I build a pipeline for this?}

\subsection{Prevalence of solutions based on macro-topics}
\vspace{-0.5em} 

\textbf{Environment Management ($\hat{R}_{06}$) stands out as the most prevalent ($23.31\%$) solution macro-topic related to ML asset management}, as illustrated in Table~\ref{rq2:macro-topic list}. This prominent prevalence underscores the intricacies and significance of overseeing environments and dependencies throughout ML asset management. Ensuring adept management of these ML environments and dependencies is paramount for achieving consistent and reproducible model training and deployment across diverse infrastructures. 

Following this, Code Development ($\hat{R}_{01}$) constitutes $15.35\%$ and is the second most prevalent solution macro-topic. The prominence of this macro-topic reflects the evolving complexities and nuances of crafting code specifically for ML applications, as opposed to traditional software development. This emphasizes the need for robust development practices, continuous integration, and testing methodologies tailored for ML. File Management ($\hat{R}_{08}$) ranks as the third most prevalent macro-topic with a proportion of $9.64\%$ of Q\&A posts. This prominence underscores the essential role of structured file storage, retrieval, and manipulation in the ML development lifecycle.

\subsection{Mapping between challenge and solution macro-topics }
\vspace{-0.5em} 

\begin{figure}
\includegraphics[width=\columnwidth]{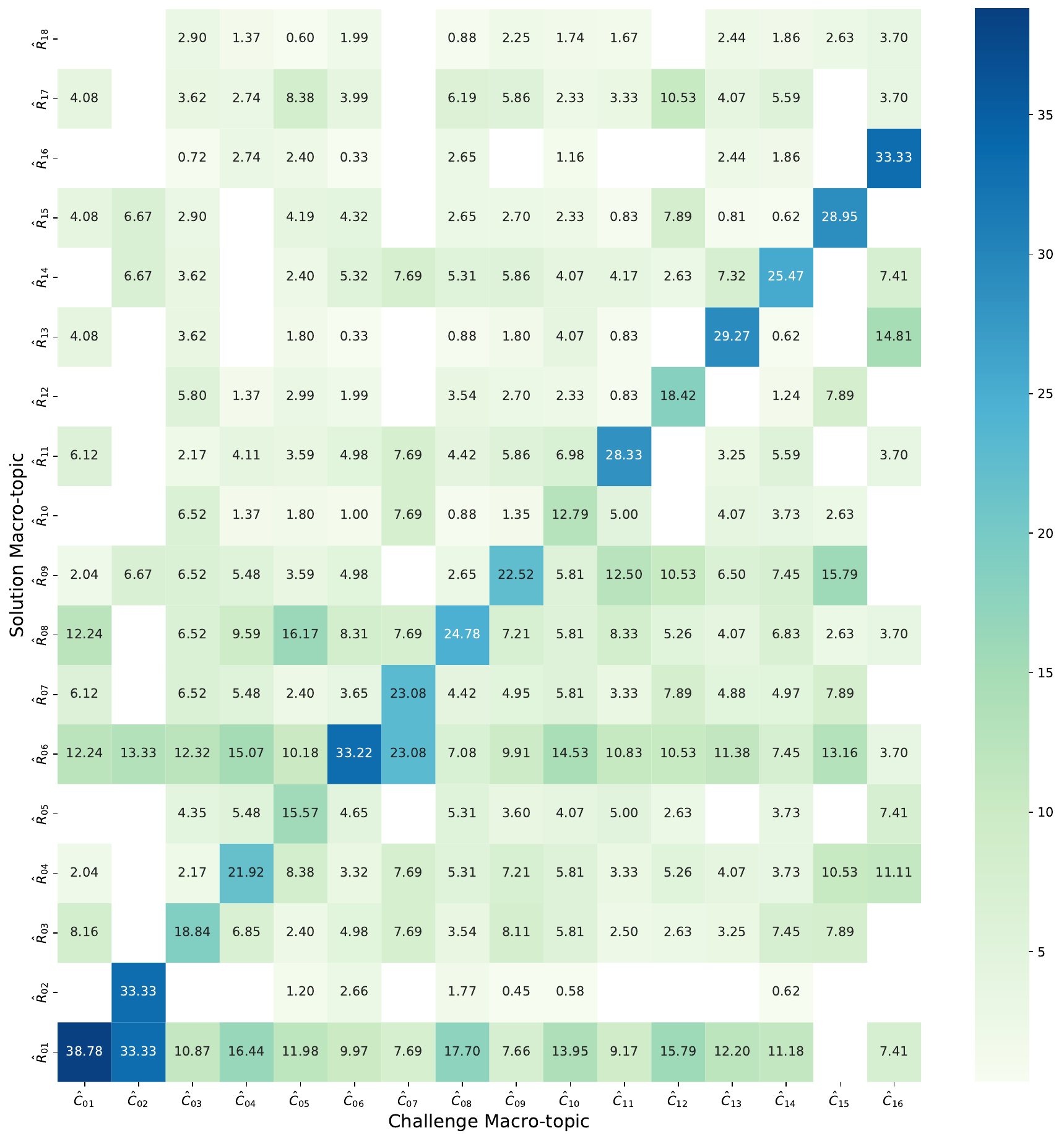}
\vspace{-2em}
\caption{Mapping of challenge marco-topics (knowledge inquiry) to solution macro-topics, showing the normalized number of posts based on post interactions.}
\label{knowledge-solution:macro-topic}
\vspace{-1.5em}
\end{figure}

\textbf{Knowledge inquiries are commonly addressed with solutions specific to their respective topics.} This correlation can be observed in Figure~\ref{knowledge-solution:macro-topic}, which shows the mapping between challenges and solution macro-topics in knowledge inquiries. Each cell in the heatmap represents the normalized number of posts corresponding to a specific challenge macro-topic across various solutions. The prominent dark blocks along the diagonal suggest that challenges in a particular domain are frequently solved in the same domain. In our study, we term this phenomenon as ``self-resolution''. For instance, inquiries related to Environment Management ($\hat{C}_{06}$) are predominantly addressed using environment-specific solutions ($\hat{R}_{06}$). To highlight the prevalence of self-resolution, we set a threshold of $25\%$ to indicate a significant rate of self-resolution. As a result, $56.25\%$ (9 out of 16) macro-topics meet this criterion, implying that the majority of knowledge inquiries rely on domain-specific knowledge from their corresponding macro-topic for solution.

Conversely, we notice a pattern in which specific types of challenge are often addressed by solutions from different areas, a phenomenon we term ``counter-self-resolution'' in our study. For example, \textbf{challenges in Code Management ($\hat{C}_{02}$) are primarily solved by methods in Code Development ($\hat{R}_{01}$).} This connection highlights that effective code management is closely related to the processes and practices in code development. Similarly, \textbf{challenges related to Experiment Management ($\hat{C}_{07}$) are primarily tackled with solutions in Environment Management ($\hat{R}_{06}$)}. This reveals that the effectiveness of experiment management depends on robust environment management practices.

% Lastly, we observe that certain solution macro-topics play a major role in addressing challenges in knowledge inquiries. Among these, \textbf{the top three most prevalent solution macro-topics are: Environment Management ($\hat{R}_{06}$) at $14.69\%$, Code Development ($\hat{R}_{01}$) at $12.15\%$, and File Management ($\hat{R}_{08}$) at $8.98\%$.} Given the findings of RQ1 and the frequent self-resolution observed in user inquiries, the prominence of Environment Management ($\hat{R}_{06}$) challenges is not unexpected. While users face different challenges managing their ML assets, addressing those in writing efficient algorithms or reorganizing their files by structures empowers them and significantly resolves their concerns. However, there is a notable disparity between the high impact of these solutions and their infrequent mention in discussions of these challenges.

\begin{figure}
\includegraphics[width=\columnwidth]{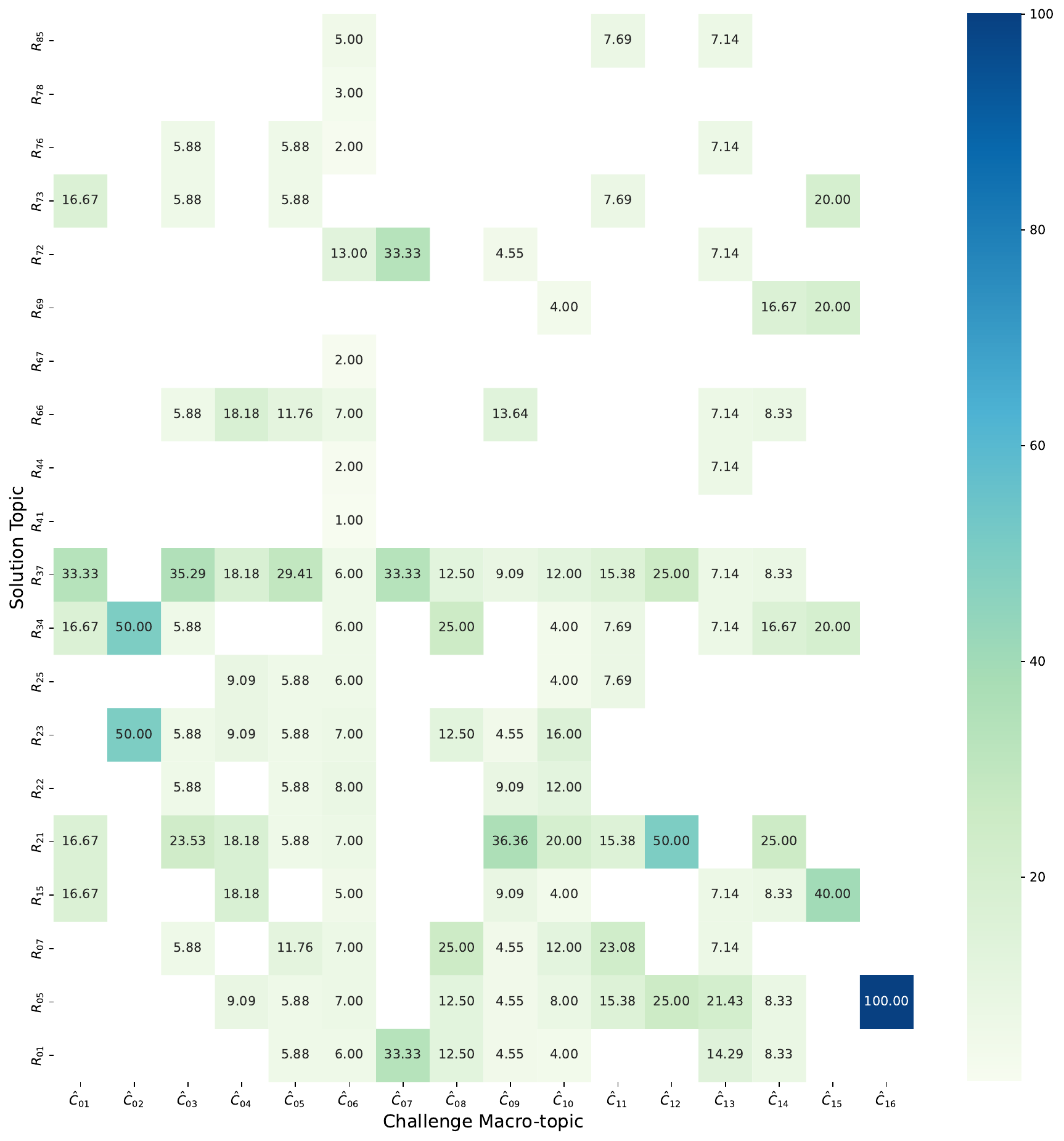}
\vspace{-2em}
\caption{Zooming down to the level of individual solution topics for Environment Management ($\hat{R}_{06}$) in Figure~\ref{knowledge-solution:macro-topic}.}
\label{knowledge-solution:topic}
\vspace{-1.5em}
\end{figure}

Lastly, since Environment Management ($\hat{R}{06}$) emerges as the most prevalent macro-topic in knowledge inquiries, we take a closer examination at the mapping of the underlying topics in this macro-topic. Figure~\ref{knowledge-solution:topic} illustrates the relationship between challenge macro-topics and the underlying solution topics in Environment Management ($\hat{R}{06}$). We observe that \textbf{Container Customization ($R_{21}$) is the most prevalent solution topic ($13.46\%$) for Environment Management ($\hat{R}_{06}$) solutions.} Specifically, it accounts for $50\%$ of the environment-oriented solutions to Network Management ($\hat{C}_{12}$) challenges. Furthermore, this figure highlights that \textbf{Package Installation ($R_{05}$) emerges as the most prevalent solution ($100\%$) for User Interface Management ($\hat{C}_{16}$).} The figure also points out that \textbf{for Code Management ($\hat{C}{02}$) challenges, Package Addition ($R_{23}$) and Environment Variable Management ($R_{34}$) are the most prevalent solutions, each accounting for $50\%$.} %The increasing emphasis on tailored solutions for environment and dependency configuration underscores the limitations of one-size-fits-all strategies. Such customized approaches better address the distinct challenges faced in various software environments, a viewpoint echoed in the prior literature~\cite{dua2014virtualization,pahl2015containerization}. 

\begin{figure}
\includegraphics[width=\columnwidth]{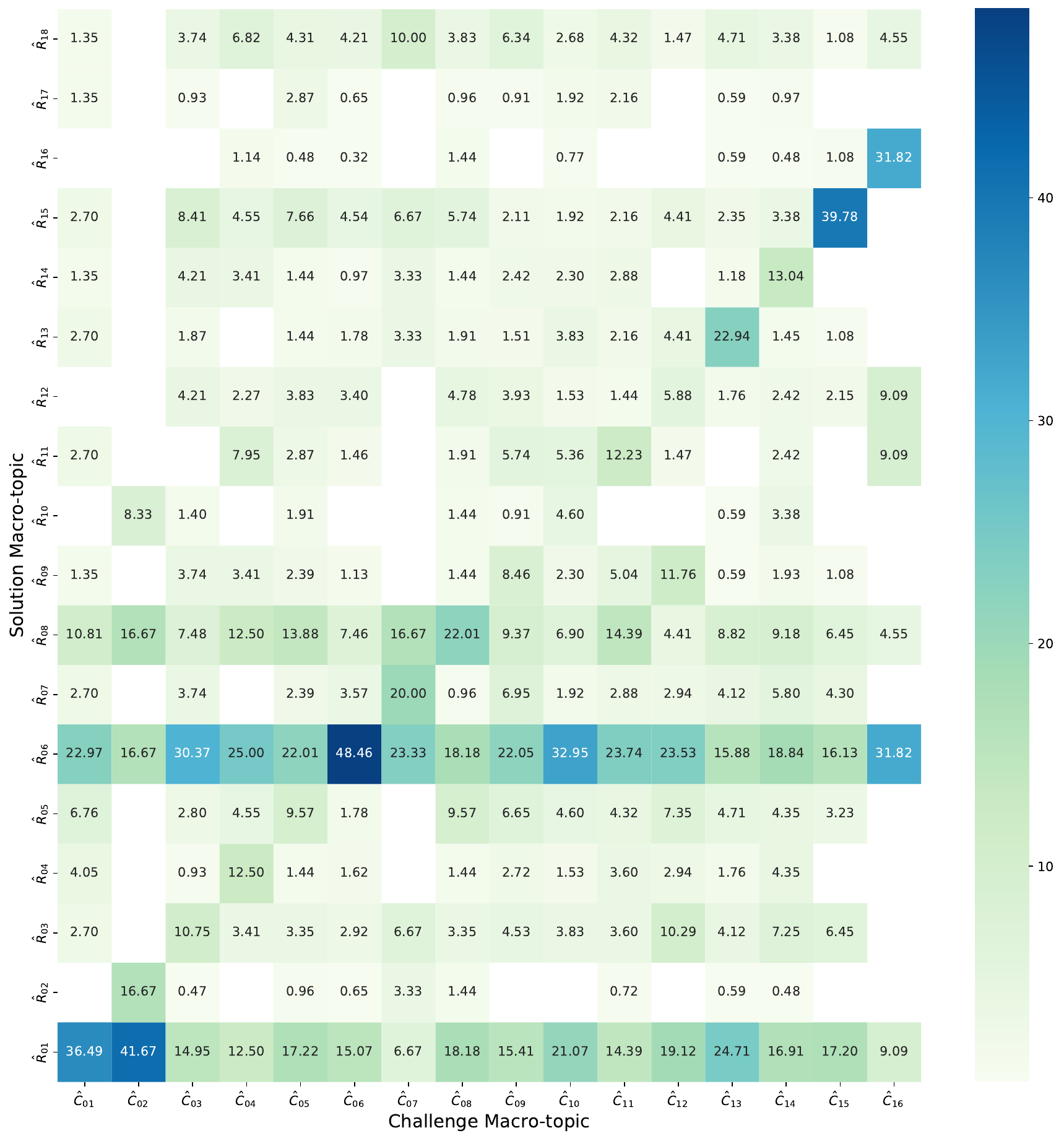}
\vspace{-2em}
\caption{Mapping of challenge marco-topics (problem inquiry) to solution macro-topics, showing the normalized number of posts based on post interactions.}
\label{problem-solution:macro-topic}
\vspace{-1.5em}
\end{figure}

\textbf{Problem inquiries are less commonly addressed through self-resolution.} This observation is evident in Figure~\ref{problem-solution:macro-topic}, where we illustrate the normalized number of posts between challenge and solution macro-topics in problem inquiries. Herein, the diagonal cells in the heatmap are lighter compared to those related to knowledge inquiries. Using a $25\%$ threshold, we observe that only $25\%$ (4 out of 16) of the macro-topics display a high level of self-resolution. This suggests a lower trend for self-resolution in problem inquiries, likely due to their inherent complexity or external, cross-domain factors influencing them. 

Conversely, we note that the counter-self-resolution trend is more pronounced in problem inquiries compared to knowledge ones. A case in point is again the resolution of Code Management ($\hat{C}_{02}$) challenges with Code Development ($\hat{R}_{01}$) methods. Furthermore, Environment Management ($\hat{R}_{06}$) emerges as the most prevalent solution macro-topic for $62.5\%$ (10 out of 16) macro-topics, including Computation Management ($\hat{C}_{03}$), Data Development ($\hat{C}_{04}$), Data Management ($\hat{C}_{05}$), Experiment Management ($\hat{C}_{07}$), Model Deployment ($\hat{C}_{09}$), Model Development ($\hat{C}_{10}$), Model Management ($\hat{C}_{11}$), Network Management ($\hat{C}_{12}$), Observability Management ($\hat{C}_{13}$) and User Interface Management ($\hat{C}_{16}$). This pattern implies that software environment and dependency management are a foundational and versatile solution, addressing a wide range of challenges on different topics. Furthermore, challenges in Observability Management ($\hat{C}_{13}$) are commonly addressed with Code Development ($\hat{R}_{01}$) solutions. This implies that the quality and effectiveness of evaluation and monitoring is significantly influenced by the methodologies and practices used in code development.

% Lastly, we similarly observe that \textbf{the three most prevalent solution macro-topics for addressing challenges in knowledge inquiries are Environment Management ($\hat{R}_{06}$) in $28.86\%$, Code Development ($\hat{R}_{01}$) in $17.42\%$, and File Development ($\hat{R}_{08}$) in $10.06\%$.} The relatively elevated percentages associated with these macro-topics in problem inquiries could signal them as primary sources of challenges. This underscores their pivotal role in both knowledge and problem inquiries. 

\begin{figure}
\includegraphics[width=\columnwidth]{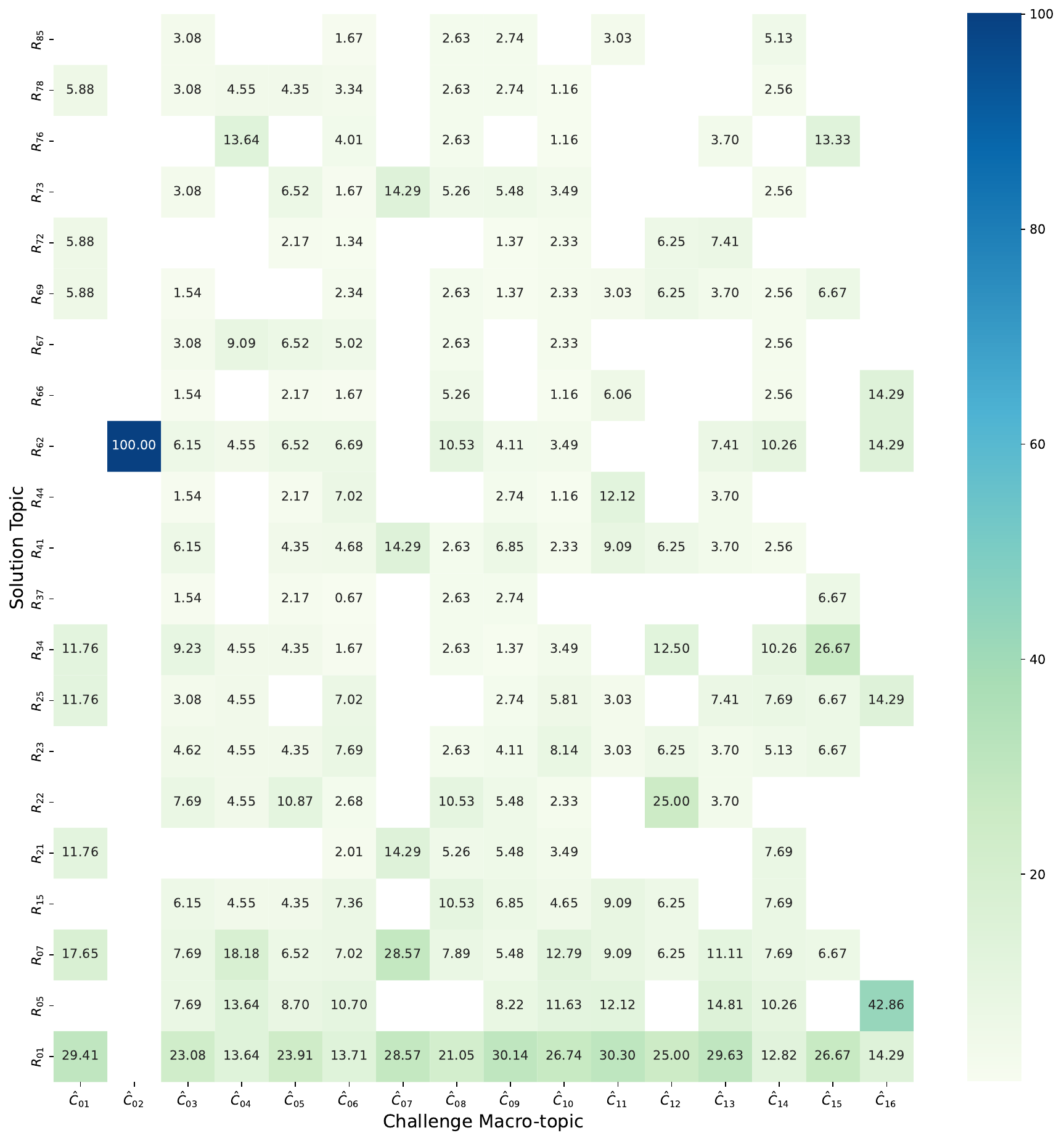}
\vspace{-2em}
\caption{Zooming down to the level of individual solution topics for Environment Management ($\hat{R}_{06}$) in Figure~\ref{problem-solution:macro-topic}.}
\label{problem-solution:topic}
\vspace{-1.5em}
\end{figure}

Lastly, since Environment Management ($\hat{R}{06}$) emerges as the most prevalent macro-topic in problem inquiries, we also examine the mapping of its underlying topics in Figure~\ref{problem-solution:topic}. In particular, \textbf{Package Upgrade ($R_{01}$), accounting for $20.45\%$ of the cases, is the most frequently encountered solution topic in the broader category of software dependency and environment, for the majority of challenge macro-topics}, with the exception of Code Management ($\hat{C}_{02}$) and User Interface Management ($\hat{C}_{16}$). This pattern indicates a significant dependence on external packages and libraries in machine learning projects. As these packages undergo updates, they may present new features, optimizations, or necessary bug fixes. Furthermore, \textbf{Package Removal ($R_{62}$) stands out as the most prevalent environment-oriented solution ($100\%$) to address the challenges related to Code Management ($\hat{C}_{02}$)}. It also reveals that \textbf{Package Installation ($R_{05}$) is the most prevalent environment-oriented solution ($42.86\%$) for User Interface Management ($\hat{C}_{16}$)}. This trend further underscores the point that package management is at the heart of environment management, maintenance, and deployment~\cite{enck2022top}. It allows developers and system administrators to define, reproduce, and share their software environments consistently and predictably.

\begin{footnotesize}
\begin{mybox}{Summary of RQ2}
\begin{itemize}
    \item Environment Management ($23.31\%$), Code Development ($15.35\%$), and File Management ($9.64\%$) are the most discussed solution topics related to machine learning asset management, indicating that these areas are the most common pain points or areas of interest for machine learning practitioners.
    \item $56.25\%$ (9 out of 16) of the challenge macro-topics in knowledge inquiries show a high-level self-resolution. In contrast, only $25\%$ (4 out of 16) of the challenge macro-topics in problem inquiries demonstrate a high self-resolution rate. This suggests that a more interconnected and cross-disciplinary approach is needed when addressing specific problems, as opposed to seeking general knowledge. 
    \item The counter-self-resolution trend is more pronounced in problem inquiries than in knowledge inquiries. This may imply that problem inquiries necessitate a broader spectrum of insights, often derived from external or cross-disciplinary domains, to devise viable solutions.
\end{itemize}
\end{mybox}
\end{footnotesize}

%% file: tables/rq2/types-of-solved-inquiry.tex
\begin{table}
\centering
\caption{Types of solved inquiries across discussion forums.}
\vspace{-0.5em}
\begin{tabular}{p{2.6cm}p{1.8cm}p{1.8cm}p{1.9cm}p{1.9cm}}
\toprule
\multirow{3}{*}{\textbf{Discussion forum}} & \parbox[t]{5cm}{\textbf{Number of \\normal \\inquiries}} & \parbox[t]{5cm}{\textbf{Number of \\ non-inquiries}} & \textbf{Number of intermittent inquiries} & \textbf{Number of non-solved inquiries} \\
\midrule
Stack Overflow & 2187 & 2 & 4 & 5 \\
\midrule
GitHub Issues & 1432 & 6 & 15 & 159 \\
\midrule
Tool-specific forums & 849 & 1 & 1 & 8 \\
\midrule
GitHub Discussions & 17 & 0 & 0 & 0 \\
\midrule
GitLab Issues & 1 & 0 & 0 & 0 \\
\bottomrule
\end{tabular}
\label{tab:types of solved inquiry}
% \vspace{-1.3em}
\end{table}

%% file: tables/rq2/rq2-macro-topic-list.tex
\begin{table}
\centering
\caption{Macro-topic list of the identified solutions.}
\vspace{-0.5em}
\begin{tabular}{lllp{10.5em}}
\toprule
\textbf{Index} & \textbf{Name} & \textbf{Prevalence (\%)} & \textbf{Topic List} \\
\midrule
\multirow{2}{*}{$\hat{R}_{01}$} & \multirow{2}{*}{Code Development} & \multirow{2}{*}{15.35} & [12, 13, 27, 28, 47, 56, 57, 58, 59, 61, 63, 84] \\
\midrule
$\hat{R}_{02}$ & Code Management & 0.80 & [43] \\
\midrule
$\hat{R}_{03}$ & Computation Management & 5.25 & [19, 30, 35, 53, 60] \\
\midrule
$\hat{R}_{04}$ & Data Development & 3.66 & [31, 33, 54] \\
\midrule
$\hat{R}_{05}$ & Data Management & 4.81 & [40, 42, 70, 75, 77] \\
\midrule
\multirow{3}{*}{$\hat{R}_{06}$} & \multirow{3}{*}{Environment Management} & \multirow{3}{*}{23.31} & [1, 5, 7, 15, 21, 22, 23, 25, 34, 37, 41, 44, 62, 66, 67, 69, 72, 73, 76, 78, 85] \\
\midrule
$\hat{R}_{07}$ & Experiment Management & 4.12 & [11, 39, 71, 83] \\
\midrule
\multirow{2}{*}{$\hat{R}_{08}$} & \multirow{2}{*}{File Management} & \multirow{2}{*}{9.64} & [16, 20, 29, 36, 49, 55, 64, 65, 68, 86] \\
\midrule
$\hat{R}_{09}$ & Model Deployment & 5.01 & [24, 26, 38, 79, 80] \\
\midrule
$\hat{R}_{10}$ & Model Development & 2.10 & [3] \\
\midrule
$\hat{R}_{11}$ & Model Management & 4.32 & [10, 32, 52] \\
\midrule
$\hat{R}_{12}$ & Network Management & 2.92 & [9, 51] \\
\midrule
$\hat{R}_{13}$ & Observability Management & 3.41 & [6, 50, 74] \\
\midrule
$\hat{R}_{14}$ & Pipeline Management & 4.08 & [14, 18, 45] \\
\midrule
$\hat{R}_{15}$ & Security Management & 4.54 & [4, 17, 81] \\
\midrule
$\hat{R}_{16}$ & User Interface Management & 1.04 & [48] \\
\midrule
$\hat{R}_{17}$ & Comparison \& Recommendation & 2.46 & [2, 82] \\
\midrule
$\hat{R}_{18}$ & Maintenance \& Support & 3.19 & [8, 46] \\
\bottomrule
\end{tabular}
\label{rq2:macro-topic list}
\vspace{-1.3em}
\end{table}

%% file: sections/6-rq3-results.tex
\section{RQ3 Results: Comparison of Discussion Forums}
\label{sec:rq3-results}
\vspace{-0.5em}

In RQ3, we explore commonalities and differences in developer discussion forums about ML asset management. Specifically, we compare Stack Overflow, repository-specific forums, and tool-specific forums. We start our analysis by examining the general prevalence metrics on various forums. This initial step gives us a broad overview of the landscape. Next, we narrow our focus to compare these metrics based on different types of inquiry in each forum. This allows us to understand how each forum performs in addressing specific types of inquiry. In the final step, we shift our attention to macro-topics in these forums, comparing the prevalence metrics once more. 

\subsection{Comparison of discussion forums in general}
\vspace{-0.5em} 

We find that \textbf{Stack Overflow is the main source of inquiries, accounting for $48.65\%$ of the total posts. Tool-specific forums are next, contributing $34.19\%$, while repository-specific forums have the fewest, at only $17.16\%$.} The prevalence of Stack Overflow posts is expected due to its popularity for developers worldwide~\cite{vasilescu2013stackoverflow}. Its highest number of posts reflects its popularity and wide reach among ML practitioners. However, the low number of posts on repository-specific forums, such as GitHub, could indicate a narrower focus of discussions, often centered on project-specific issues, bugs, or feature requests, rather than the more general questions and broad topics typically found on Stack Overflow~\cite{vadlamani2020studying,squire2015should}. %The Chi-square tests reveal no significant differences in the frequency distribution of macro-topics between any pair of forums ($p=1.0$).

\subsection{Comparison of discussion forums based on tools}
\vspace{-0.5em} 

\begin{figure}
\includegraphics[width=\columnwidth]{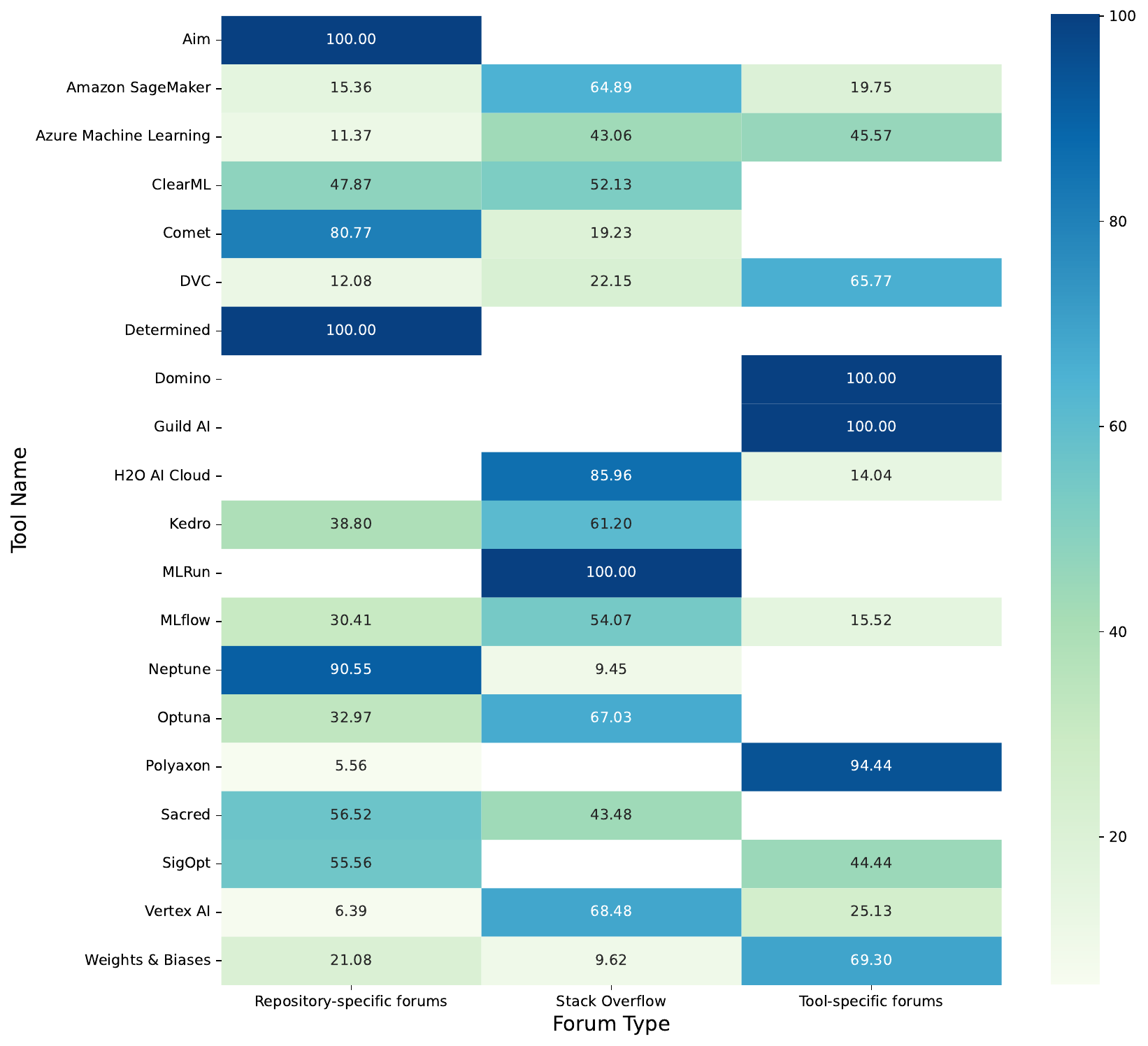}
\vspace{-2em}
\caption{Normalized distribution of Q\&A posts across different types of forum.}
\label{fig:platform tool heatmap}
\vspace{-1.5em}
\end{figure}

A distinct distribution emerges in the preferences of the types of discussion forum among users of different tools, as illustrated in Figure~\ref{fig:platform tool heatmap}. The intensity of the color in the heatmap indicates the frequency of posts for each forum-tool pairing. In particular, \textbf{$25\%$ (5 out of 20) tools}, namely DVC ($65.77\%$), Domino ($100\%$), Guild AI ($100\%$), Polyaxon ($94.44\%$), and Weights \& Biases ($69.36\%$), \textbf{have the majority of their discussions originating from tool-specific forums.} This trend could imply a positive user experience on these dedicated forums, which deserves further investigation into the factors contributing to this preference. Conversely, we observe that \textbf{$20\%$ (4 out of 20) tools}, including Aim ($100\%$), Comet ($80.77\%$), Determined ($100\%$), and Neptune ($90.55\%$), \textbf{predominantly have discussions in repository-specific forums.} This could reflect a tendency among practitioners to discuss the challenges faced in open-source projects where these tools are utilized as dependencies. Furthermore, \textbf{$35\%$ of the tools (7 out of 20)}, including Amazon SageMaker ($64.89\%$), H2O AI Cloud ($85.96\%$), Kedro ($61.20\%$), MLRun ($100\%$), MLflow ($54.07\%$), Optuna ($67.03\%$), and Vertex AI ($68.48\%$), \textbf{have the most of their discussions on Stack Overflow.} This behavior could highlight a preference to engage with a wider community or a higher popularity of Stack Overflow among developers in general. %It could also present an opportunity for these tool maintainers to improve the user experience in their forums, encouraging richer interactions.

\subsection{Comparison of discussion forums based on macro-topics}

\textbf{Environmental management ($\hat{C}_{06}$) is the most prevalent macro-topic in all forums, comprising $25.53\%$ of repository-specific issues, $18.34\%$ of Stack Overflow posts, and $16.34\%$ of discussions in tool-specific forums.} This trend reinforces the RQ1 finding that software environment and dependency management are the primary macro-topics in ML asset management. In contrast, while \textbf{Model Deployment ($\hat{C}{09}$) is the second most frequent topic on Stack Overflow ($12.49\%$) and tool-specific forums ($9.98\%$), it is ranked sixth on repository-specific forums at $6.45\%$.} Conversely, \textbf{Model Development ($\hat{C}{10}$) is second on repository-specific forums ($9.71\%$) but falls to fourth on Stack Overflow ($9.11\%$) and tool-specific forums ($8.49\%$).} 

Differences in the distribution of posts across macro-topics between any two discussion forums are significant, as evidenced by three pairwise Chi-square goodness of fit tests, conducted at a significance level of $0.05$ (Table~\ref{tab:forum comparison}). Chi-square tests are commonly used to evaluate the alignment between observed data and a hypothesized distribution~\cite{pearson1900x}. Given the multiple tests conducted, we adjust the p-values to the so-called ``q-values''~\cite{storey2002direct} using the Benjamini-Hochberg procedure to account for the False Discovery Rate (FDR)~\cite{benjamini1995controlling}. This adjustment helps mitigate overestimation of significant findings that can arise by chance and is an alternative to related approaches, such as Bonferroni correction~\cite{dunn1961multiple}, which modify the alpha value instead of the p-value.

\input{tables/rq3/forum-comparison}
\vspace{-3em}
\input{tables/rq3/cramers-v}

Following the Chi-square tests, we also calculate the Cramér's V values, which range from $0.132$ to $0.177$. Cramér's V is a metric that indicates the strength of association between two variables~\cite{cramer1999mathematical}. As depicted in Table~\ref{tab:Cramér's V}, \textbf{these values signify weak associations between the distribution of Q\&A posts and the associated forums.} Therefore, our statistical analysis implies a significant difference, but its practical implications are not pronounced. Despite this, each forum exhibits a unique pattern of macro-topics discussed. Hence, ML practitioners are still recommended to adapt their discussions to align with the distinct characteristics and preferences of each forum's audience.

\begin{footnotesize}
\begin{mybox}{Summary of RQ3}
\begin{itemize}
    \item Stack Overflow is the main source of inquiries, accounting for $48.65\%$ of the total posts. Tool-specific forums are next ($34.19\%$), while repository-specific forums have the fewest ($17.16\%$). This pattern might imply that the extensive user base and broad discussion topics make Stack Overflow the go-to platform for inquiries, while repository-specific forums, such as GitHub, are favored for project-oriented discussions.
    \item A total of $25\%$ of the analyzed tools have majority of user inquiries on tool-specific forums, $20\%$ on repository-specific forums, and $35\%$ on Stack Overflow. This pattern suggests a potential area of exploration for tool maintainers to improve user engagement on their respective forums, fostering a better collaborative community.
    \item In all types of forums, Environment Management stands out as the most prevalent topic, indicating a critical need for improved tooling, documentation, and best practices in environment setup and dependency management.
\end{itemize}
\end{mybox}
\end{footnotesize}

%% file: tables/rq3/forum-comparison.tex
\begin{table}[h]
\centering
\caption{Chi-squared test results between any two different discussion forums across macro-topics.}
\vspace{-0.7em}
\begin{tabular}{lrr}
\toprule
\textbf{Forum names} & \textbf{q-value} & \textbf{Cramér's~V} \\
\midrule
Repository-specific forums vs. Stack Overflow & 0.000 & 0.177 \\
Repository-specific forums vs. Tool-specific forums & 0.000 & 0.169 \\
Stack Overflow vs. Tool-specific forums & 0.000 & 0.132 \\
\bottomrule
\end{tabular}
\label{tab:forum comparison}
% \vspace{-1.3em}
\end{table}

%% file: tables/rq3/cramers-v.tex
\begin{table}[h]
\centering
\caption{Intepretation of Cramér's~V estimator}
% \vspace{-1em}
\begin{tabular}{cccccc}
\toprule
\textbf{Negligible} & \textbf{Weak} & \textbf{Moderate} & \textbf{Relatively strong} & \textbf{Strong} & \textbf{Very strong} \\
\midrule
$0\sim0.1$ & $0.1\sim0.2$ & $0.2\sim0.4$ & $0.4\sim0.6$ & $0.6\sim0.8$ & $0.8\sim1$ \\
\bottomrule
\end{tabular}
\label{tab:Cramér's V}
\vspace{-0.7em}
\end{table}

%% file: sections/7-implications.tex
\section{Implications}
\label{sec:implications}
\vspace{-0.5em}

Our research provides invaluable insights for a diverse group of ML practitioners, including researchers, educators, and tool/application developers in the software and machine learning domain. Through our findings, we aim to guide their future efforts toward more efficient and cohesive management of ML assets.

\subsection{Implications for Researchers}
\label{sec:implications:subsec:researchers}
\vspace{-0.5em} 

\paragraph{Prevalent Discussion Topics:} Our findings show that environment configuration, model training, and deployment are highly prevalent in asset management. Because of this, we see a great opportunity for more research on environment management systems that work well with all stages of ML development and should be flexible enough to work with different ML tools and platforms.

\paragraph{Emerging Foundation Models:} To evaluate the prevalence of discussions on foundation models~\cite{bommasani2021opportunities} (FM), we perform an analysis based on the frequency of related posts within our dataset. First, we compile a list of keywords associated with FMs from leading GitHub repositories including Awesome-LLM\footnote{\url{https://github.com/Hannibal046/Awesome-LLM}}, Open LLMs\footnote{\url{https://github.com/eugeneyan/open-llms}}, and Awesome-Multimodal-Large-Language-Models\footnote{\url{https://github.com/BradyFU/Awesome-Multimodal-Large-Language-Models}}. Our systematic search through titles and post bodies, after preprocessing and applying conditional checks to filter out irrelevant content, reveals a modest presence ($0.76\%$) of FM discussions in our dataset. Although insightful, this figure does not suffice to definitively assess the state of asset management in foundation models. The observed low frequency can be attributed to the nascent stage of local FM system deployment and the associated demand for asset management independent of FM providers, such as OpenAI\footnote{\url{https://openai.com}}. This trend began to materialize in the last half of 2023, predominantly after the cut-off date of our data collection, which ended on $26^{th} July, 2023$. In particular, the increase in local FM deployments was significantly influenced by the public release of LLaMA-2~\cite{touvron2023llama} on $18^{th} July, 2023$ and Mistral-7B~\cite{jiang2023mistral} on $27^{th} Sep, 2023$. Although there may be overlaps in asset management needs between pre-FM and FM models, future research should focus on analyzing FM-related discussion posts that have emerged since the late summer of 2023.

% \paragraph{Identifying Key Challenges:} Our study reveals that asset management inquiries have a higher unsolved rate and often take more time to resolve than those in other fields of software engineering. This finding highlights an opportunity for researchers to further explore this area. We offer 16 distinct categories that serve as a guide for this exploration. Looking into these categories, researchers identify the specific challenges in asset management. This identification enables the development of targeted solutions to tackle these challenges, improving efficiency and effectiveness in the area.

\subsection{Implications for Educators}
\label{sec:implications:subsec:educators}
\vspace{-0.5em} 

\paragraph{MLOps Curriculum Design:} Our research reveals that ML practitioners often face challenges in the software environment and dependency management, model deployment, and model development during asset management. These findings highlight the need to adapt the curriculum design. Educational institutions should prioritize course materials, hands-on experiments, and real-world case studies in the aforementioned areas. MLOps enthusiasts, not limited to computer science students, would greatly benefit from these custom curricula.

\subsection{Implications for Application Developers}
\label{sec:implications:subsec:application developers}
\vspace{-0.5em} 

\paragraph{Holistic ML Development:} Our analysis identifies 133 topics, grouped into 16 overarching macro-topics across various domains, suggesting that application developers should adopt a holistic approach to asset management. This approach includes essential steps, such as data preparation, environment setup, and model training. Much like how full-stack developers benefit from understanding every stage of development, ML developers too gain from a similar wide-ranging understanding. Adopting this holistic approach could help application developers uncover and address hidden technical debts~\cite{sculley2015hidden} in the ML development lifecycle. As a result, they face fewer obstacles and enjoy a smoother development journey.

\paragraph{Tool Selection Bootstrapping:} We find a clear pattern of association between certain tools and specific challenge macro-topics. This information is invaluable for application developers, offering a roadmap to anticipate and mitigate potential challenges during the tool selection process. AzureML and SageMaker, for instance, are associated with a broad range of macro-topics, underscoring their versatility. However, developers should also be prepared for the diverse challenges that come with such comprehensive platforms. Furthermore, the strong association of a significant number of tools with software dependency and environment management underscores a common area of challenge. Developers are recommended to pay special attention to this aspect, evaluating how the features and support of each tool can help navigate environment-related issues effectively.

% \paragraph{Optimizing Inquiry Placement:} We discover that on Stack Overflow and tool-specific forums, users often find it difficult to get answers to problem inquiries. In contrast, knowledge inquiries face similar challenges on GitHub. This observation leads to practical advice for application developers seeking help online. When developers encounter bugs, errors, or abnormal software behavior, we recommend posting these problem inquiries on GitHub, where they are more likely to receive helpful responses. On the other hand, for knowledge-based questions, such as those seeking best practices or tool comparisons, developers should turn to Stack Overflow or tool-specific forums. In this way, by choosing the appropriate platform based on the nature of their inquiries, developers can increase the chances of receiving timely and helpful responses.

\subsection{Implications for Tool Developers}
\label{sec:implications:subsec:tool developers}
\vspace{-0.5em} 

\paragraph{Feature Enhancement Prioritization:} Our study shows that ML practitioners face great challenges in solving issues related to user interface, data development, experiment management, and code management. Therefore, we recommend that tool developers prioritize the design of more intuitive interfaces, enhanced data management features, and robust tools to manage experiments and code. For emerging tool developers, tackling these identified challenges not only addresses the core issues faced by ML practitioners but also opens the door to establishing a niche market by providing tailored solutions to these distinct pain points.

\paragraph{User Engagement Analysis:} Tool developers, especially those who manage discussion forums, play a crucial role in shaping the experience of users who post their inquiries on these platforms~\cite{parra2022comparative}. Tools, such as Amazon SageMaker, H2O AI Cloud, Kedro, MLRun, MLflow, Optuna, and Vertex AI primarily have their discussions hosted on Stack Overflow. This trend may signify a preference among users to engage with a broader community or certain tool-specific forums. This suggests that maintainers of other tools could adopt certain practices from these successful forums to foster more interactions.

%% file: sections/8-threats.tex
\section{Threats to Validity}
\label{sec:threats}
\vspace{-0.5em}

We address the validity threats in our research by examining them from four perspectives: conclusion validity, construct validity, external validity, and internal validity. For each type of threat, we present the specific actions we take to mitigate them.

\subsection{Conclusion validity}
\label{sec:threats:conclusion}
\vspace{-0.5em} 

\paragraph{Macro-topic aggregation}: To mitigate the inherent risk of individual biases that might lead contributors to different conclusions about topic groupings, we have adopted a rigorous standard to aggregate topics. Initially, the first three authors take on the task independently. Following this, they conduct a collaborative review in which they cross-examine their individual categorizations. This collective approach aimes not only to harmonize their perspectives, but also to minimize any cognitive biases that might skew the results. We refer to established literature if disagreements about the categorization of a specific topic persist. By anchoring our decisions in prior research, we aim to ensure that our categorizations are not only consistent but also justifiably grounded.

\paragraph{Open card sorting}: We have manually labeled the entire dataset consisting of $4,687$ solved posts, ensuring that we incorporate every available data point for a thorough analysis. By leveraging this comprehensive sample size, we significantly enhance the statistical power of our study. Such an increase not only strengthens the reliability of our findings, but also substantially mitigates the risk of Type II errors. In the context of our study, a Type II error might lead us to inadvertently neglect a recurring category frequently used by practitioners - a misstep that could be attributed to the limitations inherent in analyzing a restricted dataset.

\subsection{Construct validity} 
\label{sec:threats:construct}
\vspace{-0.5em} 

\paragraph{Closed card sorting}: A primary threat to the construct validity of our study stems from the subjectivity inherent in the manual filtering process of tool-specific forum data. Despite implementing a multi-stage review protocol and involving several authors to minimize personal biases, the possibility of subjective interpretation in selecting and categorizing data cannot be eliminated. This subjectivity could impact the reliability and generalizability of our findings. 

\paragraph{Macro-topic aggregation}: A specific topic may relate to multiple macro-topics, leading to some uncertainty in our macro-topic aggregation choices. To address this, we conduct a manual review of a 5\% sample of posts on each topic. Through this careful examination, we identify the macro-topic that the majority of these topics align with most closely.

\paragraph{Open card sorting}: The first and second authors possess over three years of experience in software development. This expertise plays a crucial role in minimizing threats to construct validity during the open card sorting process. Before the categorization process, these contributors engage in extensive discussion to establish a clear consensus on the labeling guideline, as outlined in Step~3.1. This step guarantees that the labels accurately capture the content of the solutions.

\paragraph{Post title refinement}: Subtle nuances and the core essence of original posts may not always be accurately manifested in GPT-4 refined titles. To mitigate this, the first two authors work together to review the refined titles by examining a sample of 5\% posts. When they find discrepancies, they tweak the prompts to better align the refined titles with the content of the posts. Through this adjustment process, they observe that tweaking the prompts results in topic models with higher coherence (C\_V). Our goal is to ensure that the GPT-4 refined titles authentically encapsulate the intent of the original posts. This approach is in line with the recommendations by Sallou~et~al.~\cite{sallou2024breaking} that stress the importance of reviewing an LLM's output to ensure its applicability and validity in software engineering research.

\paragraph{Topic model selection}: In topic modeling, it is possible to encounter topics with overlapping key terms that may blur the distinctiveness of each topic. To address this, we examine the top-10 topic models, focusing on those with the highest coherence C\_V value, to better understand both challenges and solutions. We manually review a sample of 5\% of posts for each topic to assess their thematic representation and degree of overlap. Through this comparative approach, we aim to identify the topic model that best fits our study.

\subsection{External validity}
\label{sec:threats:external}
\vspace{-0.5em} 

\paragraph{Forum evolution}: We have collected Q\&A posts from multiple developer discussion forums up to $26^{th} of July, 2023$. We acknowledge that the content on these forums is dynamic and subject to constant updates and modifications. Consequently, there is a possibility that the information, challenges, and solutions we analyze may become obsolete over time. To mitigate this concern, we employ rigorous analytical methods to validate the relevance and applicability of our findings. We aim to ensure that the insights derived are not constrained by the collection period, but remain valuable and applicable in a longer timeframe.

\paragraph{Forum selection}: To ensure a comprehensive and representative dataset on ML asset management from the users' perspective, we source data from a diverse spectrum of discussion forums and tools. This strategy addresses potential threats to external validity arising from user preferences. Our dataset incorporates posts from general forums, such as Stack Overflow, repository-specific forums, such as GitHub Issues, and tool-specific forums, such as the MLflow Google Group. 

\paragraph{Post collection}: The absence of posts on certain topics does not necessarily imply the absence of such challenges for a given tool. Instead, this might reflect a potential lack of community support, which is a critical factor in tool adoption and usability. To mitigate this threat, we have broadened our scope of data collection by including data from various discussion channels, such as Stack Overflow, GitHub, and tool-specific websites. This approach aims to provide a more comprehensive understanding of the challenges within the domain.

\paragraph{Post translation}: In the open card sorting process, we utilize the built-in translator of Microsoft Edge browser\footnote{\url{https://www.microsoft.com/en-us/edge}} to convert non-English text to English text. Although such translations might not encapsulate every nuance or cultural reference inherent in the original language, software engineering inquiry terminologies generally present consistent expressions across languages. Furthermore, considering that non-English posts constitute less than $5\%$ of all entries, any potential influence on external validity remains minimal.

\paragraph{Tool selection}: The operationalization of ML encompasses numerous tools, each relevant to specific aspects of asset management. Overlooking any of these tools potentially compromises the external validity of our findings. To address this, we actively collect information on the most representative tools for ML asset management from various sources, including Google search engine, Google Scholar, GitHub, and insights provided by domain experts. Our tool selection follows the filtering criteria set by previous research~\cite{idowu2021asset}, as outlined in Step~1.1. This approach ensures that our choices remain systematic and grounded in expert knowledge.

\subsection{Internal validity}
\label{sec:threats:internal}
\vspace{-0.5em} 

\paragraph{Closed card sorting}: Including unrelated posts can introduce noise that can distort the findings of our study. To mitigate this risk, we employ the closed-open carding approach to exclude any inquiry labeled ``NA''. This step helps us filter out irrelevant data, allowing for a more precise analysis and, ultimately, more reliable results in our study.

\paragraph{Prompt tuning}: We notice that changing the way we phrase the GPT-4 prompt gives us different results. This shows that it is not always easy to figure out the best way to phrase the prompts. To tackle this issue, we use a heuristic approach to tweak the prompts. Our goal is to find the one that works best for creating refined titles. As we make these adjustments, we see a clear improvement in the coherence (C\_V) of the topic models. Most of our new prompts even perform better than our original ones. We then have two team members take a closer look at a random sample of 5\% of the posts to make sure that our new titles still match what the original poster meant. We find that for the most part, the improved titles are in line with the original intent and are similar to labels from SE experts. This step guarantees that our analysis and results are solid and takes into account the different ways that prompts can be phrased.

%% file: sections/9-conclusion.tex
\section{Conclusion}
\label{sec:conclusion}
\vspace{-0.5em}

In this study, we explore the challenges and their solutions associated with ML asset management tools through an analysis of Q\&A posts from various developer discussion forums. We employ a mixed-method approach, using BERTopic to extract and categorize the primary topics of these challenges and the corresponding solutions. Our findings reveal a comprehensive landscape of prevalent challenges and their solutions, offering a granular view of the complexities and needs in managing ML assets. We identify key areas of concern, such as the management of software environments and dependencies, as well as the training and deployment of models. It is important to note that the tools explored in our study accommodate a wide spectrum of ML models, ranging from statistical models to deep neural networks. As a result, our insights are of significant value to a diverse audience in various contexts. These insights facilitate the judicious selection of tools, the development of effective educational resources, and the guidance of future research efforts towards bridging existing gaps.

Our study suggests four promising research directions for further research on ML asset management. Firstly, we underscore the importance of focusing on the specific issues associated with the most prevalent macro-topics identified in our study. These areas, marked by the high frequency of occurrence, represent critical pain points for ML practitioners. Detailed investigations into these macro-topics can reveal underlying issues, potential solutions, and tool and resource development opportunities. 

Secondly, a more thorough analysis of the dynamics inherent in problem and knowledge inquiries across various software engineering domains is warranted. Our observations indicate a varied prevalence of inquiry types related to ML asset management. Specifically, problem inquiries are more prevalent than knowledge inquiries regarding ML asset management. This pattern aligns with previous findings in the service mesh domain~\cite{chen2023practitioners}, but diverges from general domain trends, as highlighted by earlier studies~\cite{treude2011programmers}. By exploring these patterns, we can gain a clearer understanding of the intricacies surrounding the users' intentions. 

Thirdly, our analysis shows an interesting pattern: $16$ macro-topics have the same names for both challenges and solutions. This overlap is not mere coincidence, but indicates a significant link between the challenges encountered and the solutions applied in asset management. It reveals the potential for a deeper understanding of how issues are addressed in this area. Future research should incorporate more data points and monitor the evolution of these challenges and solutions over time. Such efforts aim to establish a more comprehensive theory that accurately captures these relationships. 

Lastly, exploring the association between various types of FM (\eg, pre-trained, fine-tuned, instruction-tuned~\cite{zhang2023instruction}, proxy-tuned~\cite{liu2024tuning}, knowledge-distilled~\cite{gou2021knowledge}) and identifying challenge topics related to these types presents a promising avenue. In particular, the growing trend toward local deployment, driven by decreased privacy concerns~\cite{yao2023survey}, more cost-effective model parameters~\cite{schick2020s}, and the advent of open language models~\cite{groeneveld2024olmo}, may signal a shift in discussions toward more sophisticated asset management in FMOps. Such research could provide essential insights, fostering innovation in FM deployment strategies.

Given the pivotal role that asset management occupies in the burgeoning field of MLOps, it is essential for ongoing research, dialogue, and collaboration to shape its trajectory, ensuring that it effectively supports the diverse and dynamic needs of ML practitioners.

%% file: sections/10-conflict.tex
\section{Conflict of Interest}
\label{sec:conflict}
\vspace{-0.5em}

The authors declare that they have no conflict of interest.

%% file: sections/11-availability.tex
\section{Data Availability Statement}
\label{sec:availability}

The datasets generated and analyzed during this study are available in the replication package~\cite{replication_package}.

%% file: sections/12-appendix.tex
\section{Appendix}
\label{sec:appendix-1}

\input{tables/appendix/rq2-topic-illustration}

%% file: tables/appendix/rq2-topic-illustration.tex
Definition and illustration of solution macro-topics not discussed in section~\ref{sec:rq2-results:solution macro-topic}

\begin{longtable}{p{\linewidth}}
    \toprule
    $\hat{R}_{01}$ \textbf{Code Development} \\ 
    \midrule
    \textbf{Definition}: Same as $\hat{C}_{01}$. It integrates argument management ($R_{12}$), code modification ($R_{13}$), wait time management ($R_{27}$), command line usage ($R_{28}$), API integration ($R_{47}$), parameter update ($R_{56}$), character removal ($R_{57}$), troubleshooting guidance ($R_{58}$), function modification ($R_{59}$), syntax update ($R_{61}$), exception handling ($R_{63}$) and parameter removal ($R_{84}$) to facilitate a controlled, efficient, and error-resistant MLOps environment. \\
    \textbf{Example}: The following example\footnote{\url{https://stackoverflow.com/questions/71255132}} suggests checking the ``Action'' field when ``Use Action Name'' is selected in API methods. \\
    $E_{01}$: \textit{Accepted Answer: Check in all your API methods that you have not specified ``Use Action Name'' for any integration request and then leave the ``Action'' field blank. [TEXT]} \\
    \\
    \toprule
    $\hat{R}_{02}$ \textbf{Code Management} \\ 
    \midrule
    \textbf{Definition}: Same as $\hat{C}_{02}$. It encompasses the creation or updating of Git repositories ($R_{43}$) for optimal version control and collaboration. \\
    \textbf{Example}: The following example\footnote{\url{https://stackoverflow.com/questions/70335823}} illustrates adding the relative path to write to the output folder. \\
    $E_{02}$: \textit{Accepted Answer: [TEXT] Here is an example of an operation that reads from the relative path where your code exists: [CODE] You can then join the relative path of your git folder before your output. [TEXT]} \\
    \\
    \toprule
    $\hat{R}_{03}$ \textbf{Computation Management} \\ 
    \midrule
    \textbf{Definition}: Same as $\hat{C}_{03}$. It includes regulating resource usage through limit adjustments ($R_{19}$), facilitating functionalities through service provisioning ($R_{30}$), overseeing the creation and handling of compute instances or clusters for task execution ($R_{35}$), managing event-driven programming through lambda function management ($R_{53}$), and improving performance by increasing resource capacities ($R_{60}$). \\
    \textbf{Example}: The following example\footnote{\url{https://stackoverflow.com/questions/71398882}} suggests increasing GPU memory, decreasing batch size, and changing to a smaller model. \\
    $E_{03}$: \textit{Accepted Answer: [TEXT] Things that you can try: Provision an instance with more GPU memory; Decrease batch size; Use a different (smaller) model.} \\
    \\
    \toprule
    $\hat{R}_{04}$ \textbf{Data Development} \\ 
    \midrule
    \textbf{Definition}: Same as $\hat{C}_{04}$. It covers column manipulation ($R_{31}$), feature filtering ($R_{33}$), and data transformation ($R_{54}$) to enhance and refine data for an effective ML pipeline. \\
    \textbf{Example}: The following example\footnote{\url{https://stackoverflow.com/questions/72106030}} illustrates the usage of a feature store. \\
    $E_{04}$: \textit{Accepted Answer: [TEXT] Once a feature store is created, you will need to create an entity and then create a feature that has the labels parameter as shown in the below sample Python code. [CODE]} \\
    \\
    \toprule
    $\hat{R}_{05}$ \textbf{Data Management} \\ 
    \midrule
    \textbf{Definition}: Same as $\hat{C}_{05}$. It encompasses the conversion of data and datatypes for compatibility and processing ($R_{40}$, $R_{70}$), the creation of datasets for model training and validation ($R_{42}$), facilitating seamless data import/export for access and sharing ($R_{75}$), and the manipulation of buckets for organized storage in cloud services ($R_{77}$). \\
    \textbf{Example}: The following example\footnote{\url{https://stackoverflow.com/questions/56024351}} suggests examination of the output location in Amazon S3.  \\
    $E_{05}$: \textit{Accepted Answer: [TEXT] SageMaker places the model artifacts in a bucket that you own, check the S3 output location in the AWS SageMaker console. [TEXT]}  \\
    \\
    \toprule
    $\hat{R}_{06}$ \textbf{Environment Management} \\ 
    \midrule
    \textbf{Definition}: Same as $\hat{C}_{06}$. It encompasses package upgrades ($R_{01}$), installation ($R_{05}$), version management ($R_{07}$), SDK upgrades ($R_{15}$), container customization ($R_{21}$), Docker management ($R_{22}$), package additions ($R_{23}$), creation of environments ($R_{25}$), management of environment variables ($R_{34}$), SDK usage ($R_{37}$), package downgrades ($R_{41}$), reinstallations ($R_{44}$), removals ($R_{62}$), notebook usage ($R_{66}$), Python version management ($R_{67}$), package imports ($R_{69}$), workspace creation ($R_{72}$), region support ($R_{73}$), kernel restarts ($R_{76}$), Docker updates ($R_{78}$) and notebook instance management ($R_{85}$). \\
    \textbf{Example}: The following example\footnote{\url{https://github.com/getindata/kedro-kubeflow/issues/105}} suggests the prohibition of circular dependency. \\
    $E_{06}$: \textit{Merge Request: Fixes \#105 by not allowing circullar dependency on mlflow.}  \\
    \\
    \toprule
    $\hat{R}_{07}$ \textbf{Experiment Management} \\ 
    \midrule
    \textbf{Definition}: Same as $\hat{C}_{07}$. It encapsulates the concepts of specifying run settings ($R_{11}$), creating or updating ML experiments ($R_{39}$), providing templates for machine learning tasks ($R_{71}$) and tailoring sessions for task execution ($R_{83}$). \\
    \textbf{Example}: The following example\footnote{\url{https://github.com/Lightning-AI/lightning/issues/6745}} suggests adding information from the experiment run. \\
    $E_{07}$: \textit{Merge Request: Fix \#6745: adds additional information about the run, as in the native API. [TEXT]} \\
    \\
    \toprule
    $\hat{R}_{08}$ \textbf{File Management} \\ 
    \midrule
    \textbf{Definition}: Same as $\hat{C}_{08}$. It encompasses the processes of storage mounting($R_{16}$), directory management($R_{20}$), file deletion($R_{29}$), file download($R_{36}$), filepath update($R_{49}$), input management($R_{55}$), filepath modification($R_{64}$), file load($R_{65}$), tracking configuration($R_{68}$), and documentation update($R_{86}$). \\
    \textbf{Example}: The following example\footnote{\url{https://stackoverflow.com/questions/57126765}} suggests copying files from Amazon S3 to the local drive. \\
    $E_{08}$: \textit{Accepted Answer: [TEXT] The simplest option is to copy the files from S3 to the local drive (EBS or EFS) of the notebook instance: [CODE]}  \\
    \\
    \toprule
    $\hat{R}_{09}$ \textbf{Model Deployment} \\ 
    \midrule
    \textbf{Definition}: Same as $\hat{C}_{09}$. It involves endpoint invocation ($R_{24}$), deployment pipeline creation ($R_{26}$), model prediction ($R_{38}$), model deployment ($R_{79}$) and implementation of the inference pipeline ($R_{80}$) to efficiently deploy and serve machine learning models in the production environment. \\
    \textbf{Example}: The following example\footnote{\url{https://stackoverflow.com/questions/73811793}} suggests the usage of \texttt{undeploy\_all} function. \\
    $E_{09}$: \textit{Accepted Answer: You can undeploy all the models from an endpoint by calling the method undeploy\_all() [TEXT]} \\
    \\
    \toprule
    $\hat{R}_{10}$ \textbf{Model Development} \\ 
    \midrule
    \textbf{Definition}: Same as $\hat{C}_{10}$. It encompasses the practice of distributed training ($R_{03}$), which focuses on the implementation and configuration of parallel or continuous training processes for model creation. \\
    \textbf{Example}: The following example\footnote{\url{https://github.com/MicrosoftDocs/pipelines-azureml/issues/12}} suggests creating a custom Docker image and uploading it to Azure Container Registry. \\
    $E_{12}$: \textit{Solution Comment: Got this working by creating a custom Docker image and putting it to the ACR tied to Azure ML workspace. [TEXT]} \\
    \\
    \toprule
    $\hat{R}_{11}$ \textbf{Model Management} \\ 
    \midrule
    \textbf{Definition}: Same as $\hat{C}_{11}$. It integrates the processes of model creation ($R_{10}$), registration ($R_{32}$), and file handling ($R_{52}$) to streamline the lifecycle of machine learning models. \\
    \textbf{Example}: The following example\footnote{\url{https://stackoverflow.com/questions/72068059}} suggests replacing \texttt{load\_model} with pickle. \\
    $E_{11}$: \textit{Accepted Answer: [TEXT] In particular, this answer using load\_model instead of pickle seemed to work well for me: [CODE]} \\
    \\
    \toprule
    $\hat{R}_{12}$ \textbf{Network Management} \\ 
    \midrule
    \textbf{Definition}: Same as $\hat{C}_{12}$. It encompasses both the configuration or alteration of network settings to improve system performance ($R_{09}$) and the process of updating or configuring hyperlinks for precise navigation ($R_{51}$). \\
    \textbf{Example}: The following example\footnote{\url{https://stackoverflow.com/questions/58802366}} suggests creating an API gateway to share model as an endpoint. \\
    $E_{12}$: \textit{Accepted Answer: To share your model as an endpoint, you should use lambda and API gateway to create your API. [TEXT]} \\
    \\
    \toprule
    $\hat{R}_{13}$ \textbf{Observability Management} \\ 
    \midrule
    \textbf{Definition}: Same as $\hat{C}_{13}$. It encompasses the setup of logging systems($R_{06}$), the updating or scrutiny of performance metrics($R_{50}$), and the enhancement of log functions or levels for better debugging($R_{74}$). \\
    \textbf{Example}: The following example\footnote{\url{https://stackoverflow.com/questions/67258917}} suggests checking the metrics tab of the current step. \\
    $E_{13}$: \textit{Accepted Answer: [TEXT] In Studio, if you go to the step's Metrics tab, you will be able to see a chart/table of execution progress, including remaining items, remaining mini batches, failed items, etc. [URL]} \\
    \\
    \toprule
    $\hat{R}_{14}$ \textbf{Pipeline Management} \\ 
    \midrule
    \textbf{Definition}: Same as $\hat{C}_{14}$. It encompasses the cohesive administration and orchestration of job processing ($R_{14}$), where tasks are executed potentially in parallel or on a schedule, combined with pipeline configuration ($R_{18}$) that involves the creation, updating or modification of pipelines for efficient data and model workflow, and lifecycle configuration ($R_{45}$), which denotes the management of application states through the implementation or modification of lifecycle scripts. \\
    \textbf{Example}: The following example\footnote{\url{https://stackoverflow.com/questions/72203674}} clarifies the usage of the pipeline construct. \\
    $E_{14}$: \textit{Accepted Answer: CDK L1 Constructs correspond 1:1 to a CloudFormation resource of the same name. The construct props match the resource properties. Therefore, the go-to source is the CloudFormation documentation.} \\
    \\
    \toprule
    $\hat{R}_{15}$ \textbf{Security Management} \\ 
    \midrule
    \textbf{Definition}: Same as $\hat{C}_{15}$. It integrates the processes of establishing, allocating, or altering access permissions ($R_{04}$), updating authentication credentials ($R_{17}$), and registering or recreating user accounts ($R_{81}$) to improve access control. \\
    \textbf{Example}: The following example\footnote{\url{https://stackoverflow.com/questions/74406041}} suggests ungrading the workspace to migrate from V1 to V2. \\
    $E_{15}$: \textit{Accepted Answer: [TEXT] To migrate from Azure Machine Learning V1 to V2, you need to upgrade the az ml workspace share commands to equivalent az role assignment create commands. [TEXT]} \\
    \\
    \toprule
    $\hat{R}_{16}$ \textbf{User Interface Management} \\ 
    \midrule
    \textbf{Definition}: Same as $\hat{C}_{16}$. It includes the creation and modification of data visualizations ($R_{48}$) for improved data analysis and interpretation. \\
    \textbf{Example}: The following example\footnote{\url{https://community.wandb.ai/t/vega-code/4605}} clarifies the support for the Vega visualization language. \\
    $E_{16}$: \textit{Accepted Answer: [TEXT] currently Vega only powers custom charts and the underlying code for other panels are created in JavaScript, unfortunately.} \\
    % \bottomrule
\end{longtable}

%% file: sections/13-appendix.tex
\section{Appendix}
\label{sec:appendix-2}

\input{tables/appendix/retrieved-literature}
\input{tables/appendix/curated-tools}
\input{tables/appendix/stackoverflow-tags}
\input{tables/appendix/tool-specific-community-websites}
\input{tables/appendix/tool-usage-pattern}
\input{tables/appendix/post-title-keywords}
\input{tables/appendix/gpt-4}
\input{tables/appendix/hyperparameter-search-space}
\input{tables/appendix/rq1-topic-list}
\input{tables/appendix/rq2-topic-list}

%% file: tables/appendix/retrieved-literature.tex
\begin{table}[H]
\centering
\caption{Retrieved literature related to ML asset management tools.}
\label{tab:retrieved literature}
\begin{tabular}{C{8em}|C{6.5em}|C{6.5em}|C{6.5em}|C{6.5em}}
\toprule
\multirow{2}{*}{\textbf{Search String}} & \multicolumn{2}{c|}{\textbf{Google Scholar}} & \multicolumn{2}{c}{\textbf{GitHub Awesome Lists}} \\
\cmidrule(lr){2-3} \cmidrule(lr){4-5}
& \textbf{Literature} & \textbf{Snowballing} & \textbf{Literature} & \textbf{Snowballing} \\
\midrule
\multirow{2}{*}{MLOps tools} & 
\cite{symeonidis2022mlops,hewage2022machine,ruf2021demystifying,recupito2022multivocal,moreschi2023toward} & 
& 
\cite{mlops,awesome-LLMOPS,awesome-mlops,awesome-production-machine-learning} &
\cite{machine-learning-engineering} \\
\midrule
\parbox[t]{3cm}{machine learning \\ artifact \\ management tools} &
&
& \cite{artifact-management} & \\
\midrule
\parbox[t]{3cm}{machine learning \\ data \\ management tools} &
\cite{hartley2020dtoolai,kumar2017data} &
&
\cite{data-management} & \\
\midrule
\parbox[t]{3cm}{machine learning \\ experiment \\ management tools} & 
\cite{schlegel2023management,idowu2022emmm,idowu2021asset,gharibi2019automated,zaharia2018accelerating,sung2017nsml} &
\cite{pavao2022codalab,namaki2020vamsa,chard2019dlhub,ferenc2020deep,alberti2018deepdiva,schelter2018declarative} &
\cite{experiments,experiment-management,awesome-ml-experiment-management} &
\cite{experiment-tracking} \\
\midrule
\parbox[t]{3cm}{machine learning \\ lifecycle \\ management tools} & \cite{le2023veml,yang2021mlife,aguilar2021ease,chen2020developments,bhattacharjee2019stratum,hummer2019modelops,garcia2018context,miao2017towards,miao2017modelhub,miao2017provdb} &
&
\cite{lifecycle-management} & \\
\midrule
\parbox[t]{3cm}{machine learning \\ model \\ management tools} &
\cite{sun2020gallery,vartak2018modeldb,tsay2018runway,peili2018deep} &
&
\cite{model-management} & \\
\midrule
\parbox[t]{3cm}{machine learning \\ project \\ management tools} &
& & \cite{project-management} &\\
\midrule
\parbox[t]{3cm}{machine learning \\ workflow \\ management tools} &
\cite{bravo2022scanflow,franccoise2021marcelle,benitez2021titan,moreno2020managing,wozniak2018candle} &
&
\cite{workflow-management} & \\
\bottomrule
\end{tabular}
\end{table}

%% file: tables/appendix/curated-tools.tex
\begin{table}[H]
\centering
\caption{Information of ML asset management tools.}
\begin{tabular}{lp{5.4em}p{4.2em}p{14.8em}}
\toprule
\textbf{Tool Name} & \textbf{License Status} & \textbf{Launch Date} & \textbf{Tool Website} \\
\midrule
Aim & Open source & May 2019 & \url{https://aimstack.io}\\
\midrule
\multirow{2}{*}{Amazon SageMaker} & \multirow{2}{*}{Proprietary} & \multirow{2}{*}{Nov 2019} & \url{https://aws.amazon.com/sagemaker} \\
\midrule
\multirow{3}{*}{Azure Machine Learning} & \multirow{3}{*}{Proprietary} & \multirow{3}{*}{2015 Feb} & \url{https://azure.microsoft.com/en-us/products/machine-learning} \\
\midrule
\multirow{2}{*}{ClearML} & \multirow{2}{*}{Open source} & \multirow{2}{*}{Jun 2019} & \url{https://github.com/allegroai/clearml} \\
\midrule
cnvrg.io & Proprietary & Mar 2020 & \url{https://cnvrg.io}\\
\midrule
\multirow{2}{*}{Codalab} & \multirow{2}{*}{Open source} & \multirow{2}{*}{Nov 2014} & \url{https://github.com/codalab/codalab-worksheets} \\
\midrule
Comet & Proprietary & Jan 2017 & \url{https://www.comet.com} \\
\midrule
\multirow{2}{*}{Determined} & \multirow{2}{*}{Open source} & \multirow{2}{*}{Apr 2020}& \url{https://github.com/determined-ai/determined} \\
\midrule
Domino & Proprietary & May 2016 & \url{https://domino.ai} \\
\midrule
\multirow{2}{*}{DVC} & \multirow{2}{*}{Open source} & \multirow{2}{*}{Mar 2017} & \url{https://github.com/iterative/dvc} \\
\midrule
\multirow{2}{*}{Guild AI} & \multirow{2}{*}{Open source} & \multirow{2}{*}{Sep 2017} & \url{https://github.com/guildai/guildai} \\
\midrule
\multirow{2}{*}{H2O AI Cloud} & \multirow{2}{*}{Proprietary} & \multirow{2}{*}{Jan 2021} & \url{https://h2o.ai/platform/ai-cloud} \\
\midrule
Iterative Studio & Proprietary & May 2021 & \url{https://studio.iterative.ai} \\
\midrule
\multirow{2}{*}{Kedro} & \multirow{2}{*}{Open source} & \multirow{2}{*}{Apr 2019} & \url{https://github.com/kedro-org/kedro} \\
\midrule
\multirow{2}{*}{MLflow} & \multirow{2}{*}{Open source} & \multirow{2}{*}{Jun 2018} & \url{https://github.com/mlflow/mlflow} \\
\midrule
\multirow{2}{*}{MLRun} & \multirow{2}{*}{Open source} & \multirow{2}{*}{Sep 2019} & \url{https://github.com/mlrun/mlrun} \\
\midrule
Neptune & Proprietary & Feb 2019 & \url{https://neptune.ai} \\
\midrule
\multirow{2}{*}{Optuna} & \multirow{2}{*}{Open source} & \multirow{2}{*}{Feb 2018} & \url{https://github.com/optuna/optuna} \\
\midrule
\multirow{2}{*}{Polyaxon} & \multirow{2}{*}{Open source} & \multirow{2}{*}{Dec 2016} & \url{https://github.com/polyaxon/polyaxon} \\
\midrule
\multirow{2}{*}{Sacred} & \multirow{2}{*}{Open source} & \multirow{2}{*}{Mar 2014} & \url{https://github.com/IDSIA/sacred} \\
\midrule
SigOpt & Proprietary & Nov 2014 & \url{https://sigopt.com} \\
\midrule
Valohai & Proprietary & Feb 2017 & \url{https://valohai.com} \\
\midrule
Verta & Proprietary & Apr 2018 & \url{https://www.verta.ai} \\
\midrule
\multirow{2}{*}{Vertex AI} & \multirow{2}{*}{Proprietary} & \multirow{2}{*}{Mar 2019} & \url{https://cloud.google.com/vertex-ai} \\
\midrule
\multirow{2}{*}{Weights\&Biases} & \multirow{2}{*}{Open source} & \multirow{2}{*}{Mar 2017} & \url{https://github.com/wandb/wandb} \\
\bottomrule
\end{tabular}
\label{tab:curated tools}
\end{table}

%% file: tables/appendix/stackoverflow-tags.tex
\begin{table}[H]
\centering
\caption{Stack Overflow tag list for each curated tool.}
\begin{tabular}{lp{8cm}}
\toprule
\textbf{Tool Name} & \textbf{Stack Overflow Tags} \\
\midrule
\multirow{5}{*}{Amazon SageMaker} & amazon-sagemaker, amazon-sagemaker-experiments, amazon-sagemaker-studio, amz-sagemaker-distributed-training, amazon-sagemaker-debugger, amazon-sagemaker-clarify, amazon-sagemaker-compilers, amazon-sagemaker-experiments, amazon-sagemaker-neo \\
\midrule
\multirow{3}{*}{Azure Machine Learning} & azureml-python-sdk, azuremlsdk, azure-machine-learning-service, azure-machine-learning-studio, azure-machine-learning-workbench, azure-ml-pipelines \\
\midrule
ClearML & clearml \\
\midrule
Comet & comet-ml \\
\midrule
DVC & dvc \\
\midrule
H2O AI Cloud & h2o.ai\\
\midrule
Kedro & kedro \\
\midrule
MLflow & mlflow \\
\midrule
MLRun & mlrun \\
\midrule
Neptune & neptune, neptune-python-utils \\
\midrule
Optuna & optuna \\
\midrule
Sacred & python-sacred \\
\midrule
Vertex AI & google-cloud-vertex-ai, vertex-ai-pipeline \\
\midrule
Weights\&Biases & wandb \\
\bottomrule
\end{tabular}
\label{tab:stackoverflow tags}
\end{table}

%% file: tables/appendix/tool-specific-community-websites.tex
\begin{table}[H]
\centering
\caption{Information of tool-specific forums.}
\begin{tabular}{lp{8cm}}
\toprule
\textbf{Tool Name} & \textbf{Forum Website} \\
\midrule
\multirow{2}{*}{Amazon SageMaker} & \url{https://repost.aws/tags/TAT80swPyVRPKPcA0rsJYPuA/amazon-sage-maker} \\
\midrule
\multirow{2}{*}{Azure Machine Learning} & \url{https://learn.microsoft.com/en-us/answers/topics/25447/azure-machine-learning} \\
\midrule
\multirow{2}{*}{Domino} & \url{https://tickets.dominodatalab.com/hc/en-us/community/topics} \\
\midrule
DVC & \url{https://discuss.dvc.org/c/questions/9} \\
\midrule
Guild AI & \url{https://my.guild.ai/c/troubleshooting/6} \\
\midrule
H2O AI Cloud & \url{https://community.h2o.ai/categories/general} \\
\midrule
MLflow & \url{https://groups.google.com/g/mlflow-users} \\
\midrule
\multirow{2}{*}{Polyaxon} & \url{https://github.com/orgs/polyaxon/discussions/categories/q-a} \\
\midrule
SigOpt & \url{https://community.sigopt.com/c/general-discussion/9} \\
\midrule
\multirow{2}{*}{Vertex AI} & \url{https://www.googlecloudcommunity.com/gc/AI-ML/bd-p/cloud-ai-ml} \\
\midrule
Weights\&Biases & \url{https://community.wandb.ai/c/w-b-support/36} \\
\bottomrule
\end{tabular}
\label{tab:tool-specific community websites}
\end{table}

%% file: tables/appendix/tool-usage-pattern.tex
\begin{table}[H]
\centering
\caption{Usage pattern of curated tools.}
\footnotesize
\begin{tabular}{lp{8.1cm}}
\toprule
\textbf{Tool Name} & \textbf{Usage Pattern}\\
\midrule
Aim & \lstinline?import aim|from aim.* import ? \\
\midrule
\multirow{9}{*}{Amazon SageMaker} & \lstinline?import sagemaker|from sagemaker.* import |#include <aws\/sagemaker.*\.h>|\"github\.com\/aws\/aws-sdk-go\/service\/sagemaker\"|package com\.amazonaws\.services\.sagemaker.*;|using Amazon\.SageMaker.*;|namespace Aws\\SageMaker\.*;|import .* from (\"@aws-sdk\/client-sagemaker\"|'@aws-sdk\/client-sagemaker');|require\((\"@aws-sdk\/client-sagemaker\"|'@aws-sdk\/client-sagemaker')\);|require_relative ('aws-sdk-sagemaker\/.*'|\"aws-sdk-sagemaker\/.*\")? \\
\midrule
Azure Machine Learning & \lstinline?import azureml\.core|from azureml\.core.* import ? \\
\midrule
\multirow{2}{*}{ClearML} & \lstinline?file:clearml.conf|import clearml|from clearml.* import ? \\
\midrule
cnvrg.io & \lstinline?import cnvrgv2|from cnvrgv2 import ? \\
\midrule
Codalab & \lstinline?import codalab|from codalab.* import ? \\
\midrule
\multirow{2}{*}{Comet} & \lstinline?import comet_m(l|pm)|from comet_m(l|pm).* import |import ml\.comet\.experiment.*;|library\(cometr\)? \\
\midrule
Determined & \lstinline?import determined|from determined.* import ? \\
\midrule
\multirow{2}{*}{Domino} & \lstinline!file:domino\.y(a)?ml|import domino|from domino.* import |library\(domino\)! \\
\midrule
DVC & \lstinline!file:dvc\.y(a)?ml|import dvc|from dvc.* import ! \\
\midrule
Guild AI & \lstinline!file:guild\.y(a)?ml|import guild|from guild.* import ! \\
\midrule
\multirow{2}{*}{MLflow} & \lstinline?import mlflow|from mlflow.* import |library\(mlflow\)|import org\.mlflow.*;?\\
\midrule
\multirow{2}{*}{Neptune} & \lstinline?import neptune|from neptune.* import |library\(neptune\)? \\
\midrule
Optuna & \lstinline?import optuna|from optuna.* import ? \\
\midrule
\multirow{7}{*}{Polyaxon} & \lstinline?import polyaxon|from polyaxon.* import |\"github\.com\/polyaxon\/sdks\/go\/http_client\/v1\/.*\"|require\(('polyaxon-sdk'|\"polyaxon-sdk\")\);|require\(('@polyaxon\/sdk@1\.20\.0'|\"@polyaxon\/sdk@1\.20\.0\")\);|import .* from ('polyaxon-sdk'|\"polyaxon-sdk\");|import .* from ('@polyaxon\/sdk@1\.20\.0'|\"@polyaxon\/sdk@1\.20\.0\");? \\
\midrule
Sacred & \lstinline?import sacred|from sacred.* import ? \\
\midrule
\multirow{2}{*}{SigOpt} & \lstinline?import sigopt|from sigopt.* import |library\(SigOptR\)|import com\.sigopt.*;? \\
\midrule
\multirow{2}{*}{Valohai} & \lstinline!file:valohai\.y(a)?ml|import valohai_yaml|from valohai_yaml.* import ! \\
\midrule
Verta & \lstinline?import verta|from verta.* import ? \\
\midrule
\multirow{6}{*}{Vertex AI} & \lstinline?import google\.cloud\.aiplatform|from google\.cloud import aiplatform|import com\.google\.cloud\.aiplatform.*;|require\((\"@google-cloud\/aiplatform\"|'@google-cloud\/aiplatform')\);|import .* from (\"@google-cloud\/aiplatform\"|'@google-cloud\/aiplatform');? \\
\midrule
Weights\&Biases & \lstinline?import wandb|from wandb.* import |import com\.wandb.*;? \\
\bottomrule
\end{tabular}
\label{tab:tool usage pattern}
\end{table}

%% file: tables/appendix/post-title-keywords.tex
\begin{table}[H]
\centering
\caption{Post title keywords for each curated tool.}
\begin{tabular}{lp{8cm}}
\toprule
\textbf{Tool Name} & \textbf{Keyword List} \\
\midrule
Amazon SageMaker & sagemaker \\
\midrule
Azure Machine Learning & azure machine learning, azure ml, azure-ml, azureml \\
\midrule
ClearML & clearml \\
\midrule
cnvrg.io & cnvrg \\
\midrule
Codalab & codalab \\
\midrule
Comet & comet-ml \\
\midrule
Determined & determined \\
\midrule
Domino & domino \\
\midrule
DVC & dvc \\
\midrule
Guild AI & guild ai, guild-ai, guildai \\
\midrule
H2O AI Cloud & h2o ai, h2o.ai \\
\midrule
Kedro & kedro \\
\midrule
MLflow & mlflow \\
\midrule
MLRun & mlrun \\
\midrule
Neptune & neptune \\
\midrule
Optuna & optuna \\
\midrule
Polyaxon & polyaxon \\
\midrule
Sacred & sacred \\
\midrule
SigOpt & sigopt \\
\midrule
Valohai & valohai \\
\midrule
Verta & modeldb, verta \\
\midrule
Vertex AI & vertex ai, vertex-ai, vertexai \\
\midrule
\multirow{2}{*}{Weights\&Biases} & weights and biases, wandb, Weights\&Biases, weights\&biases, w \& b, w\&b \\
\bottomrule
\end{tabular}
\label{tab:post title keywords}
\end{table}

%% file: tables/appendix/gpt-4.tex
\begin{table}[H]
\centering
\caption{Parameters for GPT-4 prompting.}
\begin{tabular}{llp{7.5cm}}
\toprule
\textbf{Parameter} & \textbf{Value} & \textbf{Description}\\
\midrule
\multirow{2}{*}{\texttt{temperature}} & \multirow{2}{*}{0} & Temperature controls the randomness of the generated text. \\
\midrule
\multirow{3}{*}{\texttt{top\_p}} & \multirow{3}{*}{1} & Top-p sets the cumulative probability threshold to direct the sampling of words to generate coherent and grammatically correct text.\\
\midrule
\multirow{2}{*}{\texttt{frequency\_penalty}} & \multirow{2}{*}{0} & Frequency penalty controls the use of common words or phrases by the model in the generated text.\\
\midrule
\multirow{2}{*}{\texttt{presence\_penalty}} & \multirow{2}{*}{0} & Presence penalty controls the probability that a specific token is produced next in the generated text.\\
\midrule
\multirow{2}{*}{\texttt{max\_tokens}} & \multirow{2}{*}{50} & Max tokens control the maximum number of tokens to generate in the text.\\
\bottomrule
\end{tabular}
\label{tab:gpt-4}
\end{table}

%% file: tables/appendix/hyperparameter-search-space.tex
\begin{table}[H]
\centering
\caption{Hyperparameter search space for topic modeling of post inquiries.}
\begin{tabular}{p{2.5cm}|p{2.9cm}|p{2.8cm}|p{2.6cm}}
\toprule
\textbf{Hyperparameter Name} & \textbf{Range for challenge modeling} & \textbf{Range for solution modeling} & \textbf{Description} \\ 
\midrule
\multirow{2}{*}{\texttt{min\_cluster\_size}} & \multirow{2}{*}{\colorbox{green}{30},40,\ldots,100} & \multirow{2}{*}{\colorbox{green}{20},30,40} & the minimum size of a cluster. \\
\midrule
\multirow{2}{*}{\texttt{min\_samples}} & \multirow{2}{*}{\colorbox{green}{1},2,\ldots,100} & \multirow{2}{*}{1,2,\ldots,\colorbox{green}{6},\ldots,40} & the number of outliers generated.\\
\midrule
\multirow{3}{*}{\texttt{n\_components}} & \multirow{3}{*}{3,4,\colorbox{green}{5},\ldots,10} & \multirow{3}{*}{\colorbox{green}{3},4,\ldots,8} & the dimensionality of the embeddings after reduction. \\
\midrule
\multirow{3}{*}{\texttt{ngram\_range}} & \multirow{3}{*}{1,\colorbox{green}{2},3} & \multirow{3}{*}{1,\colorbox{green}{2},3} & the number of words in the topic representation. \\
\midrule
\multirow{4}{*}{\texttt{embedding\_model}} & \multirow{4}{*}{\colorbox{green}{all-mpnet-base-v2}} & \multirow{4}{*}{\colorbox{green}{all-mpnet-base-v2}} & the transformation of input documents into numerical representations. \\
\midrule
\multirow{4}{*}{\texttt{metrics}} & \multirow{4}{*}{\colorbox{green}{cosine}} & \multirow{4}{*}{\colorbox{green}{cosine}} & similarity measure between two data points in the embedding space. \\
\midrule
\multirow{3}{*}{\texttt{random\_state}} & \multirow{3}{*}{\colorbox{green}{42}} & \multirow{3}{*}{\colorbox{green}{42}} & the algorithmic randomness for reproducibility. \\
\bottomrule
\end{tabular}
\label{tab:hyperparameter search space}
\end{table} 

%% file: tables/appendix/rq1-topic-list.tex
\afterpage{
\begin{longtable}{llll}
\caption{Challenge topic information.} 
\label{rq1:topic list} \\
\toprule
\textbf{Index} & \textbf{Topic} & \textbf{Index} & \textbf{Topic} \\
\midrule
$C_{01}$ & Pipeline Step & $C_{02}$ & Notebook Instance \\\midrule
$C_{03}$ & Docker Image & $C_{04}$ & Version \\\midrule
$C_{05}$ & Log & $C_{06}$ & Plots \\\midrule
$C_{07}$ & Studio & $C_{08}$ & PyTorch \\\midrule
$C_{09}$ & Prediction & $C_{10}$ & Labeling Job \\\midrule
$C_{11}$ & Hyperparameter Tuning & $C_{12}$ & TensorFlow \\\midrule
$C_{13}$ & Spark & $C_{14}$ & RStudio \\\midrule
$C_{15}$ & Workspace & $C_{16}$ & Batch Transform \\\midrule
$C_{17}$ & Web Service & $C_{18}$ & Experiment \\\midrule
$C_{19}$ & Instance & $C_{20}$ & Sweep \\\midrule
$C_{21}$ & Deploying Model & $C_{22}$ & Columns Values \\\midrule
$C_{23}$ & Inference Endpoint & $C_{24}$ & Compute \\\midrule
$C_{25}$ & Account & $C_{26}$ & File Directory \\\midrule
$C_{27}$ & Object Attribute & $C_{28}$ & Parameters Parameter \\\midrule
$C_{29}$ & File Pandas & $C_{30}$ & Git Repo \\\midrule
$C_{31}$ & Training Job & $C_{32}$ & Save Model \\\midrule
$C_{33}$ & Bucket & $C_{34}$ & Limit Exceeded \\\midrule
$C_{35}$ & Data Factory & $C_{36}$ & Format Convert \\\midrule
$C_{37}$ & Server & $C_{38}$ & Feature Store \\\midrule
$C_{39}$ & Environment & $C_{40}$ & Initialization Init \\\midrule
$C_{41}$ & Custom Metrics & $C_{42}$ & Compute Cluster \\\midrule
$C_{43}$ & Model Endpoint & $C_{44}$ & Memory \\\midrule
$C_{45}$ & Stuck Queue & $C_{46}$ & Distributed Training \\\midrule
$C_{47}$ & File Upload & $C_{48}$ & Python \\\midrule
$C_{49}$ & TensorBoard & $C_{50}$ & Deploying Endpoint \\\midrule
$C_{51}$ & Authentication & $C_{52}$ & Module Import \\\midrule
$C_{53}$ & Deployment Model & $C_{54}$ & Artifact Download \\\midrule
$C_{55}$ & File Read & $C_{56}$ & Batch Prediction \\\midrule
$C_{57}$ & API & $C_{58}$ & API Key \\\midrule
$C_{59}$ & File Storage & $C_{60}$ & Creating Terraform \\\midrule
$C_{61}$ & CUDA Memory & $C_{62}$ & Register Model \\\midrule
$C_{63}$ & Log Metrics & $C_{64}$ & Config Configuration \\\midrule
$C_{65}$ & Inference Pipeline & $C_{66}$ & Invoke Endpoint \\\midrule
$C_{67}$ & Blob Storage & $C_{68}$ & Comparison Performance \\\midrule
$C_{69}$ & Loading Model & $C_{70}$ & Pickle File \\\midrule
$C_{71}$ & Endpoint Lambda & $C_{72}$ & File Download \\\midrule
$C_{73}$ & Broken Link & $C_{74}$ & Permissions Authorization \\\midrule
$C_{75}$ & Exporting Data & $C_{76}$ & Datasets \\\midrule
$C_{77}$ & Model Registry & $C_{78}$ & Image Read \\\midrule
$C_{79}$ & Artifacts & $C_{80}$ & Deploy Model \\\midrule
$C_{81}$ & Endpoint Content & $C_{82}$ & Creating Model \\\midrule
$C_{83}$ & Train Model & $C_{84}$ & Score Model \\\midrule
$C_{85}$ & Endpoint Ping & $C_{86}$ & Parallelization Job \\\midrule
$C_{87}$ & Module Usage & $C_{88}$ & Deployment Kubernetes \\\midrule
$C_{89}$ & Data Stores & $C_{90}$ & Permission Denied \\\midrule
$C_{91}$ & Training Reports & $C_{92}$ & Log Model \\\midrule
$C_{93}$ & Model Monitoring & $C_{94}$ & Model Deployment \\\midrule
$C_{95}$ & Training Pipeline & $C_{96}$ & Missing Module \\\midrule
$C_{97}$ & Install Package & $C_{98}$ & Logs CloudWatch \\\midrule
$C_{99}$ & Import Notebook & $C_{100}$ & Training Model \\\midrule
$C_{101}$ & Installation & $C_{102}$ & Quota Request \\\midrule
$C_{103}$ & NEO Compiling & $C_{104}$ & Connection \\\midrule
$C_{105}$ & Training Container & $C_{106}$ & Huggingface Model \\\midrule
$C_{107}$ & Container Registry & $C_{108}$ & Model ONNX \\\midrule
$C_{109}$ & Models & $C_{110}$ & Globals YAML \\\midrule
$C_{111}$ & Deployment ACI & $C_{112}$ & Trained Model \\\midrule
$C_{113}$ & Web Interface & $C_{114}$ & scikit-learn \\\midrule
$C_{115}$ & Dependency & $C_{116}$ & Nested Runs \\\midrule
$C_{117}$ & Import Data & $C_{118}$ & Endpoint Deploying \\\midrule
$C_{119}$ & Endpoint Prediction & $C_{120}$ & Lifecycle Configuration \\\midrule
$C_{121}$ & Output Inputs & $C_{122}$ & Custom Job \\\midrule
$C_{123}$ & Jupyter Notebook & $C_{124}$ & Response Endpoint \\\midrule
$C_{125}$ & Shared Cache & $C_{126}$ & Training YOLO \\\midrule
$C_{127}$ & Checkpoints Training & $C_{128}$ & Endpoint Transitioning \\\midrule
$C_{129}$ & Training File & $C_{130}$ & Inference \\\midrule
$C_{131}$ & Connecting RDS & $C_{132}$ & Model Serving \\\midrule
$C_{133}$ & Tuning Model & & \\\bottomrule
\end{longtable}
}

%% file: tables/appendix/rq2-topic-list.tex
\afterpage{
\begin{longtable}{llll}
\caption{Solution topic information.} 
\label{rq2:topic list} \\
\toprule
\textbf{Index} & \textbf{Topic} & \textbf{Index} & \textbf{Topic} \\
\midrule
$R_{01}$ & Package Upgrade & $R_{02}$ & Recommendation \\\midrule
$R_{03}$ & Distributed Training & $R_{04}$ & Permission \\\midrule
$R_{05}$ & Package Installation & $R_{06}$ & Logging Function \\\midrule
$R_{07}$ & Package Version & $R_{08}$ & Patch and Fix \\\midrule
$R_{09}$ & Network Configuration & $R_{10}$ & Model Creation \\\midrule
$R_{11}$ & Run Configuration & $R_{12}$ & Argument \\\midrule
$R_{13}$ & Code Modification & $R_{14}$ & Job Processing \\\midrule
$R_{15}$ & SDK Upgrade & $R_{16}$ & Storage Mounting \\\midrule
$R_{17}$ & Credential Update & $R_{18}$ & Pipeline Configuration \\\midrule
$R_{19}$ & Resource Limit Adjustment & $R_{20}$ & Directory \\\midrule
$R_{21}$ & Container Customization & $R_{22}$ & Docker \\\midrule
$R_{23}$ & Package Addition & $R_{24}$ & Endpoint Invocation \\\midrule
$R_{25}$ & Environment Creation & $R_{26}$ & Pipeline Deployment \\\midrule
$R_{27}$ & Wait Time & $R_{28}$ & Command Line Usage \\\midrule
$R_{29}$ & File Deletion & $R_{30}$ & Service Provisioning \\\midrule
$R_{31}$ & Column Manipulation & $R_{32}$ & Model Registration \\\midrule
$R_{33}$ & Feature Filtering & $R_{34}$ & Environment Variable \\\midrule
$R_{35}$ & Compute Instance & $R_{36}$ & File Download \\\midrule
$R_{37}$ & SDK Usage & $R_{38}$ & Model Prediction \\\midrule
$R_{39}$ & Experiment Creation & $R_{40}$ & Data Conversion \\\midrule
$R_{41}$ & Package Downgrade & $R_{42}$ & Dataset Creation \\\midrule
$R_{43}$ & Git Repository & $R_{44}$ & Package Reinstallation \\\midrule
$R_{45}$ & Lifecycle Configuration & $R_{46}$ & Support Ticket Creation \\\midrule
$R_{47}$ & API Integration & $R_{48}$ & Data Visualization \\\midrule
$R_{49}$ & Filepath Update & $R_{50}$ & Metrics \\\midrule
$R_{51}$ & Hyperlink Update & $R_{52}$ & Model File Handling \\\midrule
$R_{53}$ & Lambda Function & $R_{54}$ & Data Transformation \\\midrule
$R_{55}$ & Input Schema & $R_{56}$ & Parameter Update \\\midrule
$R_{57}$ & Character Removal & $R_{58}$ & Troubleshooting Guidance \\\midrule
$R_{59}$ & Function Modification & $R_{60}$ & Resource Increase \\\midrule
$R_{61}$ & Syntax Update & $R_{62}$ & Package Removal \\\midrule
$R_{63}$ & Exception Handling & $R_{64}$ & Filepath Modification \\\midrule
$R_{65}$ & File Load & $R_{66}$ & Notebook Usage \\\midrule
$R_{67}$ & Python Version & $R_{68}$ & Tracking Configuration \\\midrule
$R_{69}$ & Package Import & $R_{70}$ & Datatype Conversion \\\midrule
$R_{71}$ & Task Modeling & $R_{72}$ & Workspace Creation \\\midrule
$R_{73}$ & Region Support & $R_{74}$ & Logging Update \\\midrule
$R_{75}$ & Data Import/Export & $R_{76}$ & Kernel Restart \\\midrule
$R_{77}$ & Bucket Manipulation & $R_{78}$ & Docker Update \\\midrule
$R_{79}$ & Model Deployment & $R_{80}$ & Inference Pipeline \\\midrule
$R_{81}$ & Account Creation & $R_{82}$ & Comparison \\\midrule
$R_{83}$ & Session Creation & $R_{84}$ & Parameter Removal \\\midrule
$R_{85}$ & Notebook Instance & $R_{86}$ & Documentation Update \\\bottomrule
\end{longtable}
}

%% file: main.bbl
\begin{thebibliography}{100}
\providecommand{\url}[1]{{#1}}
\providecommand{\urlprefix}{URL }
\expandafter\ifx\csname urlstyle\endcsname\relax
  \providecommand{\doi}[1]{DOI~\discretionary{}{}{}#1}\else
  \providecommand{\doi}{DOI~\discretionary{}{}{}\begingroup \urlstyle{rm}\Url}\fi

\bibitem{replication_package}
\urlprefix\url{https://github.com/zhimin-z/Asset-Management-Topic-Modeling}.
\newblock \url{https://github.com/zhimin-z/MSR-Asset-Management}, \url{https://github.com/zhimin-z/QA-Asset-Management}

\bibitem{mlops}
\urlprefix\url{https://github.com/topics/mlops}

\bibitem{machine-learning-engineering}
\urlprefix\url{https://github.com/topics/machine-learning-engineering}

\bibitem{artifact-management}
\urlprefix\url{https://github.com/topics/artifact-management}

\bibitem{data-management}
\urlprefix\url{https://github.com/topics/data-management}

\bibitem{experiments}
\urlprefix\url{https://github.com/topics/experiments}

\bibitem{experiment-management}
\urlprefix\url{https://github.com/topics/experiment-management}

\bibitem{awesome-ml-experiment-management}
\urlprefix\url{https://github.com/awesome-mlops/awesome-ml-experiment-management}

\bibitem{experiment-tracking}
\urlprefix\url{https://github.com/topics/experiment-tracking}

\bibitem{lifecycle-management}
\urlprefix\url{https://github.com/topics/lifecycle-management}

\bibitem{model-management}
\urlprefix\url{https://github.com/topics/data-management}

\bibitem{project-management}
\urlprefix\url{https://github.com/topics/project-management}

\bibitem{workflow-management}
\urlprefix\url{https://github.com/topics/workflow-management}

\bibitem{agrawal2007five}
Agrawal, N., Bolosky, W.J., Douceur, J.R., Lorch, J.R.: A five-year study of file-system metadata.
\newblock ACM Transactions on Storage (TOS) \textbf{3}(3), 9--es (2007)

\bibitem{aguilar2021ease}
Aguilar~Melgar, L., Dao, D., Gan, S., G{\"u}rel, N.M., Hollenstein, N., Jiang, J., Karla{\v{s}}, B., Lemmin, T., Li, T., Li, Y., et~al.: Ease. ml: a lifecycle management system for machine learning.
\newblock In: Proceedings of the Annual Conference on Innovative Data Systems Research (CIDR), 2021. CIDR (2021)

\bibitem{Ahmed2018WhatDC}
Ahmed, S., Bagherzadeh, M.: What do concurrency developers ask about?: a large-scale study using stack overflow.
\newblock Proceedings of the 12th ACM/IEEE International Symposium on Empirical Software Engineering and Measurement  (2018)

\bibitem{alberti2018deepdiva}
Alberti, M., Pondenkandath, V., W{\"u}rsch, M., Ingold, R., Liwicki, M.: Deepdiva: a highly-functional python framework for reproducible experiments.
\newblock In: 2018 16th International Conference on Frontiers in Handwriting Recognition (ICFHR), pp. 423--428. IEEE (2018)

\bibitem{amershi2019software}
Amershi, S., Begel, A., Bird, C., DeLine, R., Gall, H., Kamar, E., Nagappan, N., Nushi, B., Zimmermann, T.: Software engineering for machine learning: A case study.
\newblock In: 2019 IEEE/ACM 41st International Conference on Software Engineering: Software Engineering in Practice (ICSE-SEIP), pp. 291--300. IEEE (2019)

\bibitem{bagherzadeh2019going}
Bagherzadeh, M., Khatchadourian, R.: Going big: a large-scale study on what big data developers ask.
\newblock In: Proceedings of the 2019 27th ACM joint meeting on european software engineering conference and symposium on the foundations of software engineering, pp. 432--442 (2019)

\bibitem{bahrampour2015comparative}
Bahrampour, S., Ramakrishnan, N., Schott, L., Shah, M.: Comparative study of deep learning software frameworks.
\newblock arXiv preprint arXiv:1511.06435  (2015)

\bibitem{baier2019challenges}
Baier, L., J{\"o}hren, F., Seebacher, S.: Challenges in the deployment and operation of machine learning in practice.
\newblock In: ECIS, vol.~1 (2019)

\bibitem{barde2017overview}
Barde, B.V., Bainwad, A.M.: An overview of topic modeling methods and tools.
\newblock In: 2017 International Conference on Intelligent Computing and Control Systems (ICICCS), pp. 745--750. IEEE (2017)

\bibitem{barrak2021co}
Barrak, A., Eghan, E.E., Adams, B.: On the co-evolution of ml pipelines and source code-empirical study of dvc projects.
\newblock In: 2021 IEEE International Conference on Software Analysis, Evolution and Reengineering (SANER), pp. 422--433. IEEE (2021)

\bibitem{belguidoum2007dependency}
Belguidoum, M., Dagnat, F.: Dependency management in software component deployment.
\newblock Electronic Notes in theoretical computer science \textbf{182}, 17--32 (2007)

\bibitem{benitez2021titan}
Ben{\'\i}tez-Hidalgo, A., Barba-Gonz{\'a}lez, C., Garc{\'\i}a-Nieto, J., Guti{\'e}rrez-Moncayo, P., Paneque, M., Nebro, A.J., del Mar Rold{\'a}n-Garc{\'\i}a, M., Aldana-Montes, J.F., Navas-Delgado, I.: Titan: A knowledge-based platform for big data workflow management.
\newblock Knowledge-Based Systems \textbf{232}, 107489 (2021)

\bibitem{benjamini1995controlling}
Benjamini, Y., Hochberg, Y.: Controlling the false discovery rate: a practical and powerful approach to multiple testing.
\newblock Journal of the Royal statistical society: series B (Methodological) \textbf{57}(1), 289--300 (1995)

\bibitem{bhattacharjee2019stratum}
Bhattacharjee, A., Barve, Y., Khare, S., Bao, S., Gokhale, A., Damiano, T.: Stratum: A serverless framework for the lifecycle management of machine learning-based data analytics tasks.
\newblock In: 2019 USENIX Conference on Operational Machine Learning (OpML 19), pp. 59--61 (2019)

\bibitem{bommasani2021opportunities}
Bommasani, R., Hudson, D.A., Adeli, E., Altman, R., Arora, S., von Arx, S., Bernstein, M.S., Bohg, J., Bosselut, A., Brunskill, E., et~al.: On the opportunities and risks of foundation models.
\newblock arXiv preprint arXiv:2108.07258  (2021)

\bibitem{borges2018s}
Borges, H., Valente, M.T.: What’s in a github star? understanding repository starring practices in a social coding platform.
\newblock Journal of Systems and Software \textbf{146}, 112--129 (2018)

\bibitem{bravo2022scanflow}
Bravo-Rocca, G., Liu, P., Guitart, J., Dholakia, A., Ellison, D., Falkanger, J., Hodak, M.: Scanflow: A multi-graph framework for machine learning workflow management, supervision, and debugging.
\newblock Expert Systems with Applications \textbf{202}, 117232 (2022)

\bibitem{campbell2013coding}
Campbell, J.L., Quincy, C., Osserman, J., Pedersen, O.K.: Coding in-depth semistructured interviews: Problems of unitization and intercoder reliability and agreement.
\newblock Sociological methods \& research \textbf{42}(3), 294--320 (2013)

\bibitem{chard2019dlhub}
Chard, R., Li, Z., Chard, K., Ward, L., Babuji, Y., Woodard, A., Tuecke, S., Blaiszik, B., Franklin, M.J., Foster, I.: Dlhub: Model and data serving for science.
\newblock In: 2019 IEEE International Parallel and Distributed Processing Symposium (IPDPS), pp. 283--292. IEEE (2019)

\bibitem{chen2020developments}
Chen, A., Chow, A., Davidson, A., DCunha, A., Ghodsi, A., Hong, S.A., Konwinski, A., Mewald, C., Murching, S., Nykodym, T., et~al.: Developments in mlflow: A system to accelerate the machine learning lifecycle.
\newblock In: Proceedings of the fourth international workshop on data management for end-to-end machine learning, pp. 1--4 (2020)

\bibitem{chen2023practitioners}
Chen, Y., Fernandes, E., Adams, B., Hassan, A.E.: On practitioners' concerns when adopting service mesh frameworks.
\newblock Empirical Software Engineering  (2023)

\bibitem{chen2020comprehensive}
Chen, Z., Cao, Y., Liu, Y., Wang, H., Xie, T., Liu, X.: A comprehensive study on challenges in deploying deep learning based software.
\newblock In: Proceedings of the 28th ACM Joint Meeting on European Software Engineering Conference and Symposium on the Foundations of Software Engineering, pp. 750--762 (2020)

\bibitem{cheng2023gpt}
Cheng, L., Li, X., Bing, L.: Is gpt-4 a good data analyst?
\newblock arXiv preprint arXiv:2305.15038  (2023)

\bibitem{coelho2017modern}
Coelho, J., Valente, M.T.: Why modern open source projects fail.
\newblock In: Proceedings of the 2017 11th Joint meeting on foundations of software engineering, pp. 186--196 (2017)

\bibitem{cramer1999mathematical}
Cram{\'e}r, H.: Mathematical methods of statistics, vol.~43.
\newblock Princeton university press (1999)

\bibitem{diamantopoulos2023semantically}
Diamantopoulos, T., Nastos, D.N., Symeonidis, A.: Semantically-enriched jira issue tracking data.
\newblock In: 2023 IEEE/ACM 20th International Conference on Mining Software Repositories (MSR), pp. 218--222. IEEE (2023)

\bibitem{dunn1961multiple}
Dunn, O.J.: Multiple comparisons among means.
\newblock Journal of the American statistical association \textbf{56}(293), 52--64 (1961)

\bibitem{enck2022top}
Enck, W., Williams, L.: Top five challenges in software supply chain security: Observations from 30 industry and government organizations.
\newblock IEEE Security \& Privacy \textbf{20}(2), 96--100 (2022)

\bibitem{esparrachiari2018tracking}
Esparrachiari, S., Reilly, T., Rentz, A.: Tracking and controlling microservice dependencies: Dependency management is a crucial part of system and software design.
\newblock Queue \textbf{16}(4), 44--65 (2018)

\bibitem{awesome-production-machine-learning}
EthicalML: awesome-production-machine-learning: A curated list of awesome open source libraries to deploy, monitor, version and scale your machine learning.
\newblock \urlprefix\url{https://github.com/EthicalML/awesome-production-machine-learning}

\bibitem{ferenc2020deep}
Ferenc, R., Viszkok, T., Aladics, T., J{\'a}sz, J., Heged{\H{u}}s, P.: Deep-water framework: The swiss army knife of humans working with machine learning models.
\newblock SoftwareX \textbf{12}, 100551 (2020)

\bibitem{franccoise2021marcelle}
Fran{\c{c}}oise, J., Caramiaux, B., Sanchez, T.: Marcelle: composing interactive machine learning workflows and interfaces.
\newblock In: The 34th Annual ACM Symposium on User Interface Software and Technology, pp. 39--53 (2021)

\bibitem{garcia2018context}
Garcia, R., Sreekanti, V., Yadwadkar, N., Crankshaw, D., Gonzalez, J.E., Hellerstein, J.M.: Context: The missing piece in the machine learning lifecycle.
\newblock In: KDD CMI Workshop, vol. 114, pp. 1--4 (2018)

\bibitem{gharibi2019automated}
Gharibi, G., Walunj, V., Alanazi, R., Rella, S., Lee, Y.: Automated management of deep learning experiments.
\newblock In: Proceedings of the 3rd International Workshop on Data Management for End-to-End Machine Learning, pp. 1--4 (2019)

\bibitem{gilardi2023chatgpt}
Gilardi, F., Alizadeh, M., Kubli, M.: Chatgpt outperforms crowd-workers for text-annotation tasks.
\newblock arXiv preprint arXiv:2303.15056  (2023)

\bibitem{giray2021software}
Giray, G.: A software engineering perspective on engineering machine learning systems: State of the art and challenges.
\newblock Journal of Systems and Software \textbf{180}, 111031 (2021)

\bibitem{goniwada2022observability}
Goniwada, S.R., Goniwada, S.R.: Observability.
\newblock Cloud Native Architecture and Design: A Handbook for Modern Day Architecture and Design with Enterprise-Grade Examples pp. 661--676 (2022)

\bibitem{gou2021knowledge}
Gou, J., Yu, B., Maybank, S.J., Tao, D.: Knowledge distillation: A survey.
\newblock International Journal of Computer Vision \textbf{129}, 1789--1819 (2021)

\bibitem{groeneveld2024olmo}
Groeneveld, D., Beltagy, I., Walsh, P., Bhagia, A., Kinney, R., Tafjord, O., Jha, A.H., Ivison, H., Magnusson, I., Wang, Y., et~al.: Olmo: Accelerating the science of language models.
\newblock arXiv preprint arXiv:2402.00838  (2024)

\bibitem{grootendorst2022bertopic}
Grootendorst, M.: Bertopic: Neural topic modeling with a class-based tf-idf procedure.
\newblock arXiv preprint arXiv:2203.05794  (2022)

\bibitem{grubb2003software}
Grubb, P., Takang, A.A.: Software maintenance: concepts and practice.
\newblock World Scientific (2003)

\bibitem{gu2023self}
Gu, H., He, H., Zhou, M.: Self-admitted library migrations in java, javascript, and python packaging ecosystems: A comparative study.
\newblock In: 2023 IEEE International Conference on Software Analysis, Evolution and Reengineering (SANER), pp. 627--638. IEEE (2023)

\bibitem{hartley2020dtoolai}
Hartley, M., Olsson, T.S.: dtoolai: Reproducibility for deep learning.
\newblock Patterns \textbf{1}(5) (2020)

\bibitem{hastie2009elements}
Hastie, T., Tibshirani, R., Friedman, J.H., Friedman, J.H.: The elements of statistical learning: data mining, inference, and prediction, vol.~2.
\newblock Springer (2009)

\bibitem{hewage2022machine}
Hewage, N., Meedeniya, D.: Machine learning operations: A survey on mlops tool support.
\newblock arXiv preprint arXiv:2202.10169  (2022)

\bibitem{hummer2019modelops}
Hummer, W., Muthusamy, V., Rausch, T., Dube, P., El~Maghraoui, K., Murthi, A., Oum, P.: Modelops: Cloud-based lifecycle management for reliable and trusted ai.
\newblock In: 2019 IEEE International Conference on Cloud Engineering (IC2E), pp. 113--120. IEEE (2019)

\bibitem{idowu2021asset}
Idowu, S., Str\"{u}ber, D., Berger, T.: Asset management in machine learning: State-of-research and state-of-practice.
\newblock ACM Comput. Surv.  (2022).
\newblock \doi{10.1145/3543847}.
\newblock \urlprefix\url{https://doi.org/10.1145/3543847}.
\newblock Just Accepted

\bibitem{idowu2022emmm}
Idowu, S., Str{\"u}ber, D., Berger, T.: Emmm: A unified meta-model for tracking machine learning experiments.
\newblock In: 2022 48th Euromicro Conference on Software Engineering and Advanced Applications (SEAA), pp. 48--55. IEEE (2022)

\bibitem{isah2019survey}
Isah, H., Abughofa, T., Mahfuz, S., Ajerla, D., Zulkernine, F., Khan, S.: A survey of distributed data stream processing frameworks.
\newblock IEEE Access \textbf{7}, 154300--154316 (2019)

\bibitem{izquierdo2017empirical}
Izquierdo, J.L.C., Cosentino, V., Cabot, J.: An empirical study on the maturity of the eclipse modeling ecosystem.
\newblock In: 2017 ACM/IEEE 20th International Conference on Model Driven Engineering Languages and Systems (MODELS), pp. 292--302. IEEE (2017)

\bibitem{jalali2012systematic}
Jalali, S., Wohlin, C.: Systematic literature studies: database searches vs. backward snowballing.
\newblock In: Proceedings of the ACM-IEEE international symposium on Empirical software engineering and measurement, pp. 29--38 (2012)

\bibitem{jiang2023mistral}
Jiang, A.Q., Sablayrolles, A., Mensch, A., Bamford, C., Chaplot, D.S., Casas, D.d.l., Bressand, F., Lengyel, G., Lample, G., Saulnier, L., et~al.: Mistral 7b.
\newblock arXiv preprint arXiv:2310.06825  (2023)

\bibitem{jiang2023empirical}
Jiang, W., Synovic, N., Hyatt, M., Schorlemmer, T.R., Sethi, R., Lu, Y.H., Thiruvathukal, G.K., Davis, J.C.: An empirical study of pre-trained model reuse in the hugging face deep learning model registry.
\newblock arXiv preprint arXiv:2303.02552  (2023)

\bibitem{awesome-mlops}
Kelvins: awesome-mlops: A curated list of awesome mlops tools.
\newblock \urlprefix\url{https://github.com/kelvins/awesome-mlops}

\bibitem{khondhu2013all}
Khondhu, J., Capiluppi, A., Stol, K.J.: Is it all lost? a study of inactive open source projects.
\newblock In: Open Source Software: Quality Verification: 9th IFIP WG 2.13 International Conference, OSS 2013, Koper-Capodistria, Slovenia, June 25-28, 2013. Proceedings 9, pp. 61--79. Springer (2013)

\bibitem{kitchenham1999towards}
Kitchenham, B.A., Travassos, G.H., Von~Mayrhauser, A., Niessink, F., Schneidewind, N.F., Singer, J., Takada, S., Vehvilainen, R., Yang, H.: Towards an ontology of software maintenance.
\newblock Journal of Software Maintenance: Research and Practice \textbf{11}(6), 365--389 (1999)

\bibitem{klaise2020monitoring}
Klaise, J., Van~Looveren, A., Cox, C., Vacanti, G., Coca, A.: Monitoring and explainability of models in production.
\newblock arXiv preprint arXiv:2007.06299  (2020)

\bibitem{kreutz2014software}
Kreutz, D., Ramos, F.M., Verissimo, P.E., Rothenberg, C.E., Azodolmolky, S., Uhlig, S.: Software-defined networking: A comprehensive survey.
\newblock Proceedings of the IEEE \textbf{103}(1), 14--76 (2014)

\bibitem{kumar2017data}
Kumar, A., Boehm, M., Yang, J.: Data management in machine learning: Challenges, techniques, and systems.
\newblock In: Proceedings of the 2017 ACM International Conference on Management of Data, pp. 1717--1722 (2017)

\bibitem{lapan2018deep}
Lapan, M.: Deep Reinforcement Learning Hands-On: Apply modern RL methods, with deep Q-networks, value iteration, policy gradients, TRPO, AlphaGo Zero and more.
\newblock Packt Publishing Ltd (2018)

\bibitem{le2023veml}
Le, V.D.: Veml: An end-to-end machine learning lifecycle for large-scale and high-dimensional data.
\newblock arXiv preprint arXiv:2304.13037  (2023)

\bibitem{liu2024tuning}
Liu, A., Han, X., Wang, Y., Tsvetkov, Y., Choi, Y., Smith, N.A.: Tuning language models by proxy.
\newblock arXiv preprint arXiv:2401.08565  (2024)

\bibitem{liu2023gpteval}
Liu, Y., Iter, D., Xu, Y., Wang, S., Xu, R., Zhu, C.: Gpteval: Nlg evaluation using gpt-4 with better human alignment.
\newblock arXiv preprint arXiv:2303.16634  (2023)

\bibitem{loeliger2012version}
Loeliger, J., McCullough, M.: Version Control with Git: Powerful tools and techniques for collaborative software development.
\newblock " O'Reilly Media, Inc." (2012)

\bibitem{lu2013study}
Lu, L., Arpaci-Dusseau, A.C., Arpaci-Dusseau, R.H., Lu, S.: A study of linux file system evolution.
\newblock In: 11th USENIX Conference on File and Storage Technologies (FAST 13), pp. 31--44 (2013)

\bibitem{manvi2014resource}
Manvi, S.S., Shyam, G.K.: Resource management for infrastructure as a service (iaas) in cloud computing: A survey.
\newblock Journal of network and computer applications \textbf{41}, 424--440 (2014)

\bibitem{mchugh2012interrater}
McHugh, M.L.: Interrater reliability: the kappa statistic.
\newblock Biochemia medica \textbf{22}(3), 276--282 (2012)

\bibitem{mcinnes2017hdbscan}
McInnes, L., Healy, J., Astels, S.: hdbscan: Hierarchical density based clustering.
\newblock J. Open Source Softw. \textbf{2}(11), 205 (2017)

\bibitem{mckinney2011pandas}
McKinney, W., et~al.: pandas: a foundational python library for data analysis and statistics.
\newblock Python for high performance and scientific computing \textbf{14}(9), 1--9 (2011)

\bibitem{melin2023tackling}
Melin, P.D.: Tackling version management and reproducibility in mlops  (2023)

\bibitem{mens2014survivability}
Mens, T., Goeminne, M., Raja, U., Serebrenik, A.: Survivability of software projects in gnome--a replication study.
\newblock In: 7th International Seminar Series on Advanced Techniques \& Tools for Software Evolution (SATToSE), pp. 79--82 (2014)

\bibitem{miao2017provdb}
Miao, H., Chavan, A., Deshpande, A.: Provdb: Lifecycle management of collaborative analysis workflows.
\newblock In: Proceedings of the 2nd Workshop on Human-in-the-Loop Data Analytics, pp. 1--6 (2017)

\bibitem{miao2017modelhub}
Miao, H., Li, A., Davis, L.S., Deshpande, A.: Modelhub: Deep learning lifecycle management.
\newblock In: 2017 IEEE 33rd International Conference on Data Engineering (ICDE), pp. 1393--1394. IEEE (2017)

\bibitem{miao2017towards}
Miao, H., Li, A., Davis, L.S., Deshpande, A.: Towards unified data and lifecycle management for deep learning.
\newblock In: 2017 IEEE 33rd International Conference on Data Engineering (ICDE), pp. 571--582. IEEE (2017)

\bibitem{miotto2018deep}
Miotto, R., Wang, F., Wang, S., Jiang, X., Dudley, J.T.: Deep learning for healthcare: review, opportunities and challenges.
\newblock Briefings in bioinformatics \textbf{19}(6), 1236--1246 (2018)

\bibitem{moreno2020managing}
Moreno, M., Louren{\c{c}}o, V., Fiorini, S.R., Costa, P., Brand{\~a}o, R., Civitarese, D., Cerqueira, R.: Managing machine learning workflow components.
\newblock International Journal of Semantic Computing \textbf{14}(02), 295--309 (2020)

\bibitem{moreschi2023toward}
Moreschi, S., Recupito, G., Lenarduzzi, V., Palomba, F., Hastbacka, D., Taibi, D.: Toward end-to-end mlops tools map: A preliminary study based on a multivocal literature review.
\newblock arXiv preprint arXiv:2304.03254  (2023)

\bibitem{munappy2022data}
Munappy, A.R., Bosch, J., Olsson, H.H., Arpteg, A., Brinne, B.: Data management for production quality deep learning models: Challenges and solutions.
\newblock Journal of Systems and Software \textbf{191}, 111359 (2022)

\bibitem{mustafa2015resource}
Mustafa, S., Nazir, B., Hayat, A., Madani, S.A., et~al.: Resource management in cloud computing: Taxonomy, prospects, and challenges.
\newblock Computers \& Electrical Engineering \textbf{47}, 186--203 (2015)

\bibitem{nagy2018survey}
Nagy, A.M., Simon, V.: Survey on traffic prediction in smart cities.
\newblock Pervasive and Mobile Computing \textbf{50}, 148--163 (2018)

\bibitem{namaki2020vamsa}
Namaki, M.H., Floratou, A., Psallidas, F., Krishnan, S., Agrawal, A., Wu, Y.: Vamsa: Tracking provenance in data science scripts.
\newblock arXiv preprint arXiv:2001.01861  (2020)

\bibitem{nguyen2019machine}
Nguyen, G., Dlugolinsky, S., Bob{\'a}k, M., Tran, V., L{\'o}pez~Garc{\'\i}a, {\'A}., Heredia, I., Mal{\'\i}k, P., Hluch{\`y}, L.: Machine learning and deep learning frameworks and libraries for large-scale data mining: a survey.
\newblock Artificial Intelligence Review \textbf{52}, 77--124 (2019)

\bibitem{openja2020analysis}
Openja, M., Adams, B., Khomh, F.: Analysis of modern release engineering topics: A large-scale study using stackoverflow.
\newblock In: Proceedings of the 36th International Conference on Software Maintenance and Evolution (ICSME), pp. 104--114 (2020)

\bibitem{paleyes2022challenges}
Paleyes, A., Urma, R.G., Lawrence, N.D.: Challenges in deploying machine learning: a survey of case studies.
\newblock ACM Computing Surveys \textbf{55}(6), 1--29 (2022)

\bibitem{parra2022comparative}
Parra, E., Alahmadi, M., Ellis, A., Haiduc, S.: A comparative study and analysis of developer communications on slack and gitter.
\newblock Empirical Software Engineering \textbf{27}(2), 40 (2022)

\bibitem{pavao2022codalab}
Pavao, A., Guyon, I., Letournel, A.C., Bar{\'o}, X., Escalante, H., Escalera, S., Thomas, T., Xu, Z.: Codalab competitions: An open source platform to organize scientific challenges.
\newblock Ph.D. thesis, Universit{\'e} Paris-Saclay, FRA. (2022)

\bibitem{pearson1900x}
Pearson, K.: X. on the criterion that a given system of deviations from the probable in the case of a correlated system of variables is such that it can be reasonably supposed to have arisen from random sampling.
\newblock The London, Edinburgh, and Dublin Philosophical Magazine and Journal of Science \textbf{50}(302), 157--175 (1900)

\bibitem{peili2018deep}
Peili, Y., Xuezhen, Y., Jian, Y., Lingfeng, Y., Hui, Z., Jimin, L.: Deep learning model management for coronary heart disease early warning research.
\newblock In: 2018 IEEE 3rd International Conference on Cloud Computing and Big Data Analysis (ICCCBDA), pp. 552--557. IEEE (2018)

\bibitem{polyzotis2018data}
Polyzotis, N., Roy, S., Whang, S.E., Zinkevich, M.: Data lifecycle challenges in production machine learning: a survey.
\newblock ACM SIGMOD Record \textbf{47}(2), 17--28 (2018)

\bibitem{recupito2022multivocal}
Recupito, G., Pecorelli, F., Catolino, G., Moreschini, S., Di~Nucci, D., Palomba, F., Tamburri, D.A.: A multivocal literature review of mlops tools and features.
\newblock In: 2022 48th Euromicro Conference on Software Engineering and Advanced Applications (SEAA), pp. 84--91. IEEE (2022)

\bibitem{rigby2009collaboration}
Rigby, P.C., Barr, E.T., Bird, C., German, D.M., Devanbu, P.: Collaboration and governance with distributed version control.
\newblock ACM Transactions on Software Engineering and Methodology, Submission number TOSEM-2009-0087 p.~33 (2009)

\bibitem{rochkind1975source}
Rochkind, M.J.: The source code control system.
\newblock IEEE transactions on Software Engineering (4), 364--370 (1975)

\bibitem{rosen2016mobile}
Rosen, C., Shihab, E.: What are mobile developers asking about? a large scale study using stack overflow.
\newblock Empirical Software Engineering \textbf{21}, 1192--1223 (2016)

\bibitem{ruf2021demystifying}
Ruf, P., Madan, M., Reich, C., Ould-Abdeslam, D.: Demystifying mlops and presenting a recipe for the selection of open-source tools.
\newblock Applied Sciences \textbf{11}(19), 8861 (2021)

\bibitem{sallou2024breaking}
Sallou, J., Durieux, T., Panichella, A.: Breaking the silence: the threats of using llms in software engineering.
\newblock In: ACM/IEEE 46th International Conference on Software Engineering. ACM/IEEE (2024)

\bibitem{schelter2015challenges}
Schelter, S., Biessmann, F., Januschowski, T., Salinas, D., Seufert, S., Szarvas, G.: On challenges in machine learning model management  (2015)

\bibitem{schelter2018declarative}
Schelter, S., B{\"o}se, J.H., Kirschnick, J., Klein, T., Seufert, S.: Declarative metadata management: A missing piece in end-to-end machine learning  (2018)

\bibitem{schick2020s}
Schick, T., Sch{\"u}tze, H.: It's not just size that matters: Small language models are also few-shot learners.
\newblock arXiv preprint arXiv:2009.07118  (2020)

\bibitem{schlegel2023management}
Schlegel, M., Sattler, K.U.: Management of machine learning lifecycle artifacts: A survey.
\newblock ACM SIGMOD Record \textbf{51}(4), 18--35 (2023)

\bibitem{sculley2015hidden}
Sculley, D., Holt, G., Golovin, D., Davydov, E., Phillips, T., Ebner, D., Chaudhary, V., Young, M., Crespo, J.F., Dennison, D.: Hidden technical debt in machine learning systems.
\newblock Advances in neural information processing systems \textbf{28} (2015)

\bibitem{soomro2016information}
Soomro, Z.A., Shah, M.H., Ahmed, J.: Information security management needs more holistic approach: A literature review.
\newblock International journal of information management \textbf{36}(2), 215--225 (2016)

\bibitem{sorokin2008utility}
Sorokin, A., Forsyth, D.: Utility data annotation with amazon mechanical turk.
\newblock In: 2008 IEEE computer society conference on computer vision and pattern recognition workshops, pp. 1--8. IEEE (2008)

\bibitem{squire2015should}
Squire, M.: "should we move to stack overflow?" measuring the utility of social media for developer support.
\newblock In: 2015 IEEE/ACM 37th IEEE International Conference on Software Engineering, vol.~2, pp. 219--228. IEEE (2015)

\bibitem{storey2002direct}
Storey, J.D.: A direct approach to false discovery rates.
\newblock Journal of the Royal Statistical Society Series B: Statistical Methodology \textbf{64}(3), 479--498 (2002)

\bibitem{sun2020gallery}
Sun, C., Azari, N., Turakhia, C.: Gallery: A machine learning model management system at uber.
\newblock In: EDBT, vol.~20, pp. 474--485 (2020)

\bibitem{sung2017nsml}
Sung, N., Kim, M., Jo, H., Yang, Y., Kim, J., Lausen, L., Kim, Y., Lee, G., Kwak, D., Ha, J.W., et~al.: Nsml: A machine learning platform that enables you to focus on your models.
\newblock arXiv preprint arXiv:1712.05902  (2017)

\bibitem{syed2017full}
Syed, S., Spruit, M.: Full-text or abstract? examining topic coherence scores using latent dirichlet allocation.
\newblock In: 2017 IEEE International conference on data science and advanced analytics (DSAA), pp. 165--174. IEEE (2017)

\bibitem{symeonidis2022mlops}
Symeonidis, G., Nerantzis, E., Kazakis, A., Papakostas, G.A.: Mlops-definitions, tools and challenges.
\newblock In: 2022 IEEE 12th Annual Computing and Communication Workshop and Conference (CCWC), pp. 0453--0460. IEEE (2022)

\bibitem{tao2023code}
Tao, L., Cazan, A.P., Ibraimoski, S., Moran, S.: Code librarian: A software package recommendation system.
\newblock In: 2023 IEEE/ACM 45th International Conference on Software Engineering: Software Engineering in Practice (ICSE-SEIP), pp. 196--198. IEEE (2023)

\bibitem{awesome-LLMOPS}
Tensorchord: awesome-llmops: An awesome curated list of best llmops tools for developers.
\newblock \urlprefix\url{https://github.com/tensorchord/Awesome-LLMOps}

\bibitem{touvron2023llama}
Touvron, H., Martin, L., Stone, K., Albert, P., Almahairi, A., Babaei, Y., Bashlykov, N., Batra, S., Bhargava, P., Bhosale, S., et~al.: Llama 2: Open foundation and fine-tuned chat models.
\newblock arXiv preprint arXiv:2307.09288  (2023)

\bibitem{treude2011programmers}
Treude, C., Barzilay, O., Storey, M.A.: How do programmers ask and answer questions on the web?(nier track).
\newblock In: Proceedings of the 33rd international conference on software engineering, pp. 804--807 (2011)

\bibitem{tsay2018runway}
Tsay, J., Mummert, T., Bobroff, N., Braz, A., Westerink, P., Hirzel, M.: Runway: machine learning model experiment management tool.
\newblock In: Conference on systems and machine learning (sysML) (2018)

\bibitem{vadlamani2020studying}
Vadlamani, S.L., Baysal, O.: Studying software developer expertise and contributions in stack overflow and github.
\newblock In: 2020 IEEE International Conference on Software Maintenance and Evolution (ICSME), pp. 312--323. IEEE (2020)

\bibitem{vartak2018modeldb}
Vartak, M., Madden, S.: Modeldb: Opportunities and challenges in managing machine learning models.
\newblock IEEE Data Eng. Bull. \textbf{41}(4), 16--25 (2018)

\bibitem{vasilescu2013stackoverflow}
Vasilescu, B., Filkov, V., Serebrenik, A.: Stackoverflow and github: Associations between software development and crowdsourced knowledge.
\newblock In: 2013 International Conference on Social Computing, pp. 188--195. IEEE (2013)

\bibitem{venkatesh2016client}
Venkatesh, P.K., Wang, S., Zhang, F., Zou, Y., Hassan, A.E.: What do client developers concern when using web apis? an empirical study on developer forums and stack overflow.
\newblock In: 2016 IEEE International Conference on Web Services (ICWS), pp. 131--138. IEEE (2016)

\bibitem{wang2019various}
Wang, Z., Liu, K., Li, J., Zhu, Y., Zhang, Y.: Various frameworks and libraries of machine learning and deep learning: a survey.
\newblock Archives of computational methods in engineering pp. 1--24 (2019)

\bibitem{werlinger2009integrated}
Werlinger, R., Hawkey, K., Beznosov, K.: An integrated view of human, organizational, and technological challenges of it security management.
\newblock Information Management \& Computer Security \textbf{17}(1), 4--19 (2009)

\bibitem{wood2008card}
Wood, J.R., Wood, L.E.: Card sorting: current practices and beyond.
\newblock Journal of Usability Studies \textbf{4}(1), 1--6 (2008)

\bibitem{wozniak2018candle}
Wozniak, J.M., Jain, R., Balaprakash, P., Ozik, J., Collier, N.T., Bauer, J., Xia, F., Brettin, T., Stevens, R., Mohd-Yusof, J., et~al.: Candle/supervisor: A workflow framework for machine learning applied to cancer research.
\newblock BMC bioinformatics \textbf{19}(18), 59--69 (2018)

\bibitem{xia2014survey}
Xia, W., Wen, Y., Foh, C.H., Niyato, D., Xie, H.: A survey on software-defined networking.
\newblock IEEE Communications Surveys \& Tutorials \textbf{17}(1), 27--51 (2014)

\bibitem{xin2021production}
Xin, D., Miao, H., Parameswaran, A., Polyzotis, N.: Production machine learning pipelines: Empirical analysis and optimization opportunities.
\newblock In: Proceedings of the 2021 International Conference on Management of Data, pp. 2639--2652 (2021)

\bibitem{xiu2020exploratory}
Xiu, M., Jiang, Z.M.J., Adams, B.: An exploratory study of machine learning model stores.
\newblock IEEE Software \textbf{38}(1), 114--122 (2020)

\bibitem{yang2021mlife}
Yang, C., Wang, W., Zhang, Y., Zhang, Z., Shen, L., Li, Y., See, J.: Mlife: A lite framework for machine learning lifecycle initialization.
\newblock Machine Learning \textbf{110}, 2993--3013 (2021)

\bibitem{Yang2016WhatSQ}
Yang, X., Lo, D., Xia, X., Wan, Z., Sun, J.: What security questions do developers ask? a large-scale study of stack overflow posts.
\newblock Journal of Computer Science and Technology \textbf{31}, 910--924 (2016)

\bibitem{yao2023survey}
Yao, Y., Duan, J., Xu, K., Cai, Y., Sun, E., Zhang, Y.: A survey on large language model (llm) security and privacy: The good, the bad, and the ugly.
\newblock arXiv preprint arXiv:2312.02003  (2023)

\bibitem{zaharia2018accelerating}
Zaharia, M., Chen, A., Davidson, A., Ghodsi, A., Hong, S.A., Konwinski, A., Murching, S., Nykodym, T., Ogilvie, P., Parkhe, M., et~al.: Accelerating the machine learning lifecycle with mlflow.
\newblock IEEE Data Eng. Bull. \textbf{41}(4), 39--45 (2018)

\bibitem{zhang2023instruction}
Zhang, S., Dong, L., Li, X., Zhang, S., Sun, X., Wang, S., Li, J., Hu, R., Zhang, T., Wu, F., et~al.: Instruction tuning for large language models: A survey.
\newblock arXiv preprint arXiv:2308.10792  (2023)

\end{thebibliography}
